\documentstyle[12pt,epsf]{article}
\newcommand{\ba}{\begin{array}}
\newcommand{\ea}{\end{array}}
\newcommand{\bi}{\begin{itemize}}
\newcommand{\ei}{\end{itemize}}
\newcommand{\no}{\nonumber}
\newcommand{\nn}{\nonumber\\}

\newcommand{\Z}{{\bf Z}}

\newcommand{\R}{{\bf R}}
\newcommand{\C}{{\bf C}}

\newcommand{\cD}{{\cal D}}

\newcommand{\cF}{{\cal F}}
\newcommand{\cG}{{\cal G}}
\newcommand{\cH}{{\cal H}}

\newcommand{\cL}{{\cal L}}
\newcommand{\cM}{{\cal M}}
\newcommand{\cN}{{\cal N}}
\newcommand{\cO}{{\cal O}}
\newcommand{\cP}{{\cal P}}

\newcommand{\cR}{{\cal R}}
\newcommand{\cS}{{\cal S}}

\newcommand{\cU}{{\cal U}}

\newcommand{\de}{\delta} 

\newcommand{\na}{\nabla}
\newcommand{\ep}{\epsilon}

\newcommand{\si}{\sigma}

\newcommand{\x}{\xi}
\newcommand{\y}{\eta}
\newcommand{\z}{\zeta}
\newcommand{\om}{\omega}
\newcommand{\Om}{\Omega}

\newcommand{\del}{\partial}
\newcommand{\vvert}{\Big{\vert}}
\newcommand{\bra}{\langle}
\newcommand{\ket}{\rangle}

\newcommand{\Bra}{\Big{\langle}}
\newcommand{\Ket}{\Big{\rangle}}

\newcommand{\diag}{\mbox{diag}\,}

\newcommand{\tr}{{\rm tr}\,}
\newcommand{\Tr}{{\rm Tr}\,}

\newcommand{\Trk}{{\rm Tr}_k}

\newcommand{\Ker}{{\rm Ker}\,}

\newcommand{\all}{{~}^{\forall}}

\newcommand{\da}{\dagger}
\newcommand{\pr}{\prime}

\newcommand{\wha}{\widehat}
\newcommand{\rar}{\rightarrow}
\newcommand{\lar}{\leftarrow}
\newcommand{\lra}{\leftrightarrow}

\newcommand{\dar}{\downarrow}

\newcommand{\Lra}{\Leftrightarrow}

\newcommand{\longr}{\longrightarrow}
\newcommand{\longl}{\longleftarrow}
\newcommand{\longlr}{\longleftrightarrow}

\newcommand{\map}{\mapsto}

\newcommand{\ti}{\times}

\newcommand{\ot}{\otimes}
\newcommand{\we}{\wedge}
\newcommand{\ap}{\approx}
\newcommand{\st}{\stackrel}
\newcommand{\unb}{\underbrace}
\newcommand{\unl}{\underline}
\newcommand{\lab}{\label}
\newcommand{\fr}{\frac}
\newcommand{\half}{\frac{1}{2}}
\newcommand{\qua}{\frac{1}{4}}
\newcommand{\scr}{\scriptsize}

\newcommand{\dis}{\displaystyle}
\newcommand{\mn}{{\mu\nu}}
\newcommand{\eb}{\bar{e}}

\newcommand{\zb}{\bar{z}}
\newcommand{\cDb}{\bar{\cD}}

\newcommand{\ah}{\hat{a}}
\newcommand{\fh}{\hat{f}}

\newcommand{\mh}{{\hat{\mu}}}
\newcommand{\nh}{{\hat{\nu}}}

\newcommand{\uh}{\hat{u}}
\newcommand{\vh}{\hat{v}}
\newcommand{\xh}{\hat{x}}
\newcommand{\zh}{\hat{z}}
\newcommand{\Ah}{\hat{A}}
\newcommand{\Bh}{\hat{B}}
\newcommand{\Dh}{\hat{D}}
\newcommand{\Fh}{\hat{F}}
\newcommand{\Ph}{\hat{P}}

\newcommand{\Th}{\hat{T}}
\newcommand{\Uh}{\hat{U}}
\newcommand{\Vh}{\hat{V}}
\newcommand{\phih}{\hat{\phi}}
\newcommand{\Phih}{\hat{\Phi}}
\newcommand{\zbh}{\hat{\bar{z}}} 
\newcommand{\xih}{\hat{\xi}}
\newcommand{\nah}{\hat{\nabla}}
\newcommand{\delh}{\hat{\partial}}

\newcommand{\whA}{\widehat{A}}

\newcommand{\wht}{\widehat{T}}
\newcommand{\whf}{\widehat{\cF}}

\newcommand{\wtps}{\widetilde{\psi}}
\newcommand{\wtc}{\widetilde{\chi}}

\newcommand{\vs}{\vspace*{0.3cm}\noindent}

\makeatletter
\@addtoreset{equation}{section}

\makeatother

\begin{document}

\begin{titlepage}
\null
\begin{flushright}
UT-03-11\\
hep-th/0303256\\
March, 2003
\end{flushright}

\vskip 1.8cm
\begin{center}
 
  {\LARGE \bf Noncommutative Solitons and D-branes}

\vskip 2.3cm
\normalsize

  {\Large Masashi Hamanaka\footnote{From 16 August, 2005
 to 15 August, 2006, the author visits
 the Mathematical Institute, University of Oxford.
 (E-mail: hamanaka@maths.ox.ac.uk)}}

\vskip 0.5cm

  {\large \it Department of Physics, University of Tokyo,\\
               Tokyo 113-0033, Japan\footnote{The
present affiliation is Graduate School of Mathematics,
 Nagoya University, Nagoya, 464-8602, Japan.
 (E-mail: hamanaka@math.nagoya-u.ac.jp)}}

\vskip 2cm

\begin{center}

A Dissertation in candidacy for \\
the degree of Doctor of Philosophy

\vskip 1.5cm

\begin{figure}[htbn]
\epsfxsize=40mm
\hspace{6cm}
\epsffile{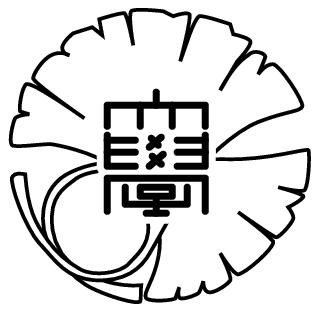}
\end{figure}

\end{center}

\newpage~

\vspace{0.5cm}

{\bf Abstract}

\end{center}

\baselineskip 7mm

D-branes are mysterious solitons in string theories
and play crucial roles in the study 
of the non-perturbative aspects.
Among many ways to analyze the properties of D-branes,
gauge theoretical analysis
often become very strong to study the dynamics
especially at the low-energy scale.
It is very interesting that 
gauge theories live on the D-branes
are useful to study the D-branes themselves
(even on non-perturbative dynamics).

Noncommutative solitons are solitons on noncommutative spaces
and have many interesting aspects.
The distinguished features on noncommutative spaces are
resolutions of singularities,
which leads to the existence of new physical objects,
such as U(1) instantons
and makes it possible to deal with
singular configurations in usual manner.

Noncommutative gauge theories have been studied
intensively for the last several years
in the context of the D-brane effective theories.
This is motivated by the fact that 
they are equivalent to the gauge theories
on D-branes in the presence of background 
NS-NS $B$-fields, or equivalently, magnetic fields.
We can examine various aspects of D-branes
from the analysis of noncommutative gauge theories
which is comparatively easier to treat.
In particular noncommutative solitons are
just the (lower-dimensional) D-branes
and successfully applied to the study of 
non-perturbative dynamics of D-branes.

In this thesis,
we discuss the noncommutative solitons in detail
with applications to D-brane dynamics.
We mainly treat noncommutative instantons
and monopoles by using Atiyah-Drinfeld-Hitchin-Manin (ADHM)
and Nahm constructions
which have the clear D-brane interpretations.
We construct various exact solutions which contain new solitons 
and discuss the corresponding D-brane dynamics.
We find that the ADHM construction potentially
possesses the ``solution generating technique,''
the strong way to confirm the Sen's conjecture related to
decays of unstable D-branes by the tachyon condensations.
We also discuss the corresponding D-brane aspects,
such as T-duality and matrix interpretations,
from gauge theoretical viewpoints.
The results are proved to be all consistent.
Finally we propose noncommutative extension
of soliton theories and integrable systems,
which, we hope, would pioneers a new study area
of integrable systems and (hopefully) string theories.  

\end{titlepage}

\clearpage

\tableofcontents

\baselineskip 6.22mm

\newpage

\section{Introduction}

D-branes are solitons in string theories
and play crucial roles in the study of the non-perturbative aspects.
Since the discovery of them by J.~Polchinski \cite{Polchinski},
there has been remarkable progress in the understanding of
string dualities, the M-theory,
the holographic principle,
microscopic origins
of the blackhole entropy, 
and so on \cite{Polchinski2}.
In the developments, D-branes have occupied central positions.

The properties of D-branes can be investigated in various ways,
for example, supergravities (SUGRA), conformal field theories (CFT),
string field theories (SFT) and so on.
In particular, the effective theories of D-branes 
are very powerful to analyze the low-energy dynamics of it.
The effective theories are described by the Born-Infeld (BI) actions
which are gauge theories on the D-branes coupled to
the bulk supergravity.
In the $\alpha^\prime\rar 0$ limit 
(called the {\it decoupling limit} or {\it zero-slope limit}),
gravities are decoupled to the theory and 
the Born-Infeld action reduces to the Yang-Mills (YM) action
which is very easy to treat.
In this thesis, we will discuss the 
D-brane dynamics from the Yang-Mills theories.

\vs

Non-Commutative (NC) gauge theories 
are gauge theories on noncommutative spaces
and have been studied intensively for the last several years
in the context of the D-brane effective theories.
NC gauge theories on D-branes are shown to be equivalent to
ordinary gauge theories on D-branes in the presence
of background magnetic fields \cite{CDS,DoHu,SeWi},
which triggers the recent explosive 
developments in noncommutative theories,
which is partly because
NC gauge theories are sometimes easier than commutative ones.

In this study, noncommutative solitons are very important 
because they can be identified with the lower-dimensional D-branes.
This makes it possible to reveal some aspects of D-brane dynamics,
such as tachyon condensations \cite{Harvey2},
by constructing exact noncommutative solitons 
and studying their properties.

\vs

\begin{figure}[htbn]
\epsfxsize=90mm
\hspace{3.5cm}
\epsffile{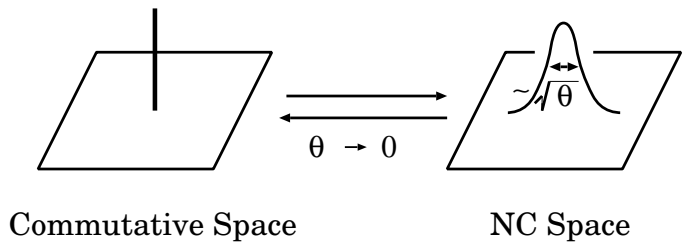}
\caption{Resolution of singularities on noncommutative spaces}
\label{resolution}
\end{figure}

Noncommutative spaces are characterized by the noncommutativity of
the spatial coordinates:
\begin{eqnarray}
\label{nc_coord}
[x^i,x^j]=i\theta^{ij}.
\end{eqnarray}
This relation looks like the canonical commutation 
relation $[q,p]=i\hbar$ in quantum mechanics
and leads to ``space-space uncertainty relation.''
Hence the singularity which exists on commutative spaces 
could resolve on noncommutative spaces
(cf. Fig. \ref{resolution}).
This is one of the distinguished features of noncommutative theories
and gives rise to various new physical objects,
for example, U(1) instantons \cite{NeSc}, 
``visible Dirac-like strings'' \cite{GrNe} and 
the fluxons \cite{Polychronakos, GrNe2}. U(1) instantons 
exist basically due to the resolution of
small instanton singularities
of the complete instanton moduli space \cite{Nakajima}.

The solitons special to noncommutative spaces 
are sometimes so simple
that we can calculate various physical quantities,
such as the energy, the fluctuation around 
the soliton configuration and so on.
This is also due to the properties on noncommutative space
that the singular configuration
becomes smooth and get suitable for the usual calculation.

\vs

In the present thesis, we discuss
noncommutative solitons with applications to the D-brane dynamics.
We mainly treat noncommutative instantons 
and noncommutative monopoles\footnote{In this thesis, 
``monopoles'' basically represents ``BPS monopoles.''} 
from section 3 to section 5.
Instantons and monopoles are stable
(anti-)self-dual configurations
in the Euclidean 4-dimensional Yang-Mills theory 
and the $(3+1)$-dimensional Yang-Mills-Higgs (YMH) theory, respectively
and actually contribute to the non-perturbative effects.
They also have the clear D-brane interpretations
such as D0-D4 brane systems
\cite{Witten3, Witten4, Douglas2}\footnote{In the D-brane picture,
instantons correspond to the static solitons on $(4+1)$-dimensional
space which the D4-branes lie on. In this sense, we consider instantons
as one of solitons in this thesis.}
and D1-D3 brane systems \cite{Diaconescu}
in type II string theories, respectively.

There are known to be strong ways
to generate exact noncommutative instantons and monopoles,
the {\it Atiyah-Drinfeld-Hitchin-Manin} (ADHM)  {\it construction} 
and the {\it Atiyah-Drinfeld-Hitchin-Manin-Nahm} (ADHMN) 
or the {\it Nahm construction}, 
respectively.\footnote{In this thesis, ``ADHM construction''
and ``Nahm construction'' are sometimes written together
as ``ADHM/Nahm construction.''}
ADHM/ Nahm construction is a wonderful application 
of the one-to-one correspondence
between the instanton/monopole moduli 
space and the moduli space of ADHM/Nahm data
and gives rise to arbitrary instantons \cite{ADHM} /
monopoles \cite{Nahm}-\cite{Nahm6}.\footnote{
In this thesis, the slash ``/'' means ``or''
and the repetition of them implies ``respectively.''}

D-branes give intuitive explanations for
various known results of field theories and
explain the reason why the instanton/monopole moduli 
spaces and the moduli space of ADHM/Nahm data correspond one-to-one.
However there still exist unknown parts of the D-brane
descriptions and 
it is expected that further study of the D-brane description of 
ADHM/Nahm construction would reveal new aspects of D-brane dynamics, 
such as Myers effect \cite{Myers}
which in fact corresponds to some boundary conditions in Nahm construction.

In section 3, we discuss the ADHM construction
of instantons focusing on new type of instantons, 
noncommutative U(1) instantons.
In the study of noncommutative U(1) instantons,
the self-duality of the noncommutative parameter is very important
and reflects on the properties of the instantons.
Usually we discuss noncommutative U(1) instantons 
which have the {\it opposite}
self-duality between the gauge field and the noncommutative parameter.
Here, in section 3.2,
we discuss noncommutative U(1) instantons which have the {\it same}
self-duality between them.
As the results, we see 
that ADHM construction of noncommutative instantons
naturally yields the essential part of 
the {\it ``solution generating technique''} (SGT) \cite{Hamanaka7}.

The ``solution generating technique''
is a transformation which leaves the equation of motion
of noncommutative gauge theories as it is
and gives rise to various new solutions from known solutions of it.
The new solutions have a clear interpretation of matrix models 
\cite{BFSS, IKKT, AIIKKT},
which concerns with the important fact 
that a D-brane can be constructed by lower-dimensional D-branes.
The ``solution generating technique'' can be also applied to 
the problem on the non-perturbative dynamics of D-branes.
One remarkable example is an exact confirmation of 
Sen's conjecture within the context of 
the effective theory of SFT that unstable D-branes decays into
the lower-dimensional D-branes by the tachyon condensation.
We discuss this technique and the applications in section 6
with a brief introduction to 
the key objects of the first breakthrough on the problem,
{\it Gopakumar-Minwalla-Strominger} (GMS) {\it solitons}.
The application of the solution generating technique 
to the noncommutative Bogomol'nyi equation
is briefly discussed in section 6.2.
This time we have to modify the technique \cite{HaTe} 
or use some trick \cite{Hashimoto}.

In section 4, we discuss Nahm constructions of monopoles.
After reviewing some typical monopoles,
we construct a special BPS configuration of 
noncommutative Yang-Mills-Higgs theory,
the {\it fluxon} \cite{Polychronakos, GrNe2} 
by Nahm procedure \cite{Hamanaka7}.
The configuration is close to the flux rather than the monopole.
The D-brane interpretation is also presented.

Monopoles can be considered as T-dualized (or Fourier-transformed)
configurations of instantons in some limit as we see in section 5.
The fluxon is also obtained by the Fourier transformation
of the noncommutative periodic instanton (caloron) 
in the zero-period limit.
The periodic solitons and 
the attempts of the Fourier-transformations
are new \cite{Hamanaka7}.
All the results are consistent with
T-duality transformation of the corresponding
D0-D4 brane systems, which is discussed in detail in section 5.

\vs

Furthermore in section 7,
we discuss noncommutative extension of
soliton theories and integrable systems
as a further direction.
We present a powerful method to generate various equations
which possess the Lax representations
on noncommutative $(1+1)$ and $(2+1)$-dimensional spaces.
The generated equations contain noncommutative integrable
equations obtained by using the bicomplex method
and by reductions of the noncommutative
(anti-)self-dual Yang-Mills equation.
This suggests that
the noncommutative Lax equations would be integrable
and be derived from reductions of the noncommutative
(anti-)self-dual Yang-Mills equations,
which implies noncommutative version of Richard Ward conjecture.

\vs

This thesis is designed for a comprehensive review
of those studies including my works
and organized as follows:
In section 2, we introduce foundation of noncommutative gauge theories and 
the commutative description briefly.
In section 3, 4 and 5, we discuss ADHM/Nahm construction of 
instantons and monopoles 
on both commutative spaces and noncommutative spaces.
In section 6, we extend the discussion to non-BPS solitons
and give a confirmation of Sen's conjecture on tachyon condensations.
In section 7, 
we discuss the noncommutative extension of soliton equations or
integrable equations as further directions.
Finally we conclude in section 8.
Appendix is devoted to an introduction to 
ADHM/Nahm construction on commutative spaces. 

\vs

The main papers contributed to the present thesis are the following:
\begin{itemize}
\item M.~Hamanaka, ``Atiyah-Drinfeld-Hitchin-Manin and Nahm constructions
of localized solitons in noncommutative gauge theories,''
Physical Review D {\bf 65} (2002) 085022 
{\sf [hep-th/0109070]} \cite{Hamanaka7} 
(Section 3.2, 3.3, 4.4, 5.2, 5.3),
\item M.~Hamanaka and K.~Toda, 
``Towards noncommutative integrable systems,''
 Physics Letters A {\bf 316} (2003) 77-83
{\sf [hep-th/0211148]} \cite{HaTo} (Section 7),
\end{itemize}
where the corresponding parts in this thesis are shown in the parenthesis.

There is another paper which is a part of this thesis:
\begin{itemize}
\item M.~Hamanaka and S.~Terashima, 
``On exact noncommutative BPS solitons,'' 
Journal of High Energy Physics {\bf 0103} (2001) 034
{\sf [hep-th/0010221]} \cite{HaTe} (The latter half of section 6.2),
\end{itemize}
though I do not consider it as a main paper for this thesis.

\newpage

\section{Non-Commutative (NC) Gauge Theories}

In this section, we introduce foundation
of noncommutative gauge theories.
Noncommutative gauge theories are equivalent to 
ordinary commutative 
gauge theories in the presence of the background magnetic fields.
This equivalence between noncommutative gauge theories and
gauge theories in magnetic fields is famous
in the area of quantum Hall effects
and recently it has been shown that 
it is also true of string theories \cite{CDS, DoHu, SeWi}.
We finally comment on the results of the equivalence 
in string theories.

\subsection{Foundation of NC Gauge Theories}

Noncommutative gauge theories 
have the following three equivalent descriptions
and are connected one-to-one by the Weyl transformation and
the Seiberg-Witten (SW) map\footnote{In this thesis,
we treat ``noncommutative Euclidean spaces'' only.
On noncommutative ``curved spaces, '' there are not in general one-to-one
correspondences between (i) and (ii).}:
\vspace{3mm}
\begin{eqnarray*}
\label{map}
\begin{array}{l}
~~~~~~~~~~~~~~~~~~~~~~~~~~~~~~\fbox{(i) 
NC Gauge theory in the star-product formalism}
\\
~~~~~~~~~~~~~~~~~~~~~~~~~~~~~~~~~~~~~~~~~~~~\uparrow\\
\langle \mbox{NC side} \rangle~~~~~~~~~~~~~~~~~~~~
\mbox{Weyl transformation}\\
~~~~~~~~~~~~~~~~~~~~~~~~~~~~~~~~~~~~~~~~~~~~\downarrow\\
~~~~~~~~~~~~~~~~~~~~~~~~~~~~~~\fbox{(ii) 
NC Gauge theory in the operator formalism}\\
~~~~~\uparrow\\
~\mbox{SW map
}\\
~~~~~\downarrow\\
\langle \mbox{Commutative side} \rangle~~~~
\fbox{(iii) Gauge theory on D-branes with magnetic fields}
\end{array}
\end{eqnarray*}
\vspace{3mm}
\noindent

In the star-product formalism (i), 
we realize the noncommutativity of 
the coordinates (\ref{nc_coord}) 
by replacing the products of the fields with the star-products.
The fields are ordinary functions.
In the commutative limit $\theta^{ij}\rightarrow 0$,
this noncommutative theories reduce to 
the ordinary commutative ones.
In the operator formalism (ii), 
we start with the noncommutativity of the coordinates (\ref{nc_coord})
and treat the coordinates and fields as operators (infinite-size
matrices).
This formalism is the most suitable to be called ``noncommutative theories,''
and has a good fit for matrix theories.
The formalism (iii) is a commutative description and 
represented as an effective theory of D-branes
in the background of $B$-fields.
The equivalence between (ii) and (iii) is clearly shown in \cite{SeWi}.

In this section, we define noncommutative gauge theories in the star-product
formalism (i) and then move to the operator formalism (ii)
by the Weyl transformation.

\noindent
\fbox{(i) The star-product formalism}
\vspace{2mm}

The star-product is defined for ordinary fields on commutative spaces 
and for Euclidean spaces, explicitly given by 
\begin{eqnarray}
f\star g(x)&:=&
\exp{\left(\frac{i}{2}\theta^{ij}\partial^{(x^\prime)}_i
\partial^{(x^{\prime\prime})}_j\right)}
f(x^\prime)g(x^{\prime\prime})\Big{\vert}_{x^{\prime}
=x^{\prime\prime}=x}\nonumber\\
&=&f(x)g(x)+\frac{i}{2}\theta^{ij}\partial_if(x)\partial_jg(x)
+{\cal O}(\theta^2).
\end{eqnarray}
This explicit representation is known 
as the {\it Groenewold-Moyal product} \cite{Groenewold,Moyal}.

The star-product has associativity: $f\star(g\star h)=(f\star g)\star h$,
and returns back to the ordinary product with $\theta^{ij}\rar 0$.
The modification of the product  makes the ordinary 
spatial coordinate ``noncommutative,'' which means :
$[x^i,x^j]_\star:=x^i\star x^j-x^j\star x^i=i\theta^{ij}$.

Noncommutative gauge theories are given by the exchange of 
ordinary products in the commutative gauge theories for the star-products
and realized as deformed theories from commutative ones.
In this context, we often call them 
the {\it NC-deformed theories}.
The equation of motion and BPS equation are also given by the same
procedure because the fields are ordinary functions 
and we can take the same steps as commutative case.

We show some examples where
all the products of the fields are the star products.

\vspace{2mm}
\noindent
\unl{4-dimensional NC-deformed Yang-Mills theory}
\vspace{2mm}

Let us consider the 4-dimensional noncommutative space
with the coordinates $x^\mu,~\mu=1,2,3,4$ where
the noncommutativity is introduced as the canonical form:
\begin{eqnarray}
\label{can_nc_coord}
\theta^{\mn}=\left(
\ba{cc|cc}
0&\theta_1&0&0\\
-\theta_1&0&0&0\\\hline
0&0&0&\theta_2\\
0&0&-\theta_2&0\\
\ea
\right).
\end{eqnarray}

The action of 4-dimensional gauge theory is given by
\begin{eqnarray}
I_{\scr\mbox{YM}}
=-\fr{1}{2g^2_{\scr\mbox{YM}}}
\int d^4x~\Tr F_\mn F^\mn.
\end{eqnarray}
The BPS equations are the ASD equations:\footnote{
When we make the distinct between ``self-dual'' or
``anti-self-dual,'' then we write ``SD'' or ``ASD'' explicitly.
For example, while ``instantons'' or ``(A)SD equations''
shows no distinction,
``ASD instantons'' or ``ASD equations'' specifies
the ASD one.}
\begin{eqnarray}
F_\mn+*F_\mn=0,
\end{eqnarray}
or equivalently,
\begin{eqnarray}
\label{star_instanton}
F_{z_1\zb_1}+F_{z_2\zb_2}=0,~~~
F_{z_1z_2}=0,
\end{eqnarray}
which are derived from the condition that the action density 
should take the minimum:
\begin{eqnarray}
I_{\scr\mbox{YM}}
&=&-\fr{1}{4g^2_{\scr\mbox{YM}}}
\int d^4x~\Tr\left(F_\mn F^\mn+*F_\mn *F^\mn\right)\no\\
&=&-\fr{1}{4g^2_{\scr\mbox{YM}}}
\int d^4x~\Tr\left((F_\mn\mp*F_{\mn})^2
\pm 2 F_\mn *F^\mn\right),
\end{eqnarray}
where the symbol $*$ is the Hodge operator defined 
by $*F_{\mn}:=(1/2) \epsilon_{\mn\rho\sigma}F^{\rho\sigma}$.

\vs
\noindent
\unl{$(3+1)$-dimensional NC-deformed Yang-Mills-Higgs theory}
\vs

Next let us consider the $(3+1)$-dimensional noncommutative space
with the coordinates $x^0,x^i,~i=1,2,3$ 
where the noncommutativity is introduced as $\theta^{12}=\theta>0$.

The action of $(3+1)$-dimensional gauge theory is given by
\begin{eqnarray}
\label{action_ymh}
I_{\scr\mbox{YMH}}
=-\fr{1}{4g^2_{\scr\mbox{YM}}}
\int d^4x~\Tr\left(F_{\mn}F^{\mn}
+2D_\mu\Phi D_\mu\Phi\right),
\end{eqnarray}
where $\Phi$ is an adjoint Higgs field.
The anti-self-dual BPS equations are 
\begin{eqnarray}
\label{star_monopole}
B_3=-D_3\Phi,~~~B_z=-D_z\Phi,
\end{eqnarray}
where $B_i$ is magnetic field 
and $B_i:=-(i/2)\epsilon_{ijk}F^{jk},~B_z:=B_1-iB_2,~D_z:=D_1-iD_2$.
These equations are usually called {\it Bogomol'nyi equations}
\cite{Bogomolnyi}
and derived from the conditions that the energy density $E$ 
should take the minimum:
\begin{eqnarray}
E&=&\frac{1}{2g^2_{\scriptsize\mbox{YM}}}
\int d^3x~{\mbox{Tr}}\left[\frac{1}{2} 
F_{ij}F^{ij}+D_i\Phi D^i\Phi\right]\no\\
&=&\frac{1}{2g^2_{\scriptsize\mbox{YM}}}
\int d^3x~{\mbox{Tr}}[(B_i\mp D_i\Phi)^2
\pm\partial_i(\epsilon_{ijk}F^{jk}\Phi)].
\end{eqnarray}

\vspace{2mm}
\noindent
\fbox{(ii) The operator formalism}
\vspace{2mm}

This time, we start with the noncommutativity of the spatial coordinates
(\ref{nc_coord}) and define noncommutative gauge theories considering 
the coordinates as operators. 
{}From now on, we write the hats on the fields 
in order to emphasize that they are operators.
For simplicity, we deal with a noncommutative plane
with the coordinates $\xh^1,\xh^2$ which satisfy $[\xh^1,\xh^2]
=i\theta,~\theta>0$.

Defining new variables $\ah,\ah^\dagger$ as
\begin{eqnarray}
\ah:=\fr{1}{\sqrt{2\theta}}\zh,~\ah^\dagger:=\fr{1}{\sqrt{2\theta}}\zbh,
\end{eqnarray}
where $\zh=\xh^1+i\xh^2,~\zbh=\xh^1-i\xh^2$,
we get the Heisenberg's commutation relation:
\begin{eqnarray}
\label{heisenberg}
{[\ah,\ah^\dagger]}=1.
\end{eqnarray}
Hence the spatial coordinates can be considered 
as the operators acting on
a Fock space $\cH$ which is spanned 
by the occupation number basis $\vert
n\ket:=\left\{(\ah^\dagger)^n/\sqrt{n!}\right\}\vert 0\ket,
~\ah\vert 0\ket=0$:
\begin{eqnarray}
\label{fock}
\cH=\oplus_{n=0}^{\infty}\C\vert n\ket.
\end{eqnarray}
Fields on the space depend on the spatial coordinates 
and are also the operators
acting on the Fock space $\cH$. 
They are represented by the occupation number basis as
\begin{eqnarray}
\fh=\sum_{m,n=0}^{\infty}f_{mn}\vert m\ket\bra n\vert.
\end{eqnarray}
If the fields have rotational symmetry on the plane,
namely, commute with the number operator $\nh:=\ah^\dagger 
\ah\sim (\xh^1)^2+(\xh^2)^2$,
they become diagonal:
\begin{eqnarray}
\fh=\sum_{n=0}^{\infty}f_{n}\vert n\ket\bra n\vert.
\end{eqnarray} 

The derivation is defined as follows:
\begin{eqnarray}
\del_i\fh :=[\delh_i,\fh] := [-i(\theta^{-1})_{ij}\xh^j, \fh],
\end{eqnarray} 
which satisfies the Leibniz rule and the desired relation:
\begin{eqnarray}
\del_i\xh^j=[-i(\theta^{-1})_{ik}\xh^k, \xh^j]=\delta_i^{~j}.
\end{eqnarray}
The operator $\delh_i$ is called the {\it derivative operator}.
The integration can also be defined as the trace of the Fock space $\cH$:
\begin{eqnarray}
\int dx^1dx^2~ \fh(\xh^1,\xh^2)&:=& 2\pi \theta\Tr_\cH \fh,
\end{eqnarray}

The covariant derivatives act on the fields which belong to
the adjoint and the fundamental representations of the gauge group as
\begin{eqnarray}
D_i\Phih^{\scr\mbox{adj.}} &:=&[\Dh_i,\Phih]:=[\delh_i+\Ah_i,\Phih],\nn
D_i\phih^{\scr\mbox{fund.}} &:=&[\delh_i,\phih]+\Ah_i \phih,
\end{eqnarray} 
respectively.
The operator $\Dh_i$ is called the {\it covariant derivative operator}.

In noncommutative gauge theories,
there are almost unitary operators $\Uh_k$ which satisfy
\begin{eqnarray}
\label{partial_iso}
\Uh_k\Uh_k^\dagger=1,~~~\Uh_k^\dagger \Uh_k=1-\Ph_k,
\end{eqnarray}
where the operator $\Ph_k$ is a projection operator whose rank is $k$.
The operator $\Uh_k$ is called the {\it partial isometry}
and plays important roles in noncommutative gauge theories
concerning the soliton charges.

The typical examples of them are
\begin{eqnarray}
\Ph_k&=&\sum_{p=0}^{k-1}\vert p \ket\bra p\vert,\\
\Uh_k&=&\sum_{n=0}^{\infty}\vert n\ket\bra n+k\vert
=\sum_{n=0}^{\infty}\vert n\ket\bra n\vert \ah^k 
\fr{1}{\sqrt{(n+k)\cdots(n+1)}},\\
\Uh_k^\dagger&=&\sum_{n=0}^{\infty}\vert n+k\ket\bra n\vert
=\sum_{n=0}^{\infty}\fr{1}{\sqrt{(n+k)\cdots(n+1)}}
(\ah^\dagger)^k\vert n\ket\bra n\vert.
\end{eqnarray}
This $\Uh_k$ is sometimes called 
the {\it shift operator}.\footnote{The shift operators can be
constructed concretely 
by applying Atiyah-Bott-Shapiro (ABS) construction \cite{ABS}
to noncommutative cases \cite{HaMo}.}

\vspace{2mm}
\noindent
{\bf [Equivalence between (i) star-product formalism and 
(ii) operator formalism]}
\vspace{2mm}

The descriptions (i) and (ii) are equivalent and connected by the Weyl
transformation. The Weyl transformation transforms the field $f(x^1,x^2)$ 
in (i) into the infinite-size matrix $\fh(\xh^1,\xh^2)$  in (ii) as
\begin{eqnarray}
\label{weyl1}
\fh(\xh^1,\xh^2)&:=&\fr{1}{(2\pi)^2}\int dk_1 dk_2~\widetilde{f}(k_1,k_2)
e^{-i(k_1\xh^1+k_2\xh^2)},
\end{eqnarray}
where
\begin{eqnarray}
\label{weyl2}
\widetilde{f}(k_1,k_2)&:=&\int dx^1dx^2~f(x^1,x^2) e^{i(k_1x^1+k_2x^2)}.
\end{eqnarray}
This map is the composite of twice Fourier transformations 
replacing the commutative coordinates $x^1,x^2$ in the exponential
with the noncommutative coordinates $\xh^1,\xh^2$ 
in the inverse transformation:
\begin{eqnarray*}
\begin{array}{ccc}
~&~&f(x^1,x^2)\\
~&~\swarrow&|\\
\widetilde{f}(k_1,k_2)&~&\mbox{Weyl transformation}\\
~&~\searrow&\dar\\
~&~&\fh(\xh^1,\xh^2).
\end{array}
\end{eqnarray*}
The Weyl transformation
preserves the product:
\begin{eqnarray}
\widehat{f\star g}=\fh\cdot\hat{g}.
\end{eqnarray}
The inverse transformation of the Weyl transformation
is given directly by
\begin{eqnarray}
f(x^1,x^2)=\int dk_2~e^{-ik_2 x^2}\Bra x^1+\fr{k_2}{2}\vvert \fh(\xh^1,\xh^2)
\vvert x^1-\fr{k_2}{2}\Ket.
\end{eqnarray}
The transformation also maps the derivation and 
the integration one-to-one. 
Hence the BPS equation and the solution are also transformed
one-to-one. The correspondences are the following:
\vspace{3mm}
\begin{eqnarray*}
\ba{ccc}
\mbox{\fbox{(i) the star-product formalism}}&\lar 
\mbox{Weyl transformation}\rar&
\mbox{\fbox{(ii) the operator formalism}}\\\\
\mbox{ordinary functions}&\mbox{[field]}&\mbox{infinite-size matrices}\\
f(x^1,x^2)&~&\dis\fh(\xh^1,\xh^2)
=\sum_{m,n=0}^{\infty}f_{mn}\vert m\ket\bra n\vert\\\\
\mbox{star-products}&
\mbox{[product]}&
\mbox{multiplications of matrices}\\
\left(f\star(g\star h)=(f\star g)\star h\right)
&\mbox{(associativity)}&\left(\hat{f}(\hat{g}\hat{h})
=(\hat{f}\hat{g})\hat{h}~\mbox{(trivial)}\right)\\\\
{[x^i,x^j]}_\star=i\theta^{ij}&\mbox{[noncommutativity]}
&[\hat{x}^i,\hat{x}^j]=i\theta^{ij}\\\\
\del_if&\mbox{[derivation]}&\del_i\fh:=
[\unb{-i(\theta^{-1})_{ij}\xh^j}_{=:~\delh_i},\fh]\\
\left(\mbox{especially,}~\del_i x^j=\delta_i^{~j}\right)&~&
\left(\mbox{especially,}~
\del_i\xh^j=\delta_i^{~j}\right)\\\\
\dis\int dx^1dx^2~f(x^1,x^2)&\mbox{[integration]}&2\pi\theta
\Tr_{\cH}\fh(\xh^1,\xh^2)\\\\
F_{ij}=\del_iA_j-\del_jA_i+[A_i,A_j]_\star&\mbox{[curvature]}
&\Fh_{ij}=\del_i\hat{A}_j-\del_j\hat{A}_i+[\hat{A}_i,\hat{A}_j]
\\
~&~&=[\hat{D}_i,\hat{D}_j]
-i(\theta^{-1})_{ij}\,\\\\
\ba{c}
\dis\sqrt{\fr{n!}{m!}}
\left(2r^2/\theta\right)^{\fr{m-n}{2}}e^{i(m-n)\varphi}\times\\
2(-1)^nL_n^{m-n}(2r^2/\theta)e^{-\fr{r^2}{\theta}}
\ea
&\mbox{[matrix element]}
&\vert n\ket\bra m\vert\\\\
\vert&\vert&\vert\\
\left(
\ba{c}
\mbox{Independent of $\varphi$}\\
\Lra m=n
\ea
\right)
&\left(
\ba{c}
\mbox{Rotational symmetry}\\
\mbox{on $x^1$-$x^2$ plane}
\ea
\right)
&
\left(
\ba{c}
\mbox{Commutes with}\\
(\xh^1)^2+(\xh^2)^2~\Lra ~m=n
\ea
\right)\\
\dar&\dar&\dar\\\\
\dis 2(-1)^nL_n(2r^2/\theta)e^{-\fr{r^2}{\theta}}
&\mbox{[some projection]}&\vert n\ket\bra n\vert
\ea
\end{eqnarray*}
\vspace{3mm}

\noindent
where $(r,\varphi)$ is the usual polar coordinate
($r=\left\{(x^1)^2+(x^2)^2\right\}^{\half}$)
and $L^{\alpha}_n(x)$ is the Laguerre polynomial:
\begin{eqnarray}
L_n^\alpha(x):=\fr{x^{-\alpha}e^{x}}{n!}\left(\fr{d}{dx}\right)^n
(e^{-x}x^{n+\alpha}).
\end{eqnarray}
(Especially $L_n(x):=L_n^0(x)$.)
 
We note that in the curvature in operator
formalism, a constant term $-i(\theta^{-1})_{ij}$ appears so that
it should cancel out the term $[\hat{\del}_i,\hat{\del_j}]
(=i(\theta^{-1})_{ij})$ 
in $[\hat{D}_i,\hat{D}_j]$.
For a review of the correspondence, see \cite{Harvey}.

We show some examples of BPS equations in operator formalism
which are simply mapped by the Weyl transformation from the BPS equations
(\ref{star_instanton}) and (\ref{star_monopole}).

\vs
\noindent
\unl{4-dimensional noncommutative Yang-Mills theory}
\vs

First we show the operator formalism on noncommutative 4-dimensional space
setting the noncommutative parameter $\theta^\mn$ anti-self-dual.
The fields on the 4-dimensional noncommutative space whose
noncommutativity is (\ref{can_nc_coord})
are operators acting on Fock space $\cH=\cH_1\ot\cH_2$ 
where $\cH_1$ and $\cH_2$ are defined by the
same steps as the previous paragraph 
on noncommutative $x^1$-$x^2$ plane
and 
on noncommutative $x^3$-$x^4$ plane
respectively.
The element in the Fock space $\cH=\cH_1\ot\cH_2$ is denoted 
by $\vert n_1\ket\ot\vert n_2\ket$ or $\vert n_1,n_2\ket$.

In order to make the noncommutative parameter anti-self-dual,
we put $\theta_1=-\theta_2=\theta>0$.
In this case, $\zh_1$ and $\zbh_2$ correspond to
annihilation operators and $\zbh_1$ and $\zh_2$ creation operators:
\begin{eqnarray}
[\zh_1,\zbh_1]=2\theta_1=2\theta,
~[\zbh_2,\zh_2]=-2\theta_2=2\theta,~\mbox{otherwise}=0.
\end{eqnarray}
We can define annihilation operators 
as $\ah_1:=(1/\sqrt{2\theta})\zh_1,~\ah_2:=(1/\sqrt{2\theta})\zbh_2$
and creation operator $\ah_1^\dagger:=(1/\sqrt{2\theta})\zbh_1,~\ah_2^\dagger
:=(1/\sqrt{2\theta})\zh_2$ 
in Fock space $\cH=\oplus_{n_1,n_2=0}^{\infty}
\C\vert n_1\ket\otimes \vert n_2\ket$ such as 
\begin{eqnarray}
[\ah_1,\ah_1^\dagger]=1,~[\ah_2,\ah_2^\dagger]=1,~\mbox{otherwise}=0,
\end{eqnarray}
where $\vert n_1\ket$ and $\vert n_2\ket$ are the occupation number basis
generated from the vacuum $\vert 0_1\ket$ and $\vert 0_2\ket$
by the action of $\ah_1^\dagger$ and $\ah_2^\dagger$,
respectively.

The anti-self-dual BPS equations in operator formalism
are transformed by Weyl transformation 
from equation (\ref{star_instanton}):
\begin{eqnarray}
\label{bps_instanton}
(\Fh_{z_1\zb_1}+\Fh_{z_2\zb_2}=)&&
-[\Dh_{z_1},\Dh_{z_1}^\dagger]-[\Dh_{z_2},\Dh_{z_2}^\dagger]
-\half\left(\fr{1}{\theta_1}+
\fr{1}{\theta_2}\right)=0,\no\\
(\Fh_{z_1z_2}=)&&[\Dh_{z_1},\Dh_{z_2}]=0,
\end{eqnarray}
The fields are represented by using the occupation number basis as
\begin{eqnarray}
\fh(\xh^\mu)&=&\sum_{m_1,m_2,n_1,n_2=0}^{\infty}f_{m_1,m_2,n_1,n_2}
\vert m_1\ket\bra n_1\vert 
\ot \vert m_2\ket\bra n_2\vert\nn
&=:&\sum_{m_1,m_2,n_1,n_2=0}^{\infty}f_{m_1,m_2,n_1,n_2}
\vert m_1,m_2\ket\bra n_1,n_2\vert.
\end{eqnarray}
We note that in the case that noncommutative parameter $\theta^{ij}$
is also anti-self-dual, the constant term $\left(1/\theta_1+
1/\theta_2\right)$ disappears.

\vs
\noindent
\unl{$(3+1)$-dimensional noncommutative Yang-Mills-Higgs theories}
\vs

The anti-self-dual BPS equations in the operator formalism 
are transformed by Weyl transformation
of equations (\ref{star_monopole}):
\begin{eqnarray}
\label{bps_monopole}
(\Bh_3=)&&[\Dh_z,\Dh_{z}^\dagger]+\fr{1}{\theta}=-[\Dh_3,\Phih],\no\\
(\Bh_z=)&&[\Dh_3,\Dh_{z}]=-[\Dh_z, \Phih].
\end{eqnarray}

The fields
are represented by using the occupation number basis as
\begin{eqnarray}
\fh(\xh^1,\xh^2, x^3)=\sum_{n=0}^{\infty}f_{mn}(x^3)\vert m\ket\bra n \vert.
\end{eqnarray}

\subsection{Seiberg-Witten Map}

Here we present the results discussed by Seiberg and Witten,
which motivates the recent explosive developments 
in noncommutative gauge theories and string theories.

Let us consider the low-energy effective theory
of open strings in the presence of
background of constant NS-NS $B$-fields.
In order to do this,
there are two ways to regularize the open-string world-sheet action
corresponding to the situation with D$p$-branes.
If we take Pauli-Villars (PV) regularization neglecting 
the derivative corrections of the field strength,
we get the ordinary (commutative) Born-Infeld action \cite{BoIn}
with $B$-field for $G=U(1)$:
\begin{eqnarray}
I_{\scr\mbox{BI}}
=\fr{1}{g_{\scr\mbox{s}} (2\pi)^p(\alpha^\prime)^{\fr{p+1}{2}}}
\int d^{p+1}x~
\sqrt{\det(g_\mn+2\pi\alpha^\prime (F_\mn+B_\mn))}
\end{eqnarray}
where $g_{\scr\mbox{s}} $ and $g_\mn$ are 
the string coupling and the closed string metric, respectively.
On the other hand, if we take the Point-Splitting (PS) regularization
neglecting the derivative corrections of the field strength,
we get the noncommutative Born-Infeld action without $B$-field
(in the star-product formalism):
\begin{eqnarray}
I_{\scr\mbox{NC BI}}
=\fr{1}{G_{\scr\mbox{s}} (2\pi)^p(\alpha^\prime)^{\fr{p+1}{2}}}
\int d^{p+1}x~
\sqrt{\det(G_\mn+2\pi\alpha^\prime F_\mn)}_\star
\end{eqnarray}
where $G_{\scr\mbox{s}}$ and $G_\mn$ are the open string coupling 
and the open string metric, respectively.

The effective theories should be independent of the ways to regularize it
and hence be equivalent to each other 
and connected by field redefinitions.
The equivalent relation between
the commutative fields $A_\mu(x), F_\mn(x)$ and
the noncommutative fields $\Ah_\mu(\xh), \Fh_\mn(\xh)$
was found by Seiberg and Witten 
as an differential equation.\footnote{
This equation is in fact not completely integrable and
has some ambiguities \cite{AsKi}.}

\begin{figure}[htbn]
\epsfxsize=130mm
\hspace{2cm}
\epsffile{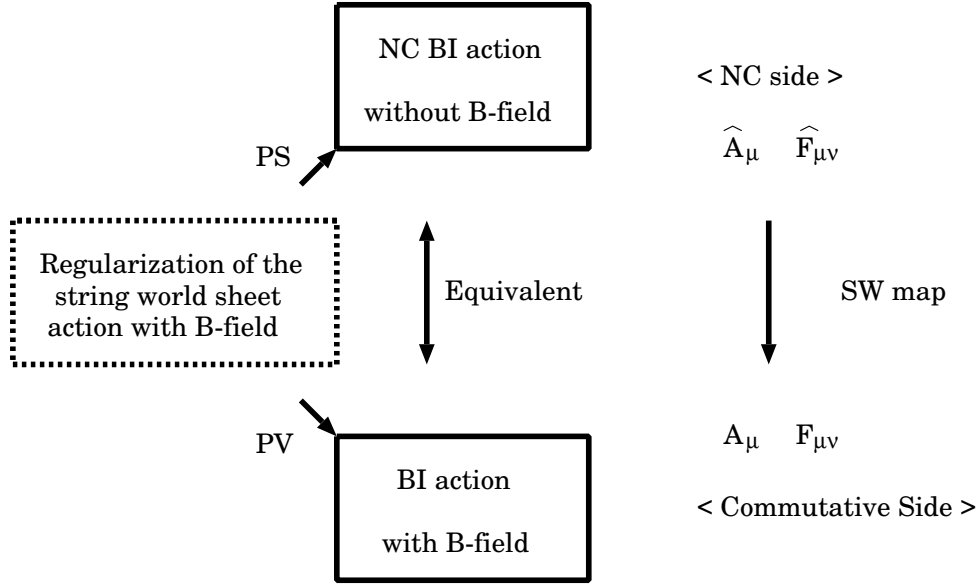}
\caption{The equivalence between NC BI action without $B$-field and 
BI action with $B$-field, and the Seiberg-Witten map}
\label{sw}
\end{figure}

A solution of it for $G=U(1)$
is obtained by \cite{OkOo, MuSu, LiMi}
and the Fourier component of the field strength of the mapped
gauge fields on commutative side is given 
in terms of the noncommutative gauge fields by
\begin{eqnarray}
&&F_{ij}(k)+(\theta^{-1})_{ij}\delta(k)\nn
&&=
\fr{1}{\mbox{Pf}(\theta)}\int dx \left[
e^{ikx}\left(\theta-\theta \fh \theta \right)_{ij}^{n-1}P 
\exp\left(i\int_0^1 \Ah(x+l\tau)l^i d\tau\right)\right],
\label{exact_sw}
\end{eqnarray}
where
\begin{eqnarray}
l^i&:=&k_j \theta^{ji},\nn
\fh_{ij}&:=&\int_0^1 \Fh_{ij}(x+l\tau)d\tau,\nn
\mbox{Pf}(\theta)&:=&\fr{1}{2^n n!}\epsilon_{i_1\ldots i_{2n}}
\theta_{i_1i_2}\cdots\theta_{i_{2n-1}i_{2n}},
\end{eqnarray}
and
\begin{eqnarray}
&&(\theta-\theta\fh\theta)_{ij}^{n-1}
=-\fr{1}{2^{n-1}(n-1)!}\epsilon_{iji_1i_2\ldots i_{2n-2}}\nn
&&\ti \int_0^1 d\tau_1 \left(\theta-\theta\Fh (x+l\tau_1)
\theta\right)^{i_1i_2}\cdots 
\int_0^1d\tau_{n-1}\left(\theta-\theta\Fh 
(x+l\tau_{n-1})\right)_.^{i_{2n-3}i_{2n-2}}
\end{eqnarray}
The exact transformation (\ref{exact_sw})
contains the {open Wilson line} \cite{IIKK}
which is gauge invariant 
in noncommutative gauge theories.
The more explicit examples of the SW map will be presented later.

{}From section 3 to section 6 except for section 6.2,
we discuss the exact solution of Yang-Mills theories as D-brane
effective theories in the zero-slope limit: $\alpha^\prime\rar 0$.
In this limit, the (NC) Born-Infeld action is reduced to the (NC) 
Super-Yang-Mills action and yields soliton solutions which are just the
(lower-dimensional) D-branes.
For example, the effective theory of $N$ D3-branes
coincides with the $G=U(N)$
Yang-Mills-Higgs action (\ref{action_ymh})
by setting the transverse Higgs fields $\Phi^4\equiv \Phi$ and 
$\Phi^\mh=0,~(\mh=5,\ldots,9)$.
We construct explicit noncommutative soliton 
solutions via ADHM/Nahm construction
and discuss the corresponding D-brane dynamics.

\newpage

\section{Instantons and D-branes}

In this section, we study noncommutative instantons in detail by
using ADHM construction.
ADHM construction is a strong method to generate all
instantons and based on a duality, that is, 
one-to-one correspondence
between the instanton moduli space and the moduli space of 
ADHM-data, which are specified by the ASD equation and ADHM equation,
respectively.
In the context of string theories,
instantons are realized as the D0-D4 brane systems
in type IIA string theory.
The numbers of D0-branes and D4-branes 
correspond to the instanton number and
the rank of the gauge group and are denoted by $k$ and $N$
in this thesis, respectively.
We will see how well ADHM construction extracts the essence
of instantons and how much it fits to the D-brane systems
in the construction of exact instanton solutions
on both commutative and noncommutative $\R^4$.

\subsection{ADHM Construction of Instantons}

In this subsection, we construct exact instanton solutions
on commutative $\R^4$.
By using ADHM procedure, we can easily construct
Belavin-Polyakov-Schwartz-Tyupkin (BPST) instanton 
solution \cite{BPST} ($G=SU(2)$ 1-instanton solution), 
't Hooft instanton solution and 
Jackiw-Nohl-Rebbi solution \cite{JNR} 
($G=SU(2)$ $k$-instanton solution). 
The concrete steps are as follows:

\begin{itemize}

\item Step (i):  Solving ADHM equation:
\begin{eqnarray}
\label{adhm_now}
&&{[B_1,B_1^\dagger]}+[B_2,B_2^\dagger]+II^\dagger-J^\dagger J
=-[z_1,\zb_1]-[z_2,\zb_2]=0,\nonumber\\
&&{[B_1,B_2]}+IJ=-[z_1,z_2]=0.
\end{eqnarray}
We note that the coordinates $z_{1,2}$ always appear in pair with
the matrices $B_{1,2}$ and that is why we see the commutator of
the coordinates in the RHS.
These terms, of course, vanish on commutative spaces, however,
they cause nontrivial contributions on noncommutative spaces,
which is seen later soon.

\item Step (ii):  Solving ``0-dimensional Dirac equation'' in the
      background of the ADHM date which satisfies ADHM eq. (\ref{adhm_now}):
\begin{eqnarray}
\label{0dirac_now}
\na^\dagger V=0,
\end{eqnarray}
with the normalization condition:
\begin{eqnarray}
\label{0norm_now}
V^\dagger V =1,
\end{eqnarray}
where the ``0-dimensional Dirac operator'' $\na$
      is defined as in Eq. (\ref{Dir_op}).

\item Step (iii): Using the solution $V$,
we can construct the corresponding instanton solution as
\begin{eqnarray}
\label{4inst_now}
A_\mu=V^\dagger \del_\mu V,
\end{eqnarray}
which actually satisfies the ASD equation:
\begin{eqnarray}
\label{asd_now}
&&F_{z_1\zb_1}+F_{z_2\zb_2}=
[D_{z_1},D_{\zb_1}]+[D_{z_2},D_{\zb_2}]=0,\nn
&&F_{z_1z_2}={[D_{z_1},D_{z_2}]}=0.
\end{eqnarray}

\end{itemize}

The detailed aspects are discussed in Appendix A.
In this subsection, we give some examples of 
the explicit instanton solutions
focusing on BPST instanton solution.

\vs
\noindent
\unl{{\it BPST instanton solution}
(1-instanton, $\dim\cM^{\scr\mbox{BPST}}_{2,1}=5$)}
\vs

This solution is the most basic and important 
and is constructed almost trivially
by ADHM procedure.

\begin{itemize}
\item Step (i): ADHM equation is a $k\ti k$ matrix-equation
and in the present $k=1$ case, is trivially solved.
The commutator part of $B_{1,2}$ is automatically dropped out
and the matrices $B_{1,2}$ can be taken as arbitrary complex
numbers.
The remaining parts $I, J$ are also easily solved:
\begin{eqnarray}
B_1=\alpha_1,~~~B_2=\alpha_2,~~~I=(\rho,0),~~~
J=\left(
\ba{c}
0\\
\rho
\ea
\right),~~~~~~~\alpha_{1, 2}\in \C,~~~\rho\in \R.
\end{eqnarray}
Here the real and imaginary parts of
$\alpha$ are denoted as $\alpha_1=b_2+ib_1,~\alpha_2=b_4+ib_3$,
respectively.

\item Step (ii): The ``0-dimensional Dirac operator'' becomes
\begin{eqnarray}
\nabla=\left(
\ba{c}
\ba{cc}
\rho&0\\
0&\rho\\
\ea\\
\ba{c}
~\\
e_\mu(x_\mu-b_\mu)\\
~
\ea
\ea
\right),~~~~~~
\nabla^\dagger=\left(
\ba{cc}
\ba{cc}
\rho&0\\
0&\rho\\
\ea&
\ba{c}
~\\
\eb_\mu(x_\mu-b_\mu)\\
~
\ea
\ea
\right),
\end{eqnarray}
and the solution of ``0-dimensional Dirac equation'' is trivially found:
\begin{eqnarray}
V=\fr{1}{\sqrt{\phi}}\left(
\ba{c}
\ba{c}
~\\
\eb_\mu(x_\mu-b_\mu)\\
~
\ea\\
\ba{cc}
-\rho&0\\
0&-\rho\\
\ea
\ea
\right),~~~\phi=\vert x-b\vert^2 +\rho^2,
\end{eqnarray}
where the normalization factor $\phi$ is determined by
the normalization condition (\ref{0norm_now}).

\item Step (iii): The instanton solution is constructed as
\begin{eqnarray}
\lab{bpstkai}
A_\mu=V^\dagger \del_\mu V
&=&\fr{i(x-b)^\nu\eta^{(-)}_\mn }{(x-b)^2+\rho^2}.
\end{eqnarray}
The field strength $F_\mn$ is calculated from this gauge field as  
\begin{eqnarray}
F_\mn=\fr{2i\rho^2}{(\vert x-b\vert^2+\rho^2)^2}\eta_{\mn}^{(-)}.
\end{eqnarray}
The distribution is just like in Fig. \ref{bpst_com}.
The dimension 5 of the instanton moduli space
corresponds to the positions $b^\mu$ and the size $\rho$
of the instanton\footnote{Here
the size of instantons is 
the full width of half maximal (FWHM) of $F_\mn$. }.

Now let us take the zero-size limit.
Then the distribution of the field strength $F_{\mn}$
converses into the singular, delta-functional configuration.
Instantons have smooth configurations by definition 
and hence the zero-size instanton does not exists,
which corresponds to the singularity of the 
(complete) instanton moduli space
which is called the {\it small instanton singularity}. 
(See Fig. \ref{bpst_com}.)\footnote{
Here the horizontal directions correspond to the degree of
global gauge transformations which act on the gauge fields
as the adjoint action.}
On noncommutative space, the singularity is resolved and
new class of instantons appear.

\begin{figure}[htbn]
\epsfxsize=110mm
\hspace{2.5cm}
\epsffile{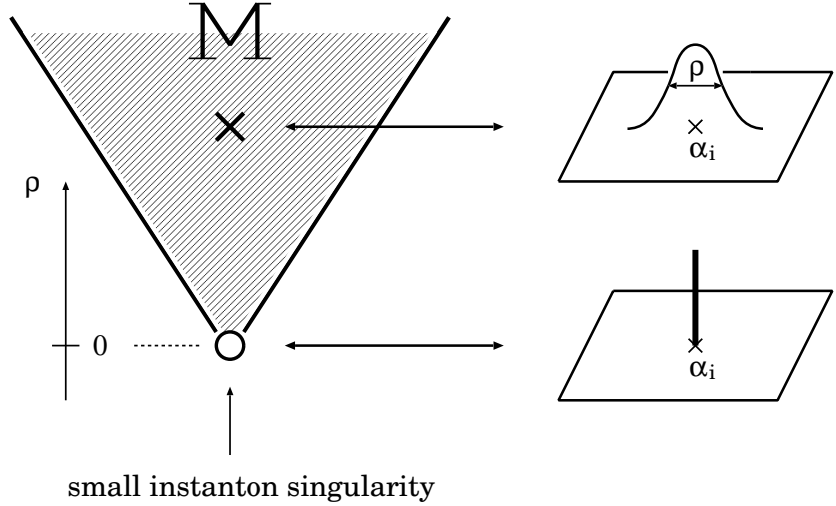}
\caption{Instanton moduli space $\cM$ and
the instanton configurations}
\label{bpst_com}
\end{figure}

\end{itemize}

\noindent
\unl{{\it 't Hooft instanton solution}
($k$-instanton, $\dim\cM^{\scr\mbox{'t Hooft}}_{2,k}=5k$)}
\vs

This solution is the most simple multi-instanton solution
without the orientation moduli parameters
and is also easily constructed by ADHM procedure.
Here we take the real representation instead of the complex
representation.

\begin{itemize}

\item  Step (i): In this case, we solve the 
ADHM equation by putting  
the matrices $B_i$ diagonal.
Then $S$ is easily solved:
\begin{eqnarray}
S&=&
\left(
\ba{ccc}
\ba{cc}
\rho_1&0\\0&\rho_1
\ea
&
\cdots
&
\ba{cc}
\rho_k&0\\0&\rho_k
\ea
\ea
\right),\nonumber\\
B_i
&=&
\left(
\ba{ccc}
\alpha_i^{(1)}&~&O\\
~&\ddots&~\\
O&~&\alpha_i^{(k)}
\ea
\right),~~~
\rho_p\in \R,~~~\alpha_i^{(p)}\in \C.
\end{eqnarray}

\item Step (ii): The solution 
of ``0-dimensional Dirac equation'' $\nabla^\dagger V=0$ is
\begin{eqnarray}
V&=&\fr{1}{\sqrt{\phi}}\left(
\ba{c}
1\\
((x^\mu-T^\mu)\ot \eb_\mu)^{-1}S^\dagger
\ea
\right),\\
{\mbox{where }}&&\phi=1+\sum_{p=1}^{k}\fr{\rho_p^2}{\vert x-b_p\vert^2},
\nonumber\\
&&((x^\mu-T^\mu) \ot \eb_\mu)^{-1}
=\diag_{p=1}^{k}
\left(\fr{(x^\mu-b_p^\mu)}{\vert x-b_p\vert^2}
\ot e^\mu\right),\nonumber
\end{eqnarray}
where $\alpha_1^{(p)}=b_p^2+ib_p^1,~\alpha_2^{(p)}=b_p^4+ib_p^3$.

\item Step (iii): The ASD gauge field is 
\begin{eqnarray}
\lab{bpstkai}
A^{(-)}_\mu=V^\dagger \del_\mu V
=-\fr{i}{\phi}\sum_{p=1}^{k}
\fr{\rho_p^2\eta_\mn^{(+)}(x_\nu-b_\nu^{(p)})}
{\vert x-b^{(p)}\vert^4}
=\fr{i}{2}\eta_\mn^{(+)}\del^{\nu}\log \phi.
\end{eqnarray}
The final form relates to {\it 't Hooft ansatz} or
{\it CFtHW ansatz} \cite{tHooft2, CoFa, Wilczek},
and originally this solution is obtained by
putting this ansatz on the ASD equation directly,
which leads to the Laplace equation of $\phi$.
This solution is singular at the centers of $k$ instantons
because a singular gauge is taken here.
In fact, in $k=1$ case,
this solution is known to be equivalent to the smooth
BPST instanton solution
up to a singular gauge transformation.
(See, for example, \cite{EGH} p. 381-383.)
The field strength is proved to be ASD though
the SD symbol $\eta_\mn^{(+)}$ is found in the gauge field (\ref{bpstkai}).
The dimension of the moduli space $5k$ consists 
of that of
the positions $b_p^\mu$ of the $k$ instantons and
the size  $\rho_p$ of them.
The diagonal components $b_p^\mu$ of ADHM date $T_\mu$
shows the positions of the instantons,
which is also seen in Eq. (\ref{adhmt})
because the constant shift of $x^\mu$ gives rise to
the shift of the date of $T^\mu$.
\end{itemize}

\subsection{ADHM Construction of NC Instantons}

In this subsection, we construct some typical noncommutative instanton
solutions by using ADHM method in the operator formalism.
In noncommutative ADHM construction, 
the self-duality of the noncommutative parameter
is important, which reflects the properties of the
instanton solutions.

The steps are all the same as commutative one:
\begin{itemize}

\item Step (i): ADHM equation is deformed by the noncommutativity of 
the coordinates as we mentioned in the previous subsection:
\begin{eqnarray}
\label{nc_adhm}
(\mu_{\scr\mbox{\R}}:=)
&&{[B_1,B_1^\dagger]}+[B_2,B_2^\dagger]+II^\dagger-J^\dagger J
=-2(\theta_{1}+\theta_{2})=:\zeta,\nonumber\\
(\mu_{\scr\mbox{\C}}:=)&&{[B_1,B_2]}+IJ=0.
\end{eqnarray}
We note that if the noncommutative parameter is ASD, 
that is, $\theta_1+\theta_2=0$,
then the RHS of the first equation of ADHM equation 
becomes zero.\footnote{When we treat SD gauge fields,
then the RHS is proportional to $(\theta_1-\theta_2)$.
Hence the relative self-duality between gauge fields and
NC parameters is important.}

\item Step (ii): Solving the noncommutative ``0-dimensional Dirac equation''
\begin{eqnarray}
\label{nc_0dirac}
\hat{\na}^\dagger \Vh=
\left(\ba{ccc}I&\zh_2-B_2&\zh_1-B_1\\
          J^\dagger&-(\zbh_1-B_1^\dagger)&\zbh_2-B^\dagger_2
          \ea\right)\Vh
=0
\end{eqnarray}
with the normalization condition.

\item Step (iii): the ASD gauge fields are
constructed from the zero-mode $V$,
\begin{eqnarray}
\label{nc_4inst}
\Ah_\mu=\Vh^\dagger \del_\mu \Vh,
\end{eqnarray}
which actually satisfies the noncommutative ASD equation:
\begin{eqnarray}
\label{nc_asd}
(\Fh_{z_1\zb_1}+\Fh_{z_2\zb_2}=)&&
[\Dh_{z_1},\Dh_{\zb_1}]+[\Dh_{z_2},\Dh_{\zb_2}]-\half\left(\fr{1}{\theta_1}
+\fr{1}{\theta_2}\right)=0,\nn
(\Fh_{z_1z_2}=)&&{[\Dh_{z_1},\Dh_{z_2}]}=0.
\end{eqnarray}
There is seen to be a beautiful duality
between Eqs. (\ref{nc_adhm}) and (\ref{nc_asd}).
We note that when the noncommutative parameter is ASD,
then the constant terms in both Eqs. (\ref{nc_adhm}) and (\ref{nc_asd})
disappear.

\end{itemize}

In this way, noncommutative instantons are actually constructed.
Here we have to take care about the inverse of the operators.

\vs
\noindent
\underline{\bf Comments on instanton moduli spaces}
\vs

\begin{figure}[htbn]
\epsfxsize=100mm
\hspace{3cm}
\epsffile{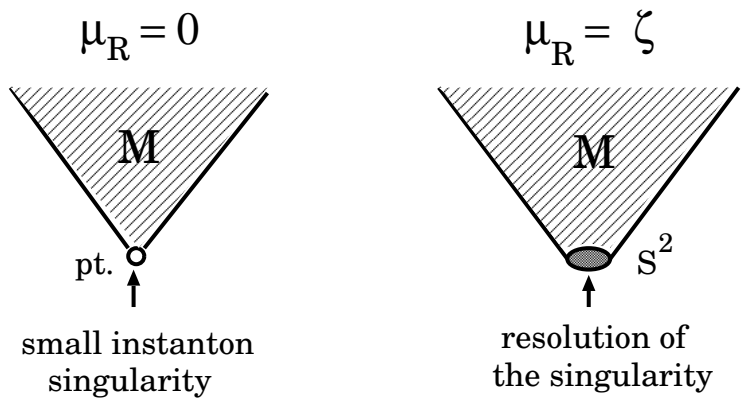}
\caption{Instanton Moduli Spaces}
\label{moduli}
\end{figure}

Instanton moduli spaces are determined by the value of $\mu_{\scr\mbox{\R}}$
\cite{Nakajima2, Nakajima3} (cf. Fig. \ref{moduli}). Namely,
\begin{itemize}
\item  In $\mu_{\scr\mbox{\R}}=0$ case, instanton moduli spaces
contain small instanton singularities,
(which is the case for  commutative $\R^4$ 
and special noncommutative $\R^4$ where
 $\theta$ : ASD).
\item In $\mu_{\scr\mbox{\R}}\neq 0=:\zeta$ case, 
small instanton singularities
are resolved and new class of smooth instantons, U(1) instantons
exist, (which is the case for general noncommutative $\R^4$)
\end{itemize}

Since $\mu_{\scr\mbox{\R}}=\zeta=-2(\theta_1+\theta_2)$ as Eq. (\ref{nc_adhm}),
the self-duality of the noncommutative parameter is important.
NC ASD instantons have the following ``phase diagram'' (Fig. \ref{phase}):

\begin{figure}[htbn]
\epsfxsize=75mm
\hspace{5.2cm}
\epsffile{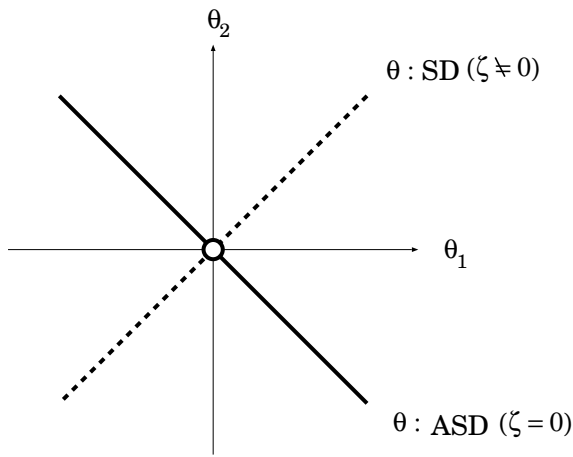}
\caption{``phase diagram'' of NC ASD instantons}
\label{phase}
\end{figure}

When the noncommutative parameter is ASD, that is, $\theta_1+\theta_2=0$,
instanton moduli space implies the singularities.
The origin of the ``phase diagram'' 
corresponds to commutative instantons.
The $\theta$-axis represents instantons on 
$\R^2_{\scr{\mbox{NC}}}\ti \R^2_{\scr{\mbox{Com}}}$.
The other instantons basically have the same properties,
and hence let us fix the noncommutative parameter $\theta$ self-dual.
This type of instantons are first discussed  
by Nekrasov and Schwarz \cite{NeSc}.\footnote{This 
Nekrasov-Schwarz type instantons (the self-duality 
of gauge field-noncommutative parameter is ASD-SD)
are discussed in \cite{Furuuchi, Furuuchi2, Furuuchi3, IKS, 
KLY, CLMS, CMS, Nekrasov, NeSc, LePo, Parvizi, TiZh, FSIv},
and the ASD-ASD instantons \cite{AGMS} are constructed by ADHM
construction in \cite{Furuuchi4, Hamanaka4}, and 
ADHM construction of instantons on $\R^2_{\scr{\mbox{NC}}}
\ti \R^2_{\scr{\mbox{Com}}}$ are discussed in \cite{KLY2}. 
For recommended articles, see \cite{CKT, Watamura}. 
Instantons on commutative side in $B$-fields are discussed in 
\cite{Moriyama2, SeWi, Terashima}.}

Now let us construct explicit noncommutative instanton solution
focusing on U(1) instantons.

\vs
\noindent
\unl{{\it$U(1),~k=1$ solution} (U(1) ASD instanton, $\theta$ : SD)}
\vs

Let us consider the ASD-SD instantons.
For simplicity, let us take $k=1$ and fix the instanton 
at the origin.
The generalization to multi-instanton is straightforward.
If we want to add the moduli 
parameters of the positions, we have only to do translations.
We note that on noncommutative space, 
translations are gauge transformations \cite{GrNe2}.

\begin{itemize}

\item Step (i):  Solving noncommutative ADHM equation

When the gauge group is U(1), the matrix $I$ or $J$
becomes zero \cite{Nakajima3}. Hence ADHM equation
is trivially solved as 
\begin{eqnarray}
\label{date_nc}
B_{1,2}=0, ~~~I=\sqrt{\zeta},~~~J=0
\end{eqnarray}

\item Step (ii):  Solving the ``0-dimensional Dirac equation''

In the background of the ADHM data (\ref{date_nc}),
the Dirac operator becomes
\begin{eqnarray}
\nah=\left(
\ba{cc}
\sqrt{\z}&0\\
\zbh_2&-\zh_1\\
\zbh_1&\zh_2
\ea
\right),~~~
\nah^\dagger=\left(
\ba{ccc}
\sqrt{\z}&\zh_2&\zh_1\\
0&-\zbh_1&\zbh_2
\ea
\right).
\end{eqnarray}
Then the inverse of $\na^\dagger\na$ exists:
\begin{eqnarray}
\fh=\sum_{n_1,n_2=0}^{\infty}
\fr{1}{n_1 +n_2+\zeta}\vert n_1,n_2\ket\bra n_1,n_2\vert.
\end{eqnarray}
In $\zeta\neq0$ case, $\fh$ always exists \cite{Furuuchi}.
One of the important points is on the Dirac zero-mode.
The solution of the ``0-dimensional Dirac equation'' is
naively obtained as follows up to the normalization factor:
\begin{eqnarray}
\label{sol_0dirac_0}
\Vh_1= \left(
\ba{cc} 
\zh_1\zbh_1 +\zh_2\zbh_2\\ 
-\sqrt{\zeta}\zbh_2\\
-\sqrt{\zeta}\zbh_1
\ea
\right),~~~ \nah^\dagger \Vh_1=0. 
\end{eqnarray}
However this does not satisfy the normalization condition
in the operator sense because $\Vh_1$ has the zero mode
$\vert 0,0\ket$ in the Fock space $\cH$ and
the inverse of $\Vh_1^\dagger\Vh_1$ does not exist in $\cH$
calculating the normalization factor.
We have to take care about this point.

K.~Furuuchi \cite{Furuuchi} shows that 
if we restrict all discussions to 
$\cH_1:=\cH -{\vert 0,0\ket\bra 0,0\vert}$, then $\Vh_1$
give the smooth ASD instanton solution in $\cH_1$.
Furthermore he transforms 
the situation in $\cH_1$ into that in $\cH$
by using shift operators and find the correctly normalized $\Vh$
and ASD instanton in $\cH$ \cite{Furuuchi2}:
\begin{eqnarray}
\Vh=\Vh_1\hat{\beta}_1 \Uh_1^\dagger,~~~ \Vh^\dagger  \Vh=1,
\end{eqnarray}
where
\begin{eqnarray}
\hat{\beta}_1&=&(1-\Ph_1) (\Vh_1^{\dagger}\Vh_1)^{-\half}(1-\Ph_1)\nn
&=&
\sum_{(n_1,n_2)\neq (0,0)}
{1\over\sqrt{(n_1+n_2)(n_1+n_2+\zeta)}}\vert n_1,n_2\ket\bra n_1,n_2\vert.
\end{eqnarray}
The projection $(1-\Ph_1)$ in the zero-mode corresponds to
the restriction to $\cH_1$ and the shift operator $\Uh_1$
transforms all the fields in $\cH_1$ to those in $\cH$.
The two prescriptions give the correct zero-mode in $\cH$.

\end{itemize}

Finally we can construct the ASD gauge field as step (iii)
and the field
strength. The instanton number is actually calculated as $-1$.

\vs
\noindent
\unl{{\it $U(2),~k=1$ solution} (NC BPST, $\theta$: SD)}
\vs

This solution is also obtained by ADHM procedure with
the ``Furuuchi's Method.''
The solution of noncommutative ADHM equation is 
\begin{eqnarray}
B_{1,2}=0,~~~I=(\sqrt{\rho^2+\zeta},0),~~~J=\left(\ba{c}0\\\rho\ea\right).
\end{eqnarray}
The date $I$ is deformed by the noncommutativity of
the coordinates, which shows that
the size of instantons becomes larger than that of commutative one
because of the existence of $\zeta$.
In fact, in the $\rho\rar 0$ limit,
the configuration is still smooth and
the U(1) part is alive.
This is essentially just the same as 
the previous $U(1),~k=1$ instanton solution.

\vs
\noindent
\unl{{\it $U(1),~k$-instanton solution} 
(Localized U(1) ASD instanton, $\theta$ : ASD)}
\vs

This time, let us consider the ASD-ASD (not ASD-SD) instanton.
In this case, there are small instanton singularities
in the instanton moduli space.
The U(1) part corresponds to this singular points.
Let us construct this solution directly. 

\begin{itemize}

\item Step (i):  The solution of ADHM equation becomes perfectly trivial:
\begin{eqnarray}
\label{sol_adhm}
&&B_i
=
\left(
\ba{ccc}
\alpha_i^{(0)}&~&O\\
~&\ddots&~\\
O&~&\alpha_i^{(k-1)}
\ea
\right),\nn
&&I=J=0,
\end{eqnarray}
where $\alpha_i^{(m)}$
should show the position of the $m$-th instanton.
The matrices $I$ and $J$ contain information of the
size of instantons and hence $I=J=0$ 
suggests that the configuration would be size-zero and singular.

\item Step (ii): \footnote{The general discussion is rather complicated.
We recommend the readers interested in the details to follow
without the moduli parameters $\alpha_i^{(m)}$ first.
Then taking the direct sum of the translation $\Th
\sim e^{\alpha_1\delh_{z_1}}\ot e^{\alpha_2\delh_{z_2}} 
\sim e^{\alpha_1\zbh_1/\theta}\ot e^{\alpha_2\zh_2/\theta}$ 
on $\nah$
and $\Vh$, we reach to the present results with the moduli parameters.
(We note $\vert \alpha_i \ket 
\sim e^{\alpha_i\ah^\dagger_i}\vert 0\ket$.)}
Next we solve ``0-dimensional Dirac equation''
in the background of the solutions (\ref{sol_adhm})
of the ADHM equation.
This is also simple.
Observing the right hand side of the completeness condition (\ref{4comp}),
we get $\vh_1^{(m)}
=\vert \alpha^{(m)}_1,\alpha^{(m)}_2\ket
\bra p_1^{(m)},p_2^{(m)}\vert$ and $\vh_2=0$,
where $\vert p_1^{(m)},p_2^{(m)}\ket$ is the
normalized orthogonal state in $\cH_1\ot \cH_2$:
\begin{eqnarray}
\bra p_1^{(m)},p_2^{(m)}\vert p_1^{(n)},p_2^{(n)}\ket=\delta_{mn},
\end{eqnarray}
and $\vert \alpha^{(m)}_1,\alpha^{(m)}_2\ket$ is the
normalized coherent state and satisfies
\begin{eqnarray}
\label{coherent}
&&\zh_1\vert \alpha^{(m)}_1,\alpha^{(m)}_2\ket
=\alpha^{(m)}_1\vert \alpha^{(m)}_1,\alpha^{(m)}_2\ket,\no\\
&&\zbh_2\vert \alpha^{(m)}_2,\alpha^{(m)}_2\ket
=\bar{\alpha}^{(m)}_2\vert \alpha^{(m)}_1,\alpha^{(m)}_2\ket,\no\\
&&\bra\alpha^{(m)}_1,\alpha^{(m)}_2\vert \alpha^{(m)}_1,\alpha^{(m)}_2\ket=1.
\end{eqnarray}
The eigen values $\alpha^{(m)}_1$ and $\alpha_2^{(m)}$
of $\zh_1$ and $\zbh_2$ are decided to be just the same as the $m$-th
diagonal components of the solutions $B_1$ and $B_2$ in Eq.
      (\ref{sol_adhm}), respectively.
Though $\uh$ is undetermined, $\Vh$ already satisfies $\na^\dagger
\Vh=0$, which comes from that
in the case that the self-dualities of gauge fields and
noncommutative parameters are the same, the coordinates in each column
of $\na^\dagger$
play the same role in the sense that they are annihilation operators or
creation operators.
Finally, the
normalization condition $\Vh^\dagger \Vh=1$
determines $\uh=\Uh_k$ where 
\begin{eqnarray}
&&\Uh_k\Uh_k^\dagger=1,\nn
&&\Uh_k^\dagger \Uh_k=1-\Ph_k=1
-\sum_{m=0}^{k-1}\vert p_1^{(m)},p_2^{(m)}\ket\bra
p_1^{(m)},p_2^{(m)}\vert.
\label{shift_4dim}
\end{eqnarray}
This is just the shift operator and naturally appears in this way.
The shift operator and $\uh$ have the same behavior
at $\vert x\vert\rar\infty$, which is consistent.

Gathering the results,
we get the Dirac zero-mode as
\begin{eqnarray}
\label{v_instanton}
\Vh=
\left(\ba{c}
\uh\\
\vh_1^{(m)}\\
\vh_2^{(m)}
\ea\right)
=
\left(
\ba{c}
\Uh_k\\
\vert \alpha^{(m)}_1,\alpha^{(m)}_2\ket\bra p_1^{(m)},p_2^{(m)}\vert\\
0
\ea\right),
\end{eqnarray}
here $\vh_i^{(m)}$ is the $m$-th low of $\vh_i$.
One example of the shift operators which satisfies (\ref{shift_4dim})
are given by
\begin{eqnarray}
\Uh_k=\sum_{n_1=1,n_2=0}^{\infty}\vert n_1,n_2\ket\bra n_1,n_2\vert
+\sum_{n_2=0}^{\infty}\vert 0,n_2\ket\bra 0,n_2+k\vert,
\end{eqnarray}
where
\begin{eqnarray}
\Ph_k=\sum_{m=0}^{k-1}\vert 0,m\ket\bra 0,m\vert.
\end{eqnarray}

We note that ASD-ASD instantons 
do not need the ``Furuuchi's method'' unlike ASD-SD instantons.

\item Step (iii): The $k$-instanton solution 
with the moduli parameters
of the positions of the instantons are:
\begin{eqnarray}
\Dh_{z_i}&=&\Vh^\dagger\delh_{z_i}\Vh
=\uh^\dagger \delh_{z_i} \uh+\vh^\dagger \delh_{z_i} \vh\no\\
&=&\Uh_k^\dagger \delh_{z_i} \Uh_k- \sum_{m=0}^{k-1}
\vert p_1^{(m)},p_2^{(m)}\ket
\bra \alpha^{(m)}_1,
\alpha^{(m)}_2\vert\fr{\zbh_i}{2\theta^i} \vert
\alpha^{(m)}_1,\alpha^{(m)}_2\ket\bra p_1^{(m)},p_2^{(m)}
\vert\no\\
&=&\Uh_k^\dagger \delh_{z_i} \Uh_k-\sum_{m=0}^{k-1}
\fr{\bar{\alpha}_{z_i}^{(m)}}{2\theta^i}
\vert p_1^{(m)},p_2^{(m)}\ket\bra p_1^{(m)},p_2^{(m)}\vert.
\end{eqnarray}

\end{itemize}

This is just the essential part of the solution generating technique.
The solution generating technique is one of the strong 
auto-B\"acklund transformation and is based on the following
transformation:
\begin{eqnarray}
\Dh_{z_i}\rar \Uh_k^\dagger \Dh_{z_i} \Uh_k 
-\sum_{m=0}^{k-1}
\fr{\bar{\alpha}_{z_i}^{(m)}}{2\theta^i}
\vert p_1^{(m)},p_2^{(m)}\ket\bra p_1^{(m)},p_2^{(m)}\vert.
\end{eqnarray}
Though this transformation looks like the gauge transformation,
it is a non-trivial transformation because $\Uh_k$ is not a unitary
operator but a shift operator.
This transformation leaves equation of motion as it is
in gauge theories and can be applied to the problems on 
tachyon condensations and Sen's conjecture,
which is discussed in section 6 in this thesis.

The field strength is calculated very easily:
\begin{eqnarray}
F_{12}=-F_{34}=i\sum_{m=0}^{k-1}\vert p_1^{(m)},p_2^{(m)}\ket
\bra p_1^{(m)},p_2^{(m)}\vert.
\end{eqnarray}
The instanton number $k$ is represented
by the dimension of the projected states $\vert p_1^{(m)},p_2^{(m)}\ket$
which appears in the relations of the shift operator $\uh=\Uh_k$
or the bra part of $\vh_1^{(m)}$
Information of the position of $k$ localized solitons
is shown in the coherent state $\vert\alpha_i^{(m)}\ket$ in the ket
part of $\vh_1^{(m)}$.

It seems to be strange
that  the field strength
contains no information of the positions $\alpha_i^{(m)}$
of the instantons.
This is due to the fact that it is hard to discuss
what is gauge invariant quantities 
in noncommutative gauge theories.
The apparent paradox is solved 
by mapping this solution to commutative side
by exact Seiberg-Witten map \cite{OkOo, MuSu, LiMi}.
The commutative description of D0-brane density $J_{D0}\sim F_\mn F_\mn$
is as follows \cite{HaOo}:
\begin{eqnarray}
J_{\scr\mbox{D0}}(k)=2\delta^{(4)}(k)
+\sum_{m=0}^{k-1}e^{ik_{z_i}\alpha_i^{(m)}},
\end{eqnarray}
that is,
\begin{eqnarray}
\label{d0density}
J_{\scr\mbox{D0}}(x)=\fr{2}{\theta^2}
+\sum_{m=0}^{k-1}
\delta^{(2)}(z_1-\alpha_1^{(m)})\delta^{(2)}(z_2-\alpha_2^{(m)}).
\end{eqnarray}
The second term shows the $k$ instantons
localized at $z_i=\alpha_i$.
The configuration is actually singular,
which is consistent with the existence of
small instanton singularities. 
The first term represents the situation that 
infinite number of D0-branes form D4-brane 
in the presence of background $B$-field,
which is consistent with interpretations in
matrix models \cite{BFSS,IKKT} (cf. section 6.2).
This D0-D4 brane system with $B$-field preserves
the original SUSY without $B$-field
and tachyon fields do not appear,
which is reflected by $\zeta=0$ (cf. section 3.3).

\vspace{2mm}
\noindent
\unl{\it localized $U(N)~k$ instantons}
\vspace{2mm}

There is an obvious generalization of the construction of $U(N)$
localized instanton, which is essentially the diagonal product
of the previous discussions.
In the solution of ADHM equations, $I,J$ can be still zero
and $B_{1,2}$ are the same as that of $N=1$ case.
The solution of ``0-dimensional Dirac equation''
is given by
\begin{eqnarray}
\Vh=
\left(\ba{c}
\uh\\
\vh_1^{(m,a)}\\
\vh_2^{(m,a)}
\ea\right)
=
\left(
\ba{c}
\Uh_k\\
\vert \alpha^{(m_a)}_1,\alpha^{(m_a)}_2\ket\bra p_1^{(m_a)},p_2^{(m_a)}\vert
\\
0
\ea\right),
\end{eqnarray}
where $m_a$ runs over some elements in
$\left\{0,1,\cdots,k-1\right\}$ whose number is $k_a$
and all $m_a$ are different. (Hence $\sum_{a=1}^N k_a=k$.)
The $N\times N$ matrix $\Uh_k$ is a partial isometry and satisfies
\begin{eqnarray}
\Uh_k\Uh_k^\dagger=1,~~~
\Uh_k^\dagger \Uh_k=1-\Ph_k,
\end{eqnarray}
where the projection $\Ph_k$ is the following diagonal sum:
\begin{eqnarray}
\Ph_k:=\diag_{a=1}^{N}\left(\diag_{m_a}\vert p_1^{(m_a)},p_2^{(m_a)}\ket
\bra p_1^{(m_a)},p_2^{(m_a)}\vert\right).
\end{eqnarray}
$\vert\alpha^{(m_a)}_i\ket$ is the normalized coherent state (\ref{coherent}).
Next
in the case of $\vert p_1^{(m_a)},p_2^{(m_a)}\ket=\vert 0,m_a\ket$,
then the shift operator
is, for example, chosen as the following diagonal sum:
\begin{eqnarray}
\Uh_k=
\diag_{a=1}^{N}\left(
\sum_{n_1=1,n_2=0}^{\infty}\vert n_1,n_2\ket\bra n_1,n_2\vert
+\sum_{n_2=0}^{\infty}\vert 0,n_2\ket\bra 0,n_2+k_a\vert\right).
\end{eqnarray}
$\vert \alpha^{(m_a)}_1,\alpha^{(m_a)}_2\ket$ is the normalized coherent state
and defined similarly as (\ref{coherent}).
We can construct another non-trivial example of a shift operator
in $U(N)$ gauge theories by using noncommutative
ABS construction \cite{ABS}.
The localized instanton solution
in \cite{Furuuchi4} is one of these generalized solutions
for $N=2$.

\vs
\noindent
\unl{{\it $U(2),~k=1$ instanton solution} (NC BPST instanton, $\theta$ : ASD)}
\vs

In the same process, we can construct exact
NC ASD-ASD BPST instanton solutions 
with the moduli parameter $\rho$ of the size and 
in the $\rho\rar 0$ limit, these solutions essentially are reduced to
the localized U(1) instantons \cite{Furuuchi4}.

\subsection{D0-D4 Brane Systems and ADHM Construction}

\begin{center}
\begin{figure}[htbn]
\epsfxsize=125mm
\hspace{2.5cm}
\epsffile{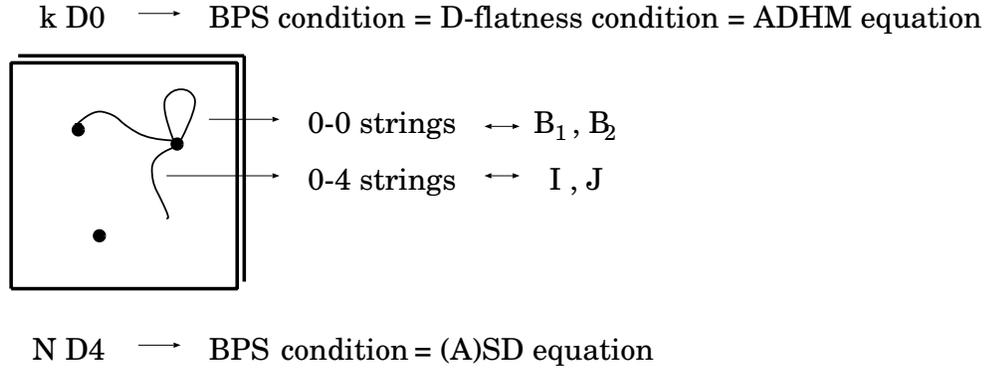}
\caption{D-brane interpretation of ADHM construction}
\label{D0D4}
\end{figure}
\end{center}

In this subsection, we discuss the
D-brane interpretation of ADHM construction of instantons.
The low-energy effective theory is
described by the Super-Yang-Mills (SYM) theory.
In particular the solitons in the SYM theory
corresponds to the lower-dimensional D-branes on the D-brane.
ADHM construction is elegantly embedded in D0-D4 systems, 
which gives the physical meaning of ADHM
construction \cite{Witten2, Douglas, Douglas2},
where the number of D0 and D4 corresponds to
the instanton number $k$ and the rank of the gauge group $N$,
respectively. (See Fig. \ref{D0D4}.)

This system preserves eight supersymmetry.
Now let us represent this SUSY condition
from two different viewpoints.

On the D4-brane, the SUSY condition 
is described as the BPS condition for
the SUSY transformation of the gaugino,
which is just the ASD Yang-Mills equation. 
On the other hand, on D0-branes,
the SUSY condition is described
as the D-flatness condition in the Higgs branch.
The D-term is an auxiliary field and
related to the massless scalar fields
which come from massless excitation modes
of 0-0 strings and 0-4 strings.
If the massless excitation modes
of 0-0 strings and 0-4 strings are denoted 
by $k\ti k$ matrices $B_{1,2}$ 
(adjoint Higgs fields) 
and $k\ti N$ matrices $I,J$ (fundamental Higgs fields), respectively,
then we get the D-flatness condition as
\begin{eqnarray}
&&{[B_1,B_1^\dagger]}+[B_2,B_2^\dagger]+II^\dagger-J^\dagger J
=0,\nonumber\\
&&{[B_1,B_2]}+IJ=0.
\end{eqnarray}
This is just the ADHM equation!
Of course, the described physical situation is unique
and hence both moduli space should be equivalent.
Furthermore the degree of freedom of the $k$ D0-branes
is apparently $4Nk$, which reproduces the results from
Atiyah-Singer index theorem.

We comment on the interpretation of $\mu_{\scr\mbox{\R}}=\zeta$
on noncommutative space from the viewpoint of effective theory of D-branes.
If $B$-field is turned on in the background of this D-brane systems,
Fayet-Illiopolous (FI) parameter appears in the D-flatness condition,
because constant expectation value of $B$-field appears
in the SUSY transformation of gaugino on D4-branes
and the constant term in the transformation equation 
is just the FI parameter.
The physical meaning of the FI parameter
is the expectation value of tachyon field which
appears first due to the unstablity of the D-brane systems
because of the presence of $B$-field.
After the tachyon condensation,
different SUSY from the original one 
is preserved again and the systems becomes stable. 
NC instanton represents such situation in general.

The interpretation of the ``0-dimensional Dirac equation''
is also discussed in \cite{Witten, Douglas2}
by D1-probe analysis of the background $k$ D5-$N$ D9
brane systems.

\newpage

\section{Monopoles and D-branes}

Monopoles are also constructed by ADHM-like procedure,
which is called {\it Nahm construction}.
This time the duality is the one-to-one
correspondence between the monopole moduli space and 
the moduli space of Nahm data.
The D-brane interpretations are also given
as D1-D3 brane systems which can be
considered as the T-dualized situation of
D0-D4 brane systems.
D-brane picture clearly explains
the equivalence between noncommutative
situation and that in the presence of the background $B$-field.

\subsection{Nahm Construction of Monopoles}

In this subsection, we construct exact BPS monopole solutions
on commutative $\R^3$.
By applying ADHM procedure to monopoles, 
we can easily construct
Dirac monopole \cite{Dirac} ($G=U(1)$ monopole solution) and
Prasad-Sommerfield (PS) solution \cite{PrSo}.
($G=SU(2)$ 1-BPS monopole solution which is the typical
example of 
't Hooft-Polyakov monopole solution \cite{tHooft0, Polyakov}.)
The concrete steps are as follows:

\begin{itemize}

\item Step (i):  Solving Nahm equation:

\begin{eqnarray}
\lab{nahm_now}
\fr{dT_i}{d\xi}&=&i\ep_{ijl}T_j T_l,
\end{eqnarray}
where $T_i(\xi)$ should satisfies the following boundary condition:
\begin{eqnarray}
\label{bc_nahm_now}
T_i(\xi)&\st{\xi\rar\pm a/2}{\longr}&\fr{\tau_i}{\dis\xi\mp \fr{a}{2}}
+(\mbox{regular terms on }\xi)\\
{\mbox{where}}&&\tau_i~:~{\mbox{ irreducible representation of }} SU(2)
~~~[\tau_i,\tau_j]=i\epsilon_{ijl}\tau_l.\nonumber
\end{eqnarray}

We note that the coordinates $x^{1,2}$ always appear in pair with
the matrices $T^{1,2}$ and that is why we see the commutator of
the coordinates in the RHS.
These terms of course vanish on commutative spaces, however,
they cause nontrivial contributions on noncommutative spaces,
which is seen later soon.

\item Step (ii):  Solving ``0-dimensional Dirac equation'' in the
      background of the Nahm date which satisfies Nahm eq. (\ref{nahm_now}):
\begin{eqnarray}
\label{1dirac}
\na^\dagger v=0,
\end{eqnarray}
with the normalization condition:
\begin{eqnarray}
\label{1norm}
\int d\xi v^\dagger v =1,
\end{eqnarray}
where the ``1-dimensional Dirac operator'' is defined by
\begin{eqnarray}
\lab{ndata}
\nabla_\xi({\bf x})=i\fr{d}{d\xi}+e_i(x^i-T^i),~~~
\nabla_\xi({\bf x})^\dagger=i\fr{d}{d\xi}+\eb_i(x^i-T^i),
\end{eqnarray}
in which $x^i$ is the coordinate of $\R^3$, 
and $\xi$ is an element of the interval $(-(a/2),a/2)$ 
for $G=SU(2)$.\footnote{The region spanned by $\xi$
depends on the gauge group, for example, 
in $G=U(2)$ case, finite interval ($a_-,a_+$), 
and in $G=U(1)$ case, semi-infinite line.} 

\item  Step (iii): Using the solution $v$,
we can construct the corresponding BPS monopole solution as
\begin{eqnarray}
\label{3inst}
\Phi=\int d\xi v^\dagger \xi v,~~~A_i=\int d\xi v^\dagger \del_\mu v,
\end{eqnarray}
which actually satisfies the Bogomol'nyi equation:
\begin{eqnarray}
B_i=-[D_i,\Phi],
\end{eqnarray}
where $B_i:=(i/2)\epsilon_{ijk}F^{jk}$ is the magnetic fields.

\end{itemize}

The detailed aspects are discussed in Appendix A.
In this subsection, we give some typical examples of 
the explicit monopole solutions.

\vs
\noindent
\unl{{\it $G=U(2)$ BPS 't Hooft-Polyakov monopole} ($k=1$)}
\vs

\begin{itemize}
\item Step (i): 

In $k=1$ case, the boundary condition (\ref{bc_nahm}) is simplified
and Nahm equation is trivially solved :
\begin{eqnarray}
T_i=b_i,
\end{eqnarray}
which shows that the monopole is located at $x_i=b_i$.
For simplicity, we set $b_i=0$.

\item Step (ii): 

In order to solve the ``1-dimensional Dirac equation,''
let us take the following ansatz on $v$
which corresponds to the gauge where the Higgs field $\Phi$
is proportional to $\sigma_3$:
\begin{eqnarray}
v=\left(\ba{c}-(x_1-ix_2)\\\del_\xi+x^3\ea\right)\beta.
\end{eqnarray}
Then the equation is reduced to 
the simple differential equation $\del_\xi^2 \beta=r^2\beta$ and we get
\begin{eqnarray}
\beta=e^{\pm r\xi},
\label{phi}
\end{eqnarray}
which says that there are two independent solutions
and the gauge group becomes $U(2)$.
From the normalization condition, the zero-mode is
\begin{eqnarray}
v=\left(
\ba{cc}
\dis -\fr{x_1-ix_2}{\sqrt{(r+x_3)(e^{2ra_+}-e^{2ra_-})}}e^{r\xi}&
\dis\fr{x_1-ix_2}{\sqrt{(r-x_3)(e^{-2ra_-}-e^{-2ra_+})}}e^{-r\xi}\\
\dis\sqrt{\fr{r+x_3}{e^{2ra_+}-e^{2ra_-}}}e^{r\xi}&
\dis \sqrt{\fr{r-x_3}{e^{2ra_+}-e^{2ra_-}}}e^{-r\xi}
\ea
\right),
\end{eqnarray}
where the integral region is $(a_-,a_+)$.

\item Step (iii): The Higgs field is calculated as follows:
\begin{eqnarray}
\label{com_tp}
\Phi&=&\left(
\ba{cc}
\dis
\fr{a_+ e^{2ra_+}-a_-e^{2ra_-}}{e^{2ra_+}-e^{2ra_-}}-\fr{1}{2r}&0\\
0&\dis\fr{a_- e^{-2ra_-}-a_+ e^{-2ra_+}}{e^{-2ra_-}-e^{-2ra_+}}+\fr{1}{2r}
\ea
\right).
\end{eqnarray}
The gauge field is also solved, however, is rather complicated.
Here if we take the integral region as $(-(a/2),a/2)$, then 
the gauge group becomes $G=SU(2)$ and the
monopole solution (\ref{com_tp}) 
coincides with Prasad-Sommerfield (PS) monopole
\cite{PrSo} up to gauge transformation\footnote{If we take
the most simple form $v\propto \exp(-x^i\sigma_i\xi)$ as the 
ansatz for $v$, this PS solution is 
directly obtained without any gauge transformation.}:
\begin{eqnarray}
\Phi=\fr{x^i\sigma_i}{2\vert\vec{x}\vert^2}
\left(\fr{a\vert\vec{x}\vert}{\tanh a\vert\vec{x}\vert}-1\right),~~~
A_i=\fr{\epsilon_{ijk}\sigma^j x^k}{2\vert\vec{x}\vert^2}
\left(\fr{a\vert\vec{x}\vert}{\sinh a\vert\vec{x}\vert}-1\right).
\end{eqnarray}
If we take the integral region $(-\infty,0)$,
then one part $e^{-r\xi}$ of the solution (\ref{phi})
becomes unnormalized and 
the gauge group becomes $G=U(1)$,
and the solution (\ref{com_tp}) is reduced to the 
Dirac monopole \cite{Dirac}
up to gauge transformation:
\begin{eqnarray}
\label{com_Dirac}
\Phi=-\frac{1}{2r},~A_r=A_\vartheta=0,~
A_\varphi=-\frac{i}{2r}\frac{1+\cos\vartheta}{\sin\vartheta},
\end{eqnarray}
where $(r,\vartheta,\varphi)$ is the ordinary polar coordinate.
The gauge fields diverse at $\vartheta=0$ and 
the magnetic fields also have the singularities at $\vartheta=0$, that is,
on the positive part of $x^3$-axis.
The string-like singularity is called Dirac string and 
can be interpreted as the infinitely-thin solenoid.
This is an unphysical object and the direction can be changed
under a gauge transformation.\footnote{For a review see, \cite{GoOl,
Harvey}.}. On the region apart from the positive part of $x^3$-axis,
the magnetic fields have the following configuration in a radial
pattern (See the left side of Fig. \ref{2dirac}):
\begin{eqnarray}
B_i=-\partial_i\Phi=-\frac{x^i}{2r^3}.
\end{eqnarray}

\end{itemize}

\subsection{Nahm Construction of NC Monopoles}

In this subsection, we construct
some typical $G=U(2)$ or $U(1)$ noncommutative 
monopole solutions by Nahm procedure.
The steps are the same as commutative one:

\begin{itemize}

\item Step (i):  Solving Nahm equation
\begin{eqnarray}
\label{nahm_monopole2}
\fr{dT_i}{d\xi}-\fr{i}{2}\epsilon_{ijk}[T_j,T_k]=-\theta\delta_{i3}
\end{eqnarray}
with the boundary condition (\ref{bc_nahm}).
There is seen to be a constant term due to
the noncommutativity of the coordinates,
which can be absorbed by a constant
shift of $T_3$ \cite{Bak, GoMa}.
In $k=1$ case, the boundary condition becomes trivial
and the solution $T_i$ is easily found.

\item Step (ii):  Solving the 1-dimensional Dirac equation
\begin{eqnarray}
\label{nc_1dirac}
\hat{\na}^\dagger \vh=0
\end{eqnarray}
with the normalization condition.

\item Step (iii):  By using the solution $\vh$ of the 
``1-dimensional Dirac equation,''
we can construct the Higgs field and gauge fields as
\begin{eqnarray}
\label{nc_monopole}
\Phih=\int d\xi~\vh^\dagger\xi \vh,~~~\Ah_i=\int d\xi~ \vh^\dagger\del_i \vh.
\end{eqnarray}

\end{itemize}

Let us construct explicit solutions.

\vs
\noindent
\unl{{\it $U(1),~k=1$ monopole solution (NC Dirac monopole)}}
\vs

For simplicity, we can set the monopole at the origin.

\begin{itemize}

\item Step (i): The solution for noncommutative Nahm equation is 
\begin{eqnarray}
T_{1,2}=0,~~~T_3=-\theta\xi,
\end{eqnarray}
where $\xi$ is an element of $(-\infty,0)$.
Here we introduce new symbols  $W,b, b^{\dagger}$ as 
\begin{eqnarray}
W(x_3,\xi)&=&x_3 \xi +  \fr{1}{2}\theta \xi^2\nonumber\\
b &=& {1\over{\sqrt{2\theta}}} \left( {\del}_\xi+  x_3 + \theta  \xi
  \right)={1\over{\sqrt{2\theta}}} e^{-W} {\del}_\xi e^{W}\nonumber\\
b^{\dagger} &=& {1\over{\sqrt{2\theta}}} \left( -
{\del}_\xi +  x_3 + \theta  \xi  \right)
=-{1\over{\sqrt{2\theta}}} e^{W} {\del}_\xi e^{-W}.
\end{eqnarray}
The operator $b$ satisfies Heisenberg' s commutation relation:
\begin{eqnarray}
[ b, b^{\dagger}]=1.
\end{eqnarray}

\item Step (ii): Now the ``1-dimensional Dirac equation'' is
\begin{eqnarray}
\left(
\ba{cc}
b&\ah^\dagger\\
\ah&-b^\dagger
\ea
\right)
\left(
\ba{c}
\vh_1\\
\vh_2
\ea
\right)
=0.
\end{eqnarray}
($\ah$ is the same as that in (\ref{heisenberg})
and satisfies $[\ah,\ah^\dagger]=1$.)
In order to solve it, let us put the following ansatz 
on $\vh$:
\begin{eqnarray}
\vh=
\left(
\ba{c}
-\ah^\dagger\\
b\ea\right)
\sum_{n=1}^{\infty}{\beta}_{n}\vert n-1 \rangle 
\bra n-1\vert\Uh_1^\dagger
+\left(
\ba{c}
-\dis{1\over{\sqrt{{\z}_0}}} e^{-W} \vert 0 \rangle \bra 0 \vert\\
0
\ea \right),
\end{eqnarray}
where ${\z}_0 = \int_{-\infty}^{0} \, d\xi \, e^{-2W}$ 
and ${\beta}_n$ satisfies
\begin{eqnarray}
\left( b^{\dagger}\, b + n \right) {\beta}_n
= 0.
\end{eqnarray}
Hence $\beta_n$ is determined by acting $b$ 
on $\beta_1$ one after another. 
The final unknown is the coefficient
which is determined by the normalization condition.
There needs to be the boundary condition
\begin{eqnarray}
{\beta}_{n} b {\beta}_{n} (0) = 1, ~~~ 
{\beta}_n(\xi) \st{\xi\rar -\infty}{\longr}0
\end{eqnarray}
and finally $\beta_n$ is obtained as
\begin{eqnarray}
{\beta}_{n}(\xi)=
{{{\zeta}_{n-1}(x_3 + \theta \xi)}\over \sqrt{{\z}_{n}(x_{3})
{\zeta}_{n-1}(x_{3})}},
~~~~~~\zeta_n(x_3):=\int_0^\infty dp~p^ne^{-\theta p^2+2px^3}.
\end{eqnarray}

\item Step (iii): The Higgs field and the gauge fields are
\begin{eqnarray}
\label{nc_dirac_mono}
\Phih&=&\sum_{n=0}^{\infty}\Phi_n\vert n \rangle\langle n \vert
    =-\sum_{n=1}^{\infty}\left(\xi_n^2-\xi_{n-1}^2\right)
       \vert n \rangle\langle n \vert-\left(\xi_0^2+\frac{x^3}{\theta}\right)
       \vert 0 \rangle\langle 0 \vert,\nonumber\\
\Dh_z&=&\frac{1}{\sqrt{2\theta}}\sum_{n=0}^{\infty}\frac{\xi_n}
       {\xi_{n+1}}a^\dagger\vert n \rangle\langle n \vert,~~~
\Ah_3=0.\\
\label{xin}
\xi_n(x_3)&:=&\sqrt{\frac{n\zeta_{n-1}}{2\theta\zeta_n}}.
\end{eqnarray}
This is smooth everywhere.
The behavior at the infinity
($r_n+x_3\rightarrow \infty,~r_n:=\sqrt{(x_3)^2+2\theta n}$)
is\footnote{The integral of $\zeta_n$ is done by the saddle point method.}:
\begin{eqnarray}
\label{asymp}
\Phi_n&\sim&\left\{
\begin{array}{ll}\displaystyle
-\frac{x_3}{\theta}&:~n=0,~x_3\rightarrow +\infty\\
\displaystyle -\frac{1}{2r_n}=-\frac{1}{2\sqrt{(x_3)^2+2\theta n}}
&{\mbox{: otherwise}}
\end{array}
\right.\\
(B_3)_n&\sim&\left\{
\begin{array}{ll}\displaystyle
\frac{1}{\theta}&~~~~~~~~~~~~~~~~~~~~~~:n=0,~x_3\rightarrow +\infty\\
\displaystyle -\frac{x_3}{2(r_n)^3}
&~~~~~~~~~~~~~~~~~~~~~~~{\mbox{: otherwise}}
\end{array}
\right.
\end{eqnarray}
This says that
the Higgs field and the magnetic field
have the special behavior at the positive part of $x_3$-axis,
that is, $n=0,~x_3\rightarrow\infty$\footnote{Here we consider $n$
as the square of the distance from the origin on the 1-2 plane.
($(x_1)^2+(x_2)^2\sim 2\theta n$).},
The distribution of the magnetic fields is roughly estimated
like the right side of Fig. \ref{2dirac}.

\begin{figure}[htbn]
\epsfxsize=120mm
\hspace{2cm}
\epsffile{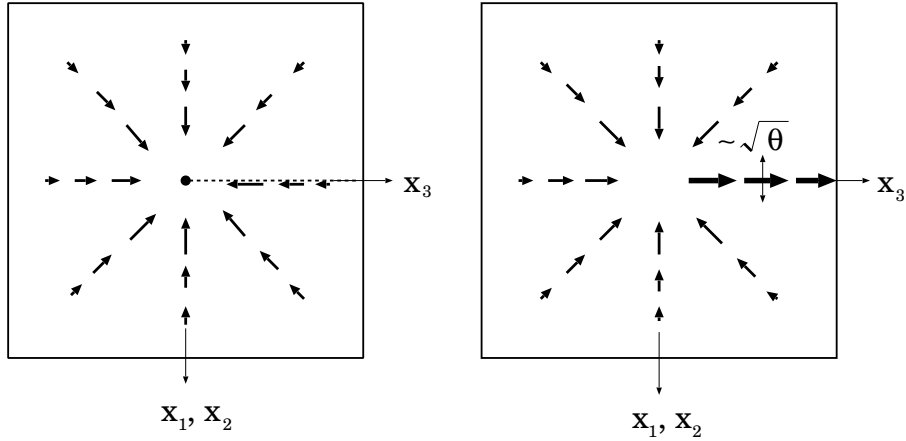}
\caption{The distribution of the magnetic fields of Dirac monopole
(On commutative space (left) V.S. On NC space (right))}
\label{2dirac}
\end{figure}

The universal magnetic field $(B_3(x_3\rightarrow +\infty))_0
\vert 0\rangle\langle 0\vert$ on the positive part of $x_3$-axis,
can be mapped into the star-product formalism 
and it has a Gaussian 
distribution $(2/\theta) \exp\left\{-((x_1)^2+(x_2)^2)/\theta\right\}$
whose width is $\sqrt{\theta}$.
Hence in the commutative limit $\theta\rightarrow 0$,
it is reduced to delta-functional distribution and   
coincides with the Dirac string.

\end{itemize}

\noindent
\unl{\bf Relation to Integrable Systems}
\vskip3mm

The solution of noncommutative 1-Dirac monopole
(\ref{nc_dirac_mono}) has an interesting form from
the integrable viewpoint.
The solution can be written as Yang's form \cite{GrNe}
(See \cite{MaWo}.):
\begin{eqnarray}
\Phi= \xih^{-1}\del_3 \xih,~A_z= \xih^{-1}[\hat{\del}_z, \xih],
\label{yang_form}
\end{eqnarray}
where
\begin{eqnarray}
\xih:=\sum_{n=0}^{\infty}\xi_n(x_3)\vert n\ket\bra n \vert.
\end{eqnarray}
This suggests that even on noncommutative spaces,
the discussion on the integrability is possible.
In fact, noncommutative Bogomol'nyi equation 
for $G=U(1)$ (\ref{bps_monopole})
can be written as the 1-dimensional 
semi-infinite Toda lattice equation \cite{GrNe}:
\begin{eqnarray}
\label{toda_lattice}
\fr{d^2q_n}{dt^2}+e^{q_{n-1}-q_n}-e^{q_n-q_{n+1}}=0,~~~(n=0,1,2,\ldots)
\end{eqnarray}
where
\begin{eqnarray}
q_n(t):=
\left\{
\begin{array}{ll}
\dis\log \left[\fr{e^{\fr{t^2}{2}}}{n!}\xi_n^2
\left(\fr{t}{2}\right)\right],
~~~t:=2x_3&n\geq 0\\
-\infty&n= -1.
\end{array}
\right.
\end{eqnarray}
The operator $\xih$ in Yang's form (\ref{yang_form}) 
is just the $\xi_n$ in (\ref{xin}).
It is interesting that discrete structure appears.

\vs
\noindent
\unl{{\it $U(2),~k=1$ monopole solution (NC Prasad-Sommerfield solution)}}
\vs

This solution is also constructed by 
Gross and Nekrasov \cite{GrNe3}.
The concrete steps are all the same as those in the 
noncommutative Dirac monopole.
The exact solution is, however, very complicated and
the properties are not yet revealed clearly.

\subsection{D1-D3 Brane Systems and Nahm Construction}

\begin{figure}[htbn]
\epsfxsize=140mm
\hspace{1cm}
\epsffile{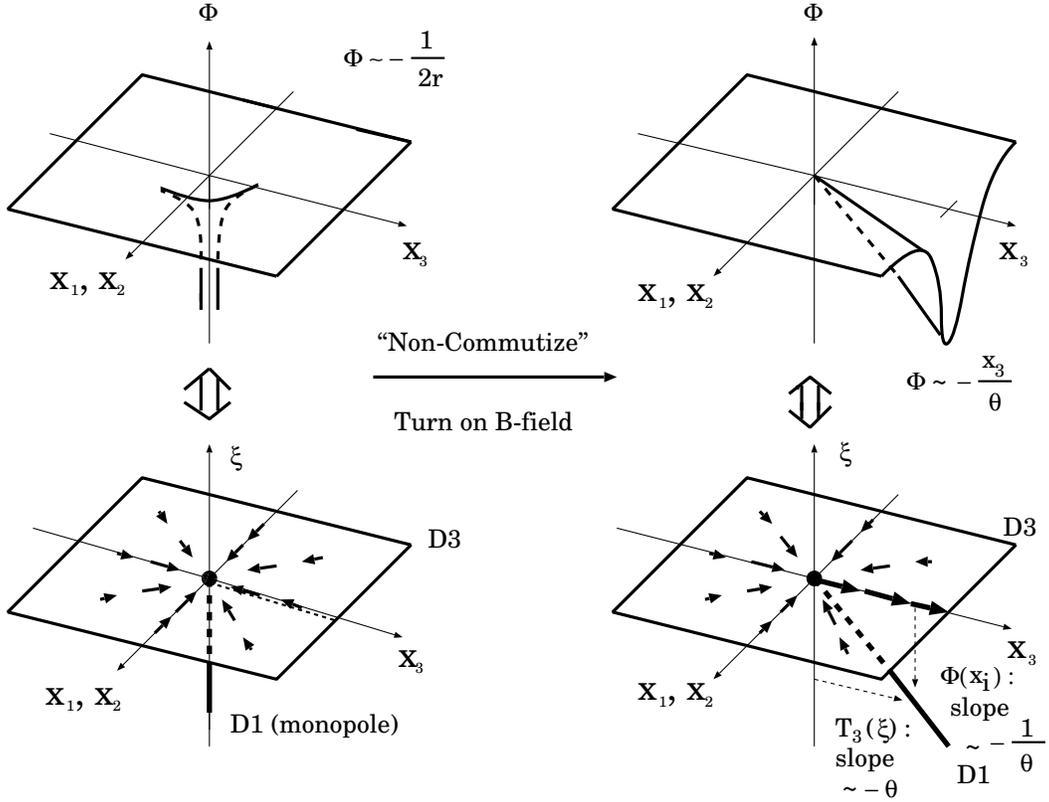}
\caption{The configuration of the Higgs field (Upper) 
and the D-brane interpretation and the magnetic field
(Lower) of
the Dirac monopole
(Left: Commutative case, Right: NC case)}
\label{higgs}
\end{figure}

The monopoles are described by 
D1-D3 brane systems.
The $G=U(N)$ Yang-Mills-Higgs theory is described by
the low-energy effective theory of $N$ D3-branes. 
Then the diagonal values of Higgs field $\Phi$ stand for
the positions of the D3-branes 
in the transverse direction of it.
For example, the Dirac monopole corresponds to
the semi-infinite D1-brane whose end attaches to
D3-brane. (See Fig. \ref{higgs}.)
This D-brane systems finally becomes stable
and then D1-branes are unified with D3-brane and 
are considered as a part of the D3-brane. 
(See the upper-left of Fig.\ref{higgs}.)
The end of D1-brane has magnetic charge
on D3-brane and is considered as magnetic monopoles.  

\begin{figure}[htbn]
\epsfxsize=100mm
\hspace{4cm}
\epsffile{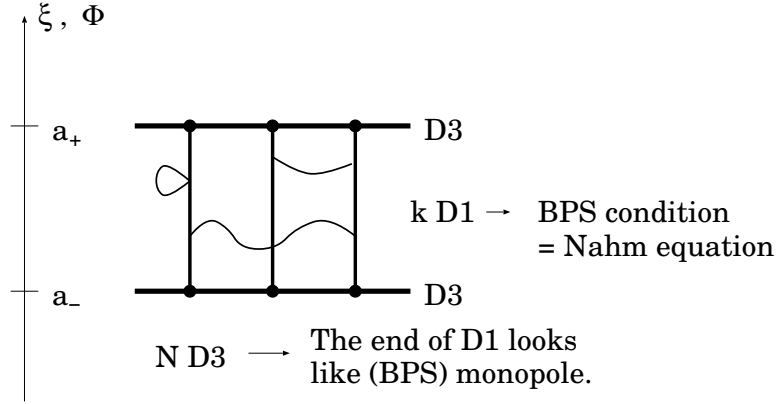}
\caption{D-brane interpretation of Nahm construction}
\label{D1D3}
\end{figure}

Nahm construction is clearly interpreted as
the D1-D3 brane systems \cite{Diaconescu}. (See Fig. \ref{D1D3}.)
The situation with $k$ D1-brane and $N$ D3-brane
represents the $G=U(N),~k$ monopoles.
As in instanton case,
Bogomol'nyi equation and Nahm equation
are described as the BPS condition on
D3-branes and D1-branes, respectively.
The physical situation is unique and 
the equivalence between two kind of moduli spaces is trivial.

\begin{center}
\begin{figure}[htbn]
\epsfxsize=65mm
\hspace{5cm}
\epsffile{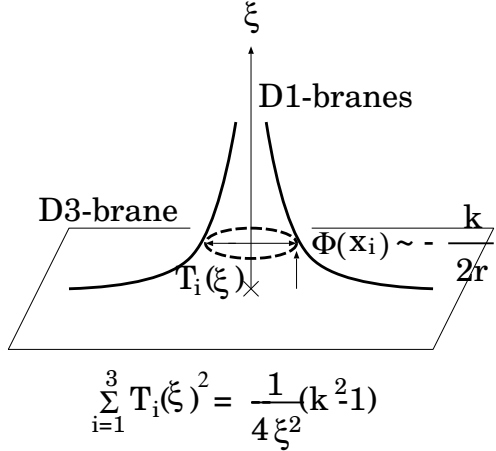}
\caption{Myers effect}
\label{myers}
\end{figure}
\end{center}

Let us consider the D-brane interpretation of
the correspondence of the 
boundary condition of the Higgs field and the Nahm data. 
On the D3-brane,
the boundary condition of the Higgs field shows
that D3-brane has a trumpet-like configuration because of
the pull-back by D1-brane.
On the other hand, on D1-brane, 
the diagonal components of $T_i$ shows the positions of the D1-branes.
However in $k>1$ case, we cannot diagonalize all $T_i$ at the same time
and cannot know all of the coordinates of D1-branes.
Instead, there is a condition for the second Casimir of $k$-dimensional
irreducible representation of $SU(2)$, 
that is, $\tau_1^2+\tau_2^2+\tau_3^2=(k^2-1)/4$ and hence
\begin{eqnarray}
 T_1^2+T_2^2+T_3^2 \st{\xi\rar \pm a/2}{\longr} \fr{1}{4\xi^2}(k^2-1).
\end{eqnarray}
This equation says that
the D1-branes have a funnel-like configuration 
near the D3-brane whose radius is $\sqrt{k^2-1}/2\xi$ (Fig. \ref{myers}).
This is in fact consistent 
with the result from 
the analysis of coincide multiple D-branes 
by using a non-abelian BI action \cite{CMT},
which strongly suggests the Myers' effect \cite{Myers}.

Next let us discuss noncommutative case.
Introducing the noncommutativity in $x^1$-$x^2$ plane
is equivalent to the presence of background $B$-field
(magnetic field) in the $x^3$ direction on the D3-brane.
Then the end of D1-brane is pulled back by the
magnetic field and finally the pulling force balances
the tension of the D1-brane and the D-brane system
becomes stable where the the slope of D1-brane
is constant \cite{HaHa, HHM, HaHi}. 
(See the lower right side of Fig. \ref{higgs}.)

The configuration of the Higgs fields 
(\ref{com_Dirac}) and (\ref{asymp}) are shown
like at the upper left and the upper right sides of 
Fig. \ref{higgs}, respectively.
Comparing the previous argument with the above D-brane interpretation
(The lower side of Fig. \ref{higgs}),
the singular behavior at the positive part of the $x_3$-axis
corresponds to the D1-brane which is considered as the
part of D3-brane.
The magnetic flux on $x_3\label{d_string}$-axis
is the ``shadow'' of the D1-brane \cite{GrNe}.
The slope of D1-brane is $- 1/\theta$ against 
``$x^i$-plane'' on the D3-brane
and $-\theta$ against $\xi$-axis,
which is very consistent (The lower right side of Fig. \ref{higgs})
and just coincides with that in commutative side
from the analysis of Born-Infeld action \cite{Moriyama, HHM2}.

\vs
\noindent
\unl{\it Nahm construction of $SU(N),~N\geq 3$ monopole 
and the D-brane interpretation}
\vs

\begin{center}
\begin{figure}[htbn]
\epsfxsize=100mm
\hspace{2cm}
\epsffile{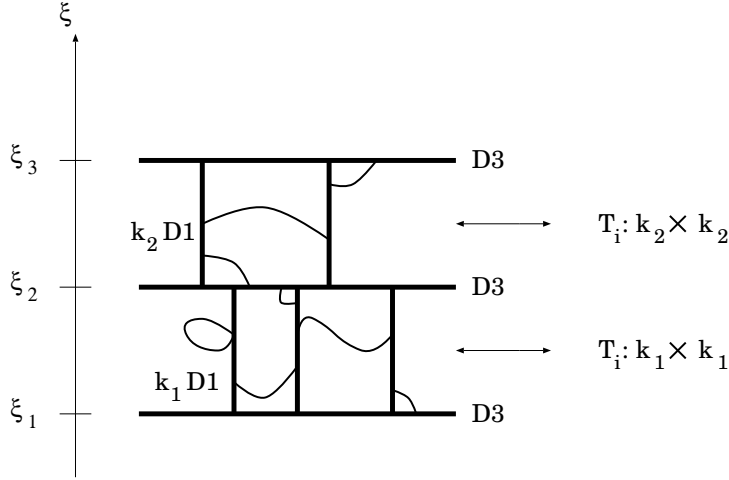} 
\caption{The D-brane interpretation of $U(3)$-monopole
(When $k_1=k_2$, the point $\xi=\xi_2$ shows the ``jumping point.'')}
\label{higher_rank}
\end{figure}
\end{center}

We give a brief introduction of Nahm construction
of $SU(N),~N\geq 3,~k$-monopole solution which corresponds to
the situation of $k$ D1-$N$ D3 brane system with $N\geq 3$ \cite{HuMu}. 
(See Fig. \ref{higher_rank}.)
The present discussion is basically commutative one,
however, also holds in noncommutative case.

Unlike $G=SU(2)$-monopole, there appear
the matrices $I,J$ in the ``0-dimensional Dirac operator''
as in ADHM construction:
\begin{eqnarray}
\label{weyl_op_caloron}
\nah:=
\left(
\ba{cc}
J&I^\dagger\\
\dis i\fr{d}{d\xi}-i(x_3-T_3)&-i(\zb_1-T_z^\dagger)\\
-i(z_1-T_z)&\dis i\fr{d}{d\xi}+i(x_3-T_3)
\ea
\right).
\end{eqnarray}
Here it is convenient to introduce the following symbols:
\begin{eqnarray}
\vec{V}\cdot\vec{V}^{\pr}&:=&\sum_{b=1}^{N_b}u_b^\dagger u_b^{\pr}
\delta(\xi-\xi_b)
+\vec{v}^\dagger\vec{v}^{\pr},\\
\bra \vec{V},\vec{V}^{\pr}\ket&:=&\int d\xi~
\vec{V}\cdot\vec{V}^{\pr}=\sum_{b=1}^{N_b}u_b^\dagger u_b^{\pr}
+\int d\xi~\vec{v}^\dagger\vec{v}^{\pr}.
\end{eqnarray}
Now Nahm data $T_i(\xi)$ are discontinuous with respect to $\xi$.
Though the size of $T_i$ is also variable at each interval of $\xi$,
here for simplicity, suppose that the size is the same.
The points $\xi=\xi_b$ where the D1-branes are attached from both side of 
the D3-brane is called ``jumping point,''
which depends on how the gauge group is broken.
(See Fig. \ref{higher_rank}.) The number $N_b$ denotes 
that of ``jumping points.''

Nahm equation is derived as the condition that
$\na\cdot\na$ commutes with Pauli matrices:
\begin{eqnarray}
\label{nahm_monopole}
&&[T_z,T_z^\dagger]+[\fr{d}{d\xi}+T_3,-\fr{d}{d\xi}+T_3]
+\sum_{b=1}^{N_b}(I_bI_b^\dagger-J_b^\dagger J_b)\delta(\xi-\xi_b)
=0,\nonumber\\
&&[T_z,\fr{d}{d\xi}+T_3]+\sum_{b=1}^{N_b}I_bJ_b\delta(\xi-\xi_b)=0. 
\end{eqnarray}
The steps are all the same as the usual Nahm construction.
Next we solve the ``1-dimensional Dirac equation''
\begin{eqnarray}
\label{weyl_monopole}
\na \cdot V
&=&
\sum_{b=1}^{N_b}
\left(
\ba{c}
J_b^\dagger\\I_b
\ea
\right)\uh_b\delta(\xi-\xi_b)\nn
&&+
\left(
\ba{cc}
\dis i\fr{d}{d\xi}+i(x_3-T_3)&i(\zb_1-T_{z}^\dagger)\\
i(z_1-T_z)&\dis i\fr{d}{d\xi}-i(x_3-T_3)
\ea
\right)
\left(\ba{c}v_1\\v_2\ea\right)
=0,\\
\label{norm_monopole}
\bra V,V\ket&=&1.
\end{eqnarray}
and construct the Higgs field and gauge fields which
satisfies the Bogomol'nyi equation
\begin{eqnarray}
\label{gauge_monopole}
\Phi=\bra V,\xi V \ket,~~~
A_i=\bra V,\del_i V \ket.
\end{eqnarray}

\vs
\noindent
{\bf Note}

\begin{itemize}
\item The boundary conditions in Nahm construction
are discussed from D-brane pictures in \cite{ChWe, KaSe, Tsimpis}
\end{itemize}

\subsection{Nahm Construction of the Fluxon}

\vs
\noindent
\unl{{\it U(1) BPS fluxon solution} ($k=1$)}
\vs

In the noncommutative Yang-Mills-Higgs theory,
there exists the special soliton corresponding to the
localized instantons.
Let construct it for $k=1$ for simplicity.

From the suggestion of caloron solutions,
this solution is considered as 
the noncommutative version of the monopole
with $\rho=\zeta=0$, that is, $D=0$.
Hence $\xi$ runs all real number and
there are ``jumping points.''
(Suppose $\xi_b=0$.)

\begin{itemize}

\item Step (i): The solution of Nahm equation is 
\begin{eqnarray}
\label{sol_nahm_mono}
I=J=0,~~~
T_i(\xi)=-\theta\delta_{i3}\xi.
\end{eqnarray}

\item Step (ii): The solution of ``1-dimensional Dirac equation'' is
\begin{eqnarray}
\label{v_monopole}
\Vh=
\left(
\ba{c}
\uh\\\vh_1\\\vh_2
\ea
\right)
=
\left(
\ba{c}
\Uh_k\\
f(\xi,x_3)
\vert 0\ket\bra 0\vert\\
0
\ea
\right),
\end{eqnarray}
where
\begin{eqnarray}
\label{f}
f(\xi,x_3)=\left(\fr{\pi}{\theta}\right)^{\qua}
\exp\left[-\fr{\theta}{2}\left(\xi+\fr{x_3}{\theta}\right)^2
\right].
\end{eqnarray}

\item Step (iii):  Substituting this to (\ref{gauge_monopole}),
we get the Higgs field and the gauge fields
which satisfies noncommutative Bogomol'nyi equation \cite{Hamanaka7}:
\begin{eqnarray}
\Phih&=&
\xi_1\Uh_1^\dagger \Uh_1+\left(\fr{\theta}{\pi}\right)^{\half}
\int_{-\infty}^{\infty}d\xi~
\left(\xi-\fr{x_3}{\theta}
\right)e^{-\theta\xi^2}
\vert 0\ket\bra 0\vert=-
\fr{x_3}{\theta}\vert 0\ket\bra 0\vert,\nonumber\\
\Ah_3&=&
\int_{-\infty}^{\infty}d\xi~\vh^\dagger 
\left(-\fr{x_3}{\theta}-\xi\right)\vh
=\left(-\fr{x_3}{\theta}-\Phih\right)
\vert 0\ket\bra 0\vert=0,\nonumber\\
\Dh_{z}&=&
\Uh_1^\dagger\hat{\del}_{z}\Uh_1.
\end{eqnarray}
This is a special soliton on noncommutative space which is 
called the {\it BPS fluxon} \cite{Polychronakos, GrNe2}.
The magnetic field is easily calculated as 
\begin{eqnarray}
\Bh_3&=&\frac{1}{\theta}\Ph_1,
~~~\Bh_1
=\Bh_2=0.
\label{fluxon_sol}
\end{eqnarray}
We can also take the Seiberg-Witten map to the configuration.
The D1-brane current density is calculated \cite{HaOo} as
\begin{eqnarray}
J_{\mbox{\scr{D1}}}(x)=\fr{1}{\theta}+\delta^{(2)}(z)
\delta\left(\Phi+\fr{x_3}{\theta}\right).
\end{eqnarray}
The configuration of the Higgs field and the distribution
of the magnetic field are as like in Fig. \ref{flux_fluxon}.

\begin{figure}[htbn]
\begin{center}
\epsfxsize=120mm
\hspace{-1cm}
\epsffile{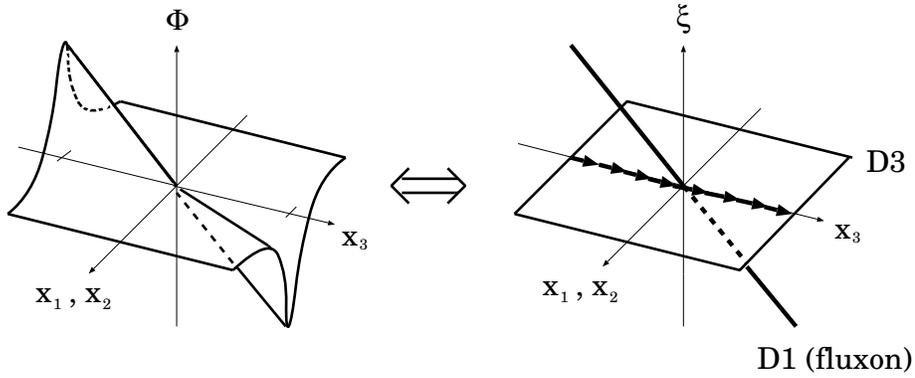}
\caption{The Higgs field of 1 fluxon (Left)
and the D-brane interpretation and the magnetic field (Right)}
\label{flux_fluxon}
\end{center}
\end{figure}  

The fluxon can be interpreted as the infinite
magnetic flux which appears on the positive part of
$x_3$-axis in noncommutative Dirac monopole
and is close to a flux rather than a monopole.
The tension of the flux is calculated as
$2\pi/g_{\scriptsize\mbox{YM}}^2\theta$ \cite{GrNe2}.

The generalization to $k$-fluxon solution with the moduli parameters
which show the positions of the fluxons 
are straightforwardly made \cite{Hamanaka7} as follows.

The Dirac zero-mode is
\begin{eqnarray}
\Vh=
\left(
\ba{c}
\uh\\\vh_1^{(m)}\\\vh_2^{(m)}
\ea
\right)
=
\left(
\ba{c}
\Uh_k\\
f^{(m)}(\xi,x_3)
\vert \alpha_z^{(m)}\ket\bra m\vert\\
0
\ea
\right),
\end{eqnarray}
where
\begin{eqnarray}
f^{(m)}(\xi,x_3)=\left(\fr{\pi}{\theta}\right)^{\qua}
\exp\left[-\fr{\theta}{2}\left(\xi+\fr{x_3-b_3^{(m)}}{\theta}\right)^2
\right].
\end{eqnarray}

The $k$-fluxon solution with the moduli parameters is
\begin{eqnarray}
\label{sol_monopole}
\Phih&=&
\xi_1\Uh_k^\dagger \Uh_k+\left(\fr{\theta}{\pi}\right)^{\half}
\sum_{m=0}^{k-1}\int_{-\infty}^{\infty}d\xi~
\left(\xi-\fr{x_3-b_3^{(m)}}{\theta}
\right)e^{-\theta\xi^2}
\vert m\ket\bra m\vert\no\\
&=&-\sum_{m=0}^{k-1}\left(
\fr{x_3-b_3^{(m)}}{\theta}\right)\vert m\ket\bra m\vert\no\\
\Ah_3&=&\bra\Vh,\del_3 \Vh\ket
=\int_{-\infty}^{\infty}d\xi~\vh^\dagger
\left(-\fr{x_3-b_3^{(m)}}{\theta}-\xi\right)\vh
=\sum_{m=0}^{k-1}\left(-\fr{x_3-b_3^{(m)}}{\theta}-\Phi^{(m)}\right)
\vert m\ket\bra m\vert\no\\
&=&0,\no\\
\Ah_{z}&=&\bra\Vh,\del_{z}\Vh\ket
=\Uh_k^\dagger\hat{\del}_{z}\Uh_k-\hat{\del}_{z}
-\sum_{m=0}^{k-1}\fr{\bar{\alpha}_z^{(m)}}{2\theta}
\vert m\ket\bra m\vert.
\end{eqnarray}

The D1-brane current density is calculated \cite{HaOo} as
\begin{eqnarray}
J_{\mbox{\scr{D1}}}(x)=\fr{1}{\theta}
+\sum_{m=0}^{k-1}\delta^{(2)}(z-\alpha_z^{(m)})
\delta\left(\Phi+\fr{x_3-b_3^{(m)}}{\theta}\right).
\end{eqnarray}

When we apply the ``solution generating technique'' 
to Bogomol'nyi equation, 
we have to find a modification or a trick 
on the transformation of 
the Higgs field \cite{HaTe, Hashimoto} (cf. section 6.2).
On the other hand, Nahm construction naturally 
shows the modification part as in (\ref{sol_monopole}).

\end{itemize}

\newpage

\section{Calorons and D-branes}

In section 3 and 4, we treat instantons and monopoles separately.
In fact, monopoles are considered as the Fourier-transformed
instantons in some sense,
which is clearly understood 
from the T-duality transformation of D0-D4 brane systems.
In this section,
we discuss the reasons introducing
periodic instantons which corresponds to
D0-D4 brane systems on $\R^3\ti S^1$ which is called {\it calorons}.
We do not examine the detailed properties but
just give the D-brane interpretation of it and 
take the T-duality transformation.

\subsection{Instantons on $\R^3\ti S^1$ (=Calorons) and T-duality}

Calorons are periodic instantons in one direction, 
that is, instantons on $\R^3\times S^1$.
They were first constructed explicitly in \cite{HaSh}
as infinite number of 't Hooft instantons 
periodic in one direction
and used for the discussion on non-perturbative aspects of
finite-temperature field theories \cite{HaSh, GPY}.
Calorons can intermediate between instantons and monopoles 
and coincide with them in the limits
of $\beta\rar \infty$ and $\beta\rar 0$ respectively 
where $\beta$ is the perimeter of $S^1$ \cite{Rossi}.
Hence calorons also can be 
reinterpreted clearly from D-brane picture \cite{LeYi}
and constructed by Nahm construction 
\cite{Nahm6, KrvB, LeLu, BrvB}. 

The D-brane pictures of them are the following. (See Fig. \ref{caloron}.)
Instantons and monopoles are represented as D0-branes on D4-branes
and D-strings ending to D3-branes respectively.
Hence calorons are represented as D0-branes on D4-branes
lying on $\R^3\ti S^1$.

\begin{center}
\begin{figure}[htbn]
\epsfxsize=105mm
\hspace{3cm} 
\epsffile{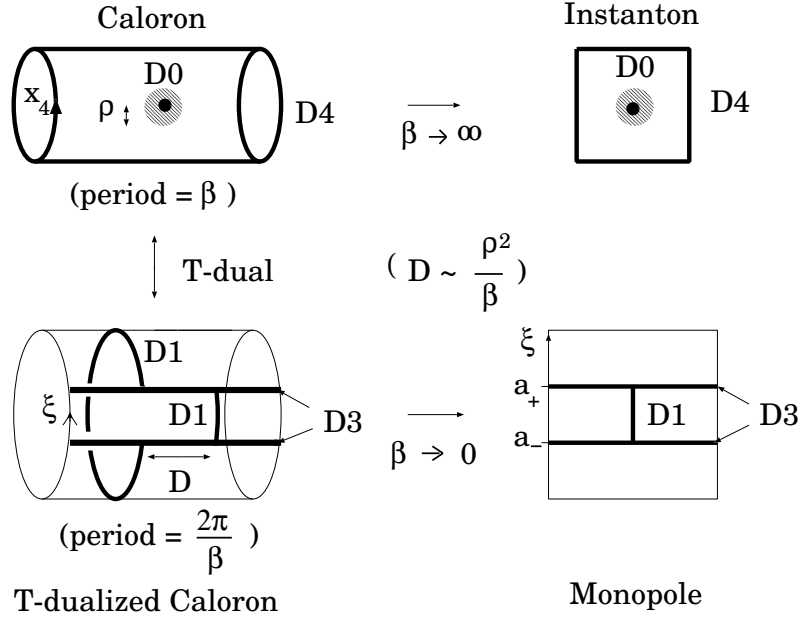}
\caption{The D-brane description of $U(2)$ 1 caloron.}
\label{caloron}
\end{figure}
\end{center}

In the T-dualized picture, $U(N)$ 1 caloron can be interpreted
as $N-1$ fundamental monopoles and the $N$-th monopole which appears 
from the Kaluza-Klein sector \cite{LeYi}.
The value of the fourth component of the gauge field
at spatial infinity
on D4-brane determines the positions of the D3-branes
which denote the Higgs expectation values of the monopole.
The positions of the D3-branes are called the jumping points
because at these points, the D1-brane is generally separated.
In $N=2$ case, the separation interval 
(see Fig. \ref{caloron}) $D$ satisfies $D\sim \rho^2/\beta$ \cite{LeYi, LeLu},
and if the size $\rho$ of periodic instanton is fixed and
the period $\beta$ goes to zero,
then one monopole decouples and 
the situation exactly coincides with that of PS-monopole \cite{PrSo}.
BPS fluxons are represented as infinite D-strings piercing D3-branes
in the background constant $B$-field 
and considered to be the T-dualized noncommutative calorons 
in the limit with the period $\beta\rar0$
and the interval $D\rar 0$, which suggests $\rho =0$. 

\subsection{NC Calorons and T-duality}

In this subsection, 
we construct the noncommutative caloron solution by putting
infinite number of localized instantons in the one direction
at regular intervals.

\vspace{2mm}
\noindent
\unl{\it localized U(1) 1 caloron}
\vspace{2mm}

Now let us construct a localized caloron solution as commutative
caloron solution in section 3.1, that is, we take the instanton number
$k\rar \infty$ and put
infinite number of localized instantons in the $x_4$ direction
at regular intervals. We have to find an appropriate shift operator
so that it gives rise to an infinite-dimensional projection operator
and put the moduli parameter $b_4$ periodic.

The solution is found as:
\begin{eqnarray}
\label{loc_caloron}
\Ah_{z_1}&=&\Uh_{k\ti\infty}^\dagger \delh_{z_1} \Uh_{k\ti\infty}-\delh_{z_1}
-\sum_{m=0}^{k-1}
\fr{\bar{\alpha}_1^{(m)}}{2\theta}\vert m\ket\bra m\vert\ot 1_{\cH_2},
\no\\
\Ah_{z_2}&=&\Uh_{k\ti\infty}^\dagger \delh_{z_2} \Uh_{k\ti\infty}-\delh_{z_2}
+\sum_{m=0}^{k-1}\sum_{n=-\infty}^{\infty}
\fr{\bar{\alpha}_2^{(m)}-in\beta}
{2\theta}\vert m\ket\bra m\vert\ot \vert n\ket\bra n\vert,
\end{eqnarray}
where the shift operator is defined as
\begin{eqnarray}
\Uh_{k\times \infty}
=\sum_{n_1=0}^{\infty}\vert n_1\ket\bra n_1+k\vert\ot1_{\cH_2}.
\end{eqnarray}
The field strength is calculated as
\begin{eqnarray}
\label{f_cal}
\Fh_{12}=-\Fh_{34}=i\fr{1}{\theta}\Ph_k\ot 1_{\cH_2},
\end{eqnarray}
which is trivially periodic in the $x_4$ direction.
It seems to be strange
that  this contains no information of the period $\beta$.
Hence one may wonder if this solution is the charge-one caloron
solution on $\R^3\ti S^1$ whose perimeter is $\beta$.
Furthermore one may doubt if
this suggests that this soliton represents D2-brane
not infinite number of D0-branes.

The apparent paradox is solved by mapping this solution to commutative side
by exact Seiberg-Witten map.
The commutative description of D0-brane density
is as follows
\begin{eqnarray}
\label{d0density2}
J_{\scr\mbox{D0}}(x)=\fr{2}{\theta^2}
+\sum_{m=0}^{k-1}\sum_{n=-\infty}^{\infty}
\delta^{(2)}(z_1-\alpha_1^{(m)})
\delta^{(2)}(z_2-\alpha_2^{(m)}-in\beta).
\end{eqnarray}
The information of the period has appeared
and the solution (\ref{loc_caloron}) is
shown to be an appropriate charge-one caloron solution
with the period $\beta$.
The above paradox is due to the fact that in noncommutative gauge theories,
there is no local observable and the period becomes obscure.
And as is pointed out in \cite{HaOo},
the D2-brane density is exactly zero.
Hence the paradox has been solved clearly.

This soliton can be interpreted as a localized instanton
on noncommutative $\R^3\ti S^1$.
It is interesting to study the relationship between our
solution and that in \cite{DDRS}.

\vspace{2mm}
\noindent
\unl{\it localized U(1) 1 doubly-periodic instantons}
\vspace{2mm}

In similar way, we can construct doubly-periodic (in the $x_3$ and
$x_4$ directions) instanton solution:
\begin{eqnarray}
\label{double}
\Ah_{z_1}&=&\Uh_{k\ti\infty}^\dagger \delh_{z_1} \Uh_{k\ti\infty}-\delh_{z_1}
-\sum_{m=0}^{k-1}
\fr{\bar{\alpha}_1^{(m)}}{2\theta}\vert m\ket\bra m\vert\ot 1_{\cH_2},
\no\\
\Ah_{z_2}&=&\Uh_{k\ti\infty}^\dagger \delh_{z_2} \Uh_{k\ti\infty}-\delh_{z_2}
\no\\
&&+\sum_{m=0}^{k-1}\sum_{n_1,n_2=-\infty}^{\infty}
\fr{\bar{\alpha}_2^{(m)}+\beta_1n_1-i\beta_2n_2}
{2\theta}\vert m\ket\bra m\vert\ot \vert
\widetilde{\alpha}_{n_1n_2}^{(l_1,l_2)}
\ket\bra \widetilde{\alpha}_{n_1n_2}^{(l_1,l_2)}\vert,
\end{eqnarray}
where the system
$\left\{\vert \widetilde{\alpha}^{(l_1,l_2)}_{n_1,n_2}
\ket\right\}_{n_1,n_2\in\scr\Z}$
is von Neumann lattice \cite{vonNeumann} and an orthonormal and
complete set \cite{Perelov, BBGK}\footnote{
To make this system complete, the sum over the labels $(n_1,n_2)$
of von Neumann lattice is taken removing some one pair.
We apply this summation rule
to the doubly-periodic instanton solution (\ref{double}).}.
Von Neumann lattice is the complete subsystem of the set of
the coherent states which is over-complete,
and generated by $e^{l_1\delh_3}$ and $e^{l_2\delh_4}$,
where the periods of the lattice $l_1,l_2 \in \R$
satisfies $l_1 l_2=2\pi\theta$.
(See also \cite{BGZ, GHS}.)
This complete system has two kind of labels and suitable to
doubly-periodic instanton.
Of course, another complete system can be available if one label the
system appropriately.

The field strength in the noncommutative side
is the same as (\ref{f_cal}) and
the commutative description of D0-brane density becomes
\begin{eqnarray}
\label{d0density3}
J_{\scr\mbox{D0}}(x)&=&\fr{2}{\theta^2}
+\sum_{m=0}^{k-1}\sum_{n_1,n_2=-\infty}^{\infty}
\delta^{(1)}(z_1-\alpha_1^{(m)})
\delta^{(2)}(z_2-\alpha_2^{(m)}-n_1\beta_1-in_2\beta_2),
\end{eqnarray}
which
guarantees that this is an appropriate charge-one doubly-periodic
instanton solution
with the period $\beta_1,\beta_2$.

This soliton can be interpreted as a localized instanton on
noncommutative $\R^2\ti T^2$.
The exact known solitons on noncommutative torus are very
refined or abstract as is found in
\cite{GHS, Boca, KrSc, KMT}.
It is therefore notable that our simple solution (\ref{double}) is
indeed doubly-periodic. The point is that
we treat noncommutative $\R^4$ not noncommutative torus
and apply ``solution generating technique'' to $\cH_1$ side
only.

\subsection{Fourier Transformation of Localized Calorons}

Now we discuss the Fourier transformation
of the gauge fields of localized caloron
and show that the transformed configuration exactly coincides with
the BPS fluxon in the $\beta\rar 0$ limit.
This discussion is similar to that
the commutative caloron exactly
coincides with PS monopole in the $\beta\rar 0$ limit
up to gauge transformation as in the end of section 3.1,.

\begin{center}
\begin{figure}[htbn]
\epsfxsize=105mm
\hspace{3cm}
\epsffile{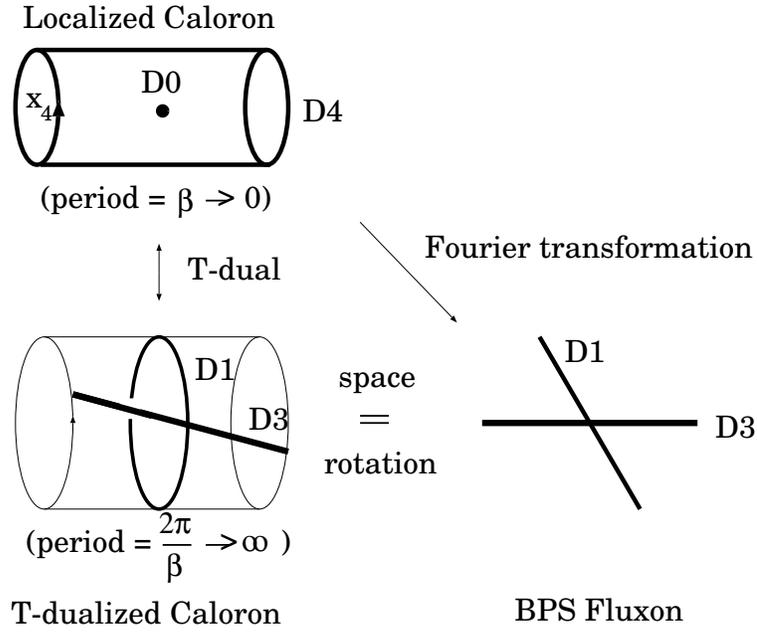}
\caption{Localized U(1) 1 caloron
and the relation to BPS fluxon}
\label{t-dual}
\end{figure}
\end{center}

The Fourier transformation can be defined by
\begin{eqnarray}
\label{fourier}
\hat{1}_{\cH_2}&\rar& 1,~~~\xh_{3,4}\hat{1}_{\cH_2}\rar x_{3,4},\no\\
\Ah_\mu &\rar&\widetilde{\Ah_\mu^{[l]}}=\lim_{\beta\rar 0}\fr{1}{\beta}
\int_{-\fr{\beta}{2}}^{\fr{\beta}{2}}dx_4~e^{2\pi il\fr{x_4}{\beta}}\Ah_\mu.
\end{eqnarray}
In the  $\beta\rar 0$ limit, only $l=0$ mode survives and the
Fourier transformation (\ref{fourier}) becomes trivial.
Then we rewrite these zero modes $\widetilde{\Ah_i^{[0]}}$
and $i\widetilde{\Ah_4^{[0]}}$ as
$\Ah_i$ and $\Phih$ in $(3+1)$-dimensional
noncommutative gauge theory respectively.
Noting that in the localized caloron solution (\ref{loc_caloron}),
$\Uh_{k\ti\infty}^\dagger \delh_{z_2} \Uh_{k\ti\infty}-\delh_{z_2}=
\Ph_k\ot\hat{1}_{\cH_2}(\zbh_2/2\theta_2)$, where the $\Ph_k$
is the same as the projection in (\ref{proj}),
the transformed fields are easily calculated as follows:
\begin{eqnarray}
\Ah_{z_1}&=&\Uh^\dagger_k \delh_{z_1}\Uh_k-\delh_{z_1}-\sum_{m=0}^{k-1}
\fr{\bar{\alpha}_{z_1}^{(m)}}{2\theta_1}\vert m\ket\bra m\vert,\no\\
\Ah_3&=&i\sum_{m=0}^{k-1}\fr{b_4^{(m)}}{\theta_2}
\vert m\ket\bra m\vert,\no\\
\Phih&=&\sum_{m=0}^{k-1}\fr{x_3-b_3^{(m)}}{\theta_2}
\vert m\ket\bra m\vert.
\end{eqnarray}
The Fourier transformation (\ref{fourier})
also reproduces the anti-self-dual
BPS fluxon
rewriting $\theta_1$, $\theta_2$ and $z_1$
as $\theta$, $-\theta$ and $z$ respectively.
We note that the anti-self-dual condition of the noncommutative parameter
$\theta_1+\theta_2=0$ in the localized caloron would
correspond to the anti-self-dual condition of the BPS fluxon.
In the D-brane picture,
the Fourier transformation (\ref{fourier}) can be considered
as the composite
of T-duality in the $x_4$ direction and the space rotation
in $x_3$-$\Phi$ plane \cite{HaHi,Moriyama2, HHM2}.
(cf. Fig. \ref{t-dual})

\newpage

\section{NC Solitons and D-branes}

So far, we have discussed mainly the 
Yang-Mills(-Higgs) theories which correspond to
the gauge theories on D-branes in the decoupling limit.
{}From now on, we treat other noncommutative theories.
In this section,
we discuss the applications of noncommutative
solitons to the problems on tachyon condensations,
which was a breakthrough in the understanding of
non-perturbative aspects of D-branes.

\subsection{Gopakumar-Minwalla-Strominger (GMS) Solitons}

In this subsection, we briefly review
the Gopakumar-Minwalla-Strominger (GMS) solitons
which are the special scalar solitons
in the $\theta\rar \infty$ limit. 
The structure is very simple and 
easy to be applied to tachyon condensations.

Let us consider the Yang-Mills-Higgs theory 
on the noncommutative $(2+1)$-dimensional space-time:
\begin{eqnarray}
\label{2ymh}
I=\int dt d^2 x \left(-\fr{1}{4}F_\mn F^\mn
+\half D_\mu \Phi D^\mu\Phi+V(\Phi)\right),
\end{eqnarray}
where the Higgs field $\Phi$ belongs to the adjoint representation of 
the gauge group and the potential term $V(\Phi)$ is a polynomial 
in $\Phi$:
\begin{eqnarray}
V(\Phi)=\fr{m^2}{2}\Phi^2+c_1 \Phi^3+\cdots
\end{eqnarray}
Now let us take the scale 
transformation $x^i\rar \sqrt{\theta}x^i, 
A_\mu \rar \sqrt{\theta^{-1}} A_\mu$
and the $\theta\rar \infty$ limit,
then the kinetic terms in the action drop out and
the action (\ref{2ymh}) is reduced to the simple one:
\begin{eqnarray}
I=\int dt d^2 x~V(\Phi).
\end{eqnarray} 
The equation of motion is easily obtained:
\begin{eqnarray}
\fr{dV}{d\Phi}=c\Phi(\Phi-\lambda_1)\cdots(\Phi-\lambda_n)=0.
\end{eqnarray}
On commutative spaces, the solution is trivial: $\Phi=\lambda_i$.
However, on noncommutative space, there is a simple, but non-trivial
solution:
\begin{eqnarray}
\Phi=\lambda_i P
\end{eqnarray}
where $P$ is a projection.
The typical example is found in operator formalism:
\begin{eqnarray}
\Phi=\lambda_i \vert 0\ket\bra 0\vert.
\end{eqnarray}
This solution has the Gaussian distribution in star-product formalism
and hence has a localized energy.
This configuration is stable as far as $\theta\rar \infty$,
which guarantees this is a soliton solution called
the {\it GMS solitons}.

The action (\ref{2ymh}) is equivalent to the
effective action of D2-brane in the decoupling limit.
Hence the solitons are considered as the D0-branes
on the D2-branes.
This is confirmed by the coincidence of the energy and
the spectrum of the fluctuation around the soliton configuration,
which makes the studies of 
noncommutative solitons and tachyon condensations
joined. (For other discussion on noncommutative solitons and D-branes,
see e.g. \cite{AcSo, CIMM, FIO, Itoyama}.)

\subsection{The Solution Generating Technique}

The ``solution generating technique'' 
is a transformation which leaves an equation 
as it is, that is, one of the auto-B\"acklund
transformations.
The transformation is almost a gauge transformation
and defined as follows:
\begin{eqnarray}
\label{hkl}
\Dh_z\rar \Uh^\dagger \Dh_z \Uh,
\end{eqnarray}
where $\Uh$ is an almost unitary operator and satisfies
\begin{eqnarray}
\label{isometry}
\Uh \Uh^\dagger=1.
\end{eqnarray}
We note that we don't put $\Uh^\dagger\Uh=1$.
If $\Uh$ is finite-size, $\Uh\Uh^\dagger=1$
implies $\Uh^\dagger \Uh=1$ and then $\Uh$ and
the transformation (\ref{hkl}) become a unitary operator and
just a gauge transformation respectively.
Now, however, $\Uh$ is infinite-size
and we only claim that $\Uh^\dagger\Uh$ is a projection
because $(\Uh^\dagger\Uh)^2=\Uh^\dagger(\Uh
\Uh^\dagger)\Uh=\Uh^\dagger\Uh$.
Hence the operator $\Uh$ is the partial isometry (\ref{partial_iso}).

The transformation (\ref{hkl}) generally
leaves an equation of motion as it is \cite{HKL}:
\begin{eqnarray}
\fr{\delta I}{\delta\cO}\rar\Uh^\dagger\fr{\delta I}{\delta\cO}\Uh,
\end{eqnarray}
where $I$ and $\cO$ are the Lagrangian and the field in the
Lagrangian.
Hence if one prepares a known solution of
the equation of motion $\delta I/\delta\cO=0$, then we can get various
new solution of it by applying the transformation (\ref{hkl}) to
the known solution.
The new soliton solutions from vacuum solutions
are called {\it localized solitons}.
The dimension of the projection $\Ph_k$ in fact
represents the charge of the localized solitons.
In general, the new solitons generated from known solitons
by the ``solution generating technique'' are
the composite of known solitons and localized solitons.

The ``solution generating technique'' (\ref{hkl}) can be generalized
so as to include moduli parameters.
In U(1) gauge theory,
the generalized transformation becomes as follows:
\begin{eqnarray}
\label{hkl2}
\Dh_z\rar \Uh_k^\dagger \Dh_z \Uh_k-\sum_{m=0}^{k-1}
\fr{\bar{\alpha}_z^{(m)}}{2\theta}\vert m\ket\bra m\vert,
\end{eqnarray}
where $\alpha_z^{(m)}$ is an complex number
and represents the position of the $m$-th localized soliton.

This technique is all found by hand.
However as we saw in section 3.2,
ADHM construction naturally gives rise to
all elements in the solution generating technique 
including moduli parameters.
Next we will see how strong the solution generating technique 
is to generate new soliton solution,
how simple the solution is to be calculated,
and how well it fits to D-brane interpretation including
matrix models. 

\vs
\noindent
\unl{\bf Application to Sen's Conjecture on Tachyon Condensations}
\vs

\begin{figure}[htbn]
\begin{center}
\epsfxsize=70mm
\hspace{2cm}
\epsffile{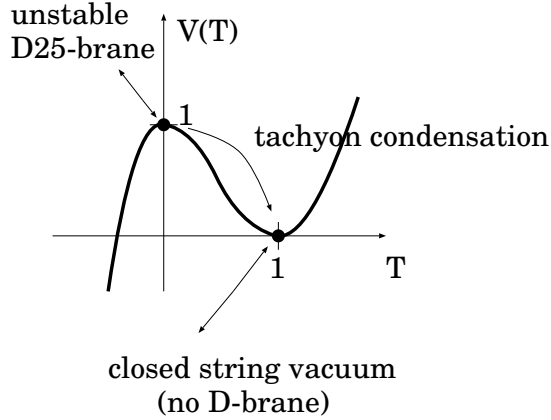}
\caption{The tachyon potential}
\label{tachyon}
\end{center}
\end{figure}

For simplicity, let us consider the bosonic
effective theory of a D25-brane 
in the background constant $B$-field 
whose non-zero component is $B_{24,25}~(=:-b<0$)\footnote{
Here 24-25 plane is supposed to be noncommutative. 
The variable $z$ is 
the complex coordinate of this plane.}:
\begin{eqnarray}
\label{eff}
I&=&\frac{T_{\scriptsize\mbox{D25}}g_s}{G_s}
\int d^{24}x~(2\pi \theta {\mbox{Tr}}_{\cal H})~{\cal L},\\
{\cal L}&=&-V(T-1)\sqrt{-\det(G_{\mu\nu}+2\pi\alpha^{\prime}
(F+\Phi)_{\mu\nu})}\nonumber\\
&&+\frac{1}{2}\sqrt{G}f(T-1)[D^\mu, T][D_\mu, T]
+(\mbox{higher derivative terms of }F),
\end{eqnarray}
where
\begin{eqnarray}
&&G_{\mu\nu}
={\mbox{diag}}(1,-1,\cdots,-1,-(2\pi\alpha^{\prime}b)^{2},
-(2\pi\alpha^{\prime}b)^{2}),~~~
G_s=g_s(2\pi\alpha^{\prime}),\nonumber\\
&&\theta^{24,25}=:\theta=\fr{1}{b},~~~
F_{24,25}+\Phi_{24,25}=\fr{1}{\theta}[D_z,D_z^\dagger]\nonumber,
\end{eqnarray}
where $T_{\scriptsize\mbox{Dp}}$ denotes the tension of the D$p$-brane
and $\mu,\nu=0,\ldots,25$.
This effective action is obtained by
remaining massless tachyon fields $T$ and gauge fields $A_\mu$,
integrating out the other massive fields,
and imposing the ordinary gauge symmetry\footnote{
We have to make further discussion on the Born-Infeld part.
However we do not need the details here.}.
Let us suppose that the tachyon potential $V(T)$
has the following shape like Fig. \ref{tachyon}

Following Sen's conjecture
\footnote{For a review see \cite{Sen}.},
the part at the valley ($T=1$) corresponds to
the closed string vacuum where there is no D-branes
because there is no open strings.
The important point here is that
the ``solution generating technique''
leaves the equation of motion as it is
independent of the details of the action (\ref{eff})
because of the gauge invariance of it.

That is why we can get non-trivial exact solution of the
effective theory of SFT very easily
by applying the solution generating technique (\ref{hkl}) to
the vacuum solution $\Th=1,~\hat{D}_z=\hat{\partial}_z,
~\Ah_i=0,~(i=0,\ldots,23)$
even if we know {\it no} parts of the action:
\begin{eqnarray}
\Th=\Uh_k^\dagger 1 \Uh_k=1-\Ph_k,
~\hat{D}_z=\Uh_k^\dagger \hat{\partial}_z \Uh_k,
~\Ah_i=0.
\end{eqnarray}
The tension of this solution is easily calculated
\begin{eqnarray}
(\mbox{Tension})=(2\pi)^2\alpha^\prime k T_{\scriptsize\mbox{D25}}
=kT_{\scriptsize\mbox{D23}},
\end{eqnarray}
which shows that this localized configuration
have the same tension as that of $k$ D23-branes!
The fluctuation spectrum is also coincident with
that of D23-brane, which are both evidences that
this noncommutative soliton solution is just D23-branes!
This implies an {\it exact} confirmation of Sen's conjecture
that an unstable D25-brane decays into D23-brane
by the tachyon condensation
in the context of the effective theory of SFT.

\vs
\noindent
\unl{\bf Application to NC Bogomol'nyi Equation}
\vs

Now let us apply the solution generating technique to BPS equations.
Unlike EOM, BPS equations contain constants in general
and therefore do not be transformed covariantly under the 
transformation (\ref{hkl}).

Here we introduce some results on this problems.
focusing on the noncommutative Bogomol'nyi equation for $G=U(1)$ here.
The following modified transformation
leaves the noncommutative Bogomol'nyi equation 
as it is \cite{HaTe, Hashimoto}:
\begin{eqnarray}
\label{ours}
\Phi&\rightarrow& \Uh_k^\dagger\Phi \Uh_k-\sum_{m=0}^{k-1}
\frac{x_3-b_3^{(m)}}{\theta}\vert m\rangle\langle m \vert,\nonumber\\
D_3&\rightarrow&\partial_3+\Uh_k^\dagger A_3 \Uh_k
+i\sum_{m=0}^{k-1}\frac{b^{(m)}_4}{\theta}
\vert m\rangle\langle m \vert,\nonumber\\
D_z&\rightarrow& \Uh_k^\dagger D_z \Uh_k-\sum_{m=0}^{k-1}
\frac{\bar{\alpha}^{(m)}_z}{2\theta}\vert m\rangle\langle m \vert.
\end{eqnarray}
The crucial modification part appears in the 
transformation law of the Higgs field, which is,
interestingly, seen naturally in the fluxon solutions (\ref{fluxon_sol})
by Nahm construction.

\begin{figure}[htbn]
\epsfxsize=90mm
\hspace{3.5cm}
\epsffile{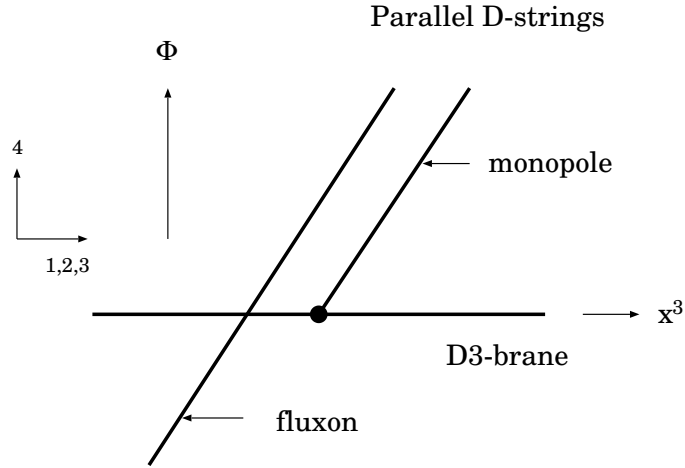}
\caption{Bound state at threshold of an
Abelian monopole and $k$ fluxons ($k$=1).}
\label{composite}
\end{figure}

We can generate various new BPS soliton solutions 
from known solutions.
For example, from the noncommutative 1-Dirac monopole solution 
(\ref{nc_monopole}),
we get the following new solution by the BPS solution generating 
technique (\ref{ours}) \cite{HaTe}:
\begin{eqnarray}
\label{new_mono}
\Phi^{\scriptsize\mbox{new}}&=&
-\left\{\sum_{n=k+1}^{\infty}(\xi_{n-k}^2-\xi_{n-k-1}^2)
\vert n\rangle\langle n \vert
+\left(\xi_0^2+\frac{x_3}{\theta}\right)\vert k\rangle\langle k \vert
+\sum_{m=0}^{k-1}
\left(\frac{x_3-b^{(m)}_3}{\theta}\right)
\vert m\rangle\langle m \vert\right\}, \nonumber\\
D_z^{\scriptsize\mbox{new}}
&=&\frac{1}{\sqrt{2\theta}}\sum_{n=k}^{\infty}\sqrt{\frac{n+1-k}{n+1}}
\frac{\xi_{n-k}}{\xi_{n+1-k}}a^\dagger\vert n\rangle\langle n \vert
-\sum_{m=0}^{k-1}\fr{\bar{\alpha}^{(m)}_z}{2\theta}
\vert m\rangle\langle m \vert,\nonumber\\
A_3^{\scriptsize\mbox{new}}&=&i\sum_{m=0}^{k-1}\fr{b_4^{(m)}}{\theta}
\vert m\rangle\langle m \vert.
\end{eqnarray}
This is the composite of a noncommutative Dirac monopole
and $k$ fluxons (See Fig. \ref{composite}).

Here let us interpret this transformation
from the viewpoints of matrix models
\cite{BFSS, IKKT, AGMS, AIIKKT}.
The new solution has the following matrix representation
setting $b_3=0$: 
\begin{eqnarray}
\label{new_matrix}
~~~D_z^{\scriptsize\mbox{new}}&=&\Uh^\dagger_k\left(\sum_{m,n=0}^{\infty}
(\Dh_z^{\scriptsize\mbox{original}})_{m,n}
\vert m\rangle\langle n\vert\right)
\Uh_k
-\sum_{m=0}^{k-1}\frac{\bar{\alpha}^{(m)}_z}{2\theta}
\vert m\rangle\langle m \vert,\nonumber\\
&=&\sum_{m,n=k}^{\infty}
(\Dh_z^{\scriptsize\mbox{original}})_{m-k,n-k}
\vert m\rangle\langle n\vert
-\sum_{m=0}^{k-1}\frac{\bar{\alpha}^{(m)}_z}{2\theta}
\vert m\rangle\langle m \vert,\nonumber\\
&=&
\left(
\begin{array}{ccc|ccc}
\displaystyle-\frac{\bar{\alpha}_z^{(0)}}{2\theta}&~&O&~&~&~\\
~&\ddots&~&~&O&~\\
O&~&\displaystyle-\frac{\bar{\alpha}_z^{(k-1)}}{2\theta}&~&~&~\\\hline
~&~&~&~&~&~\\
~&O&~&~&\Dh^{\scriptsize\mbox{original}}_z&~\\
~&~&~&~&~&~
\end{array}
\right)\nn
\Phih^{\scriptsize\mbox{new}}
&=&
\left(
\begin{array}{ccc|ccc}
\displaystyle -\frac{x_3}{\theta}&~&O&~&~&~\\
~&\ddots&~&~&O&~\\
O&~&\displaystyle-\frac{x_3}{\theta}&~&~&~\\\hline
~&~&~&~&~&~\\
~&O&~&~&\Phih^{\scriptsize\mbox{original}}&~\\
~&~&~&~&~&~
\end{array}
\right).
\end{eqnarray}
The transformed configuration can be
interpreted as the composite
of the original configuration (basically a D3-brane)and
the additional $k$ fluxons 
(unbounded $k$ D1-branes).
The upper-left $k\times k$ part and 
the lower-right part correspond to
the additional independent $k$ D1-branes and 
the original D3-brane as the bound state of
infinite D1-branes, respectively. 
The zero components in the off-diagonal parts
show the no-bound between the original configuration
and the $k$ fluxons. (See the right side of Fig. \ref{flux_fluxon}.)
The diagonal elements of the upper-left $k\times k$ part
are the Higgs vacuum expectation value (VEV) on the D1-branes
and represents the positions of $k$ D1-branes.
That is why the parameters $\bar{\alpha}_z^{(m)}$
are the moduli parameters which shows the positions of the 
fluxons, which is consistent with the previous results.
The linear terms $-x_3/\theta$ in the Higgs field $\Phi$
gives rise to the slope $-1/\theta$ of D1-brane
in the $x_3$ direction
and play the crucial role so that 
the transformed configuration should be BPS.

\begin{figure}[htbn]
\epsfxsize=100mm
\hspace{3cm}
\epsffile{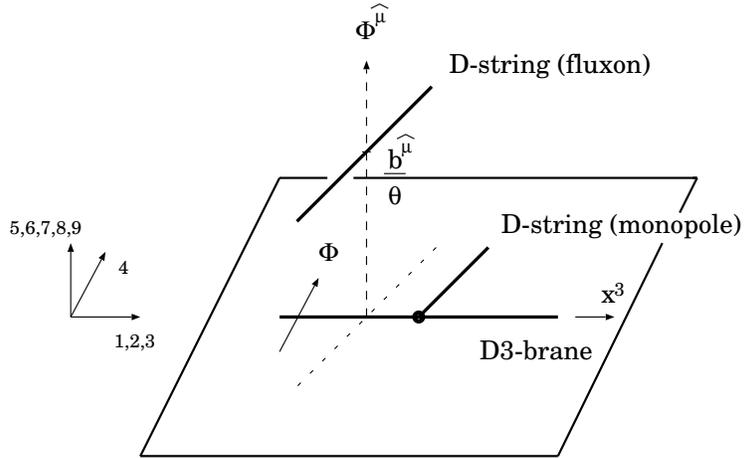}
\caption{Fluxons can escape from the D3-brane.}
\label{small2}
\end{figure}

We have set the transverse coordinates $\Phi^\mh=0$
in the last paragraph in section 2.
After the transformation, however, we can take $\Phi^\mh\neq 0$
keeping the BPS condition.
For example, to the general solutions
(\ref{new_mono}), we can set
\begin{eqnarray}
\label{transverse}
\Phi_\mh=\sum_{m=0}^{k-1}\fr{b^{(m)}_\mh}{\theta} 
\vert m\ket\bra m\vert=\left(
\begin{array}{ccc|ccc}
\dis \fr{b^{(0)}_\mh}{\theta}&~&O&~&~\\
~&\ddots&~&~&O&~\\
O&~&\dis \fr{b^{(k-1)}_\mh}{\theta}&~\\\hline
~&~&~&~&~&~\\
~&O&~&~&O&~\\
~&~&~&~&~&~
\end{array}\right),
\end{eqnarray}
where $b^{(m)}_\mh,~\mh=5,\ldots,9$ are real constants and
denote the $\mh$-th transverse
coordinates of the $m$-th fluxon. This shows that the $k$ fluxons can
escape from the D3-brane. (See Fig. \ref{small2}.)

\newpage

\section{Towards NC Soliton Theories and NC Integrable Systems}

In this section, we discuss noncommutative extension
of integrable systems.
We believe that this study would pioneer 
new area of integrable systems.

In star-product formalism,
noncommutative theories are considered as deformed theories
from commutative ones.
Under the NC-deformation,
the (anti-)self-dual (ASD) Yang-Mills equations
could be considered to preserve the integrability
in the same sense as in commutative cases \cite{KKO, Nekrasov}.
On the other hand, with regard to typical integrable equations
such as the Korteweg-de Vries (KdV) equation \cite{KdV} and
the Kadomtsev-Petviasfvili (KP) equation \cite{KaPe},
naive noncommutative extension generally destroys the integrability.
There is known to be a method, the {\it bicomplex method}, to yield
noncommutative integrable equations which have many conserved quantities
\cite{Bicomplex, DiMH, DiMH2, DiMH3, GrPe}.
There are many other works on noncommutative integrable systems,
for example, \cite{Bieling, CCMo, CLMS, CFZ, FSIv, FIY, Hannabuss, HLW, 
LePo, LePo2, LPS, Legare, Paniak, Takasaki, Wolf}.

\begin{figure}[htbn]
\epsfxsize=110mm
\hspace{2.5cm}
\epsffile{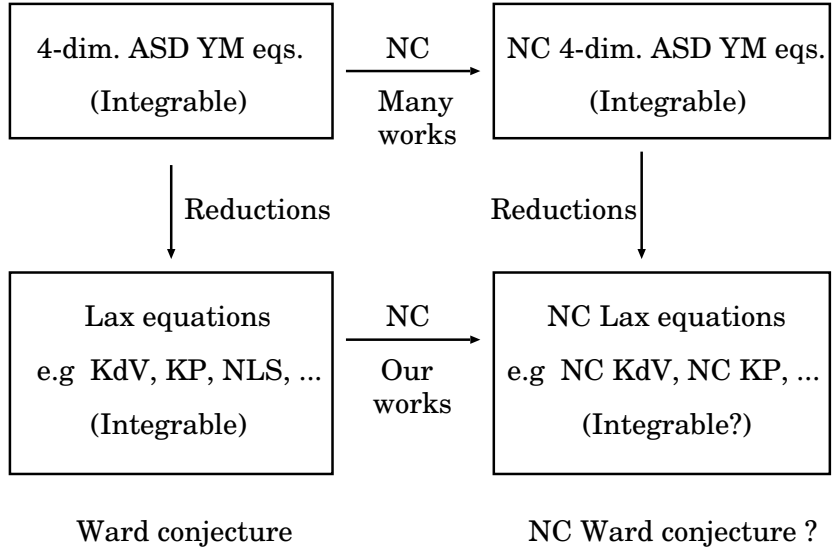}
\caption{NC Ward Conjecture}
\label{ward}
\end{figure}

In this section, we discuss
noncommutative extension of wider class of integrable equations
which are expected to preserve the integrability.
First, we present a strong method 
to give rise to noncommutative Lax pairs 
and construct various noncommutative Lax equations.
Then we discuss the relationship between
the generated equations and the noncommutative integrable equations 
obtained from the bicomplex method
and from reductions of the noncommutative ASD Yang-Mills equations.
All the results are consistent and we can expect that 
the noncommutative Lax equations would be integrable. 
Hence it is natural to
propose the following conjecture which contains 
the noncommutative version of Ward conjecture:
{\it many of noncommutative Lax equations would be integrable
and be obtained from reductions 
of the noncommutative ASD Yang-Mills equations}.
(See Fig. \ref{ward}.)

\subsection{The Lax-Pair Generating Technique}

In commutative cases, 
Lax representations \cite{Lax} are
common in many known integrable equations
and fit well to the discussion of reductions 
of the ASD Yang-Mills equations.
Here we look for the Lax representations on noncommutative spaces.
First we introduce how to find Lax representations on commutative spaces.

An integrable equation which possesses the Lax representation
can be rewritten as the following equation:
\begin{eqnarray}
[L,T+\partial_t]=0,
\label{lax}
\end{eqnarray}
where $\partial_t:=\partial/\partial t$.
This equation and the pair of operators $(L,T)$ are 
called the {\it Lax equation} and the {\it Lax pair}, respectively.

The noncommutative version of the Lax equation (\ref{lax}),
the {\it noncommutative Lax equation}, is easily defined
just by replacing the product of $L$ and $T$ with the star product.

In this subsection, we look for the noncommutative Lax equation
whose operator $L$ is a differential operator.
In order to make this study systematic,
we set up the following problem :

\vspace{3mm}
\noindent
{\bf Problem :}
For a given operator $L$, 
find the corresponding operator $T$
which satisfies the Lax equation (\ref{lax}).
\vspace{3mm}

This is in general very difficult to solve.
However if we put an ansatz on the operator $T$,
then we can get the answer 
for wide class of Lax pairs
including noncommutative case.
The ansatz for the operator $T$ is of the following type:

\vspace{3mm}
\noindent
{\bf Ansatz for the operator $T$ :}
\begin{eqnarray}
\label{ansatz}
T=\partial_i^n L+T^\prime.
\end{eqnarray}
\noindent
Then the problem for $T$ is reduced to that for $T^\prime$.
This ansatz is very simple, however, very strong
to determine the unknown operator $T^\prime$.
In this way, we can get the Lax pair $(L,T)$,
which is called, in this paper, the
{\it Lax-pair generating technique}.

In order to explain it more concretely,
let us consider the Korteweg-de-Vries (KdV) 
equation on commutative $(1+1)$-dimensional space
where the operator $L$ is given 
by $L_{\scriptsize\mbox{KdV}}:=\partial_x^2+u(t,x)$.

The ansatz for the operator $T$ is given by
\begin{eqnarray}
T=\partial_x L_{\scriptsize\mbox{KdV}} +T^\prime,
\label{ansatz_KdV}
\end{eqnarray}
which corresponds to $n=1$ and $\partial_i=\partial_x$ in
the general ansatz (\ref{ansatz}).
This factorization 
was first used to find wider class of
Lax pairs in higher dimensional case \cite{ToYu}.

The Lax equation (\ref{lax}) leads to the equation for 
the unknown operator $T^\prime$:
\begin{eqnarray}
[\partial_x^2+u,T^\prime]=u_x\partial_x^2+u_t+uu_x,
\label{Tpr}
\end{eqnarray}
where $u_x:=\partial u/\partial x$ and so on.
Here we would like to 
delete the term $u_x\partial_x^2$ in the RHS of (\ref{Tpr})
so that this equation
finally is reduced to a differential equation.
Therefore the operator $T^\prime$ could be taken as
\begin{eqnarray}
T^\prime=A \partial_x+B,
\end{eqnarray}
where $A, B$ are polynomials of $u, u_x, u_t, u_{xx},$ etc.
Then the Lax equation becomes $f\partial_x^2+g\partial_x+h=0$.
{}From $f=0, g=0$, we get\footnote{Exactly speaking, 
an integral constant should appear in $A$ as $A=u/2+\alpha$.
This constant $\alpha$ is unphysical
and can be absorbed by the scale transformation $u\rightarrow u+2\alpha/3$. 
Hence we can take $\alpha=0$ without loss of generality.
{}From now on, we always omit such kind of integral constants.}
\begin{eqnarray}
A=\frac{u}{2},~~~B=-\frac{1}{4}u_x+\beta,
\end{eqnarray}
that is,
\begin{eqnarray}
T=\del_x L_{\scr\mbox{KdV}}+A\del_x+B
=\partial_x^3+\frac{3}{2}u\partial_x+\frac{3}{4}u_x.
\end{eqnarray}
Finally $h=0$ yields the Lax equation, the KdV equation:
\begin{eqnarray}
u_t+\frac{3}{2}uu_x+\frac{1}{4}u_{xxx}=0.
\end{eqnarray}

In this way, we can generate a wide class of Lax equations 
including higher dimensional integrable equations \cite{ToYu}.
For example, $L_{\scriptsize\mbox{mKdV}}:=\partial_x^2+v(t,x)\partial_x$ and 
$L_{\scriptsize\mbox{KP}}:=\partial_x^2+u(t,x,y)+\partial_y$ give rise to 
the modified KdV equation and the KP equation, respectively 
by the same ansatz (\ref{ansatz_KdV}) for $T$.
If we take $L_{\scriptsize\mbox{BCS}}:=\partial_x^2+u(t,x,y)$ and 
the modified ansatz $T=\partial_y L_{\scriptsize\mbox{BCS}}+T^\prime$,
then we get the Bogoyavlenskii-Calogero-Schiff (BCS) 
equation \cite{Bogoyavlenskii, Calogero, 
Schiff}.\footnote{The multi-soliton solution 
is found in \cite{FSTY,FTY}.}

Good news here is that this technique is also applicable to
noncommutative cases.

\subsection{NC Lax Equations}

We present some results by using the Lax-pair generating technique.
First we focus on noncommutative $(2+1)$-dimensional
Lax equations.
Let us suppose that the noncommutativity is basically
introduced in the space directions.  
\begin{itemize}

\item The NC KP equation \cite{Paniak}:

The Lax operator is given by
\begin{eqnarray}
L_{\scriptsize\mbox{KP}}=\partial_x^2+u(t,x,y)+\partial_y
=:L_{\scriptsize\mbox{KP}}^\prime+\partial_y.
\end{eqnarray}
The ansatz for the operator $T$ is the same as commutative case:
\begin{eqnarray}
T=\partial_x L_{\scriptsize\mbox{KP}}^\prime+T^\prime.
\end{eqnarray}
Then we find
\begin{eqnarray}
T^\prime=A\del_x+B
=\frac{1}{2}u\partial_x-\frac{1}{4}u_x-\frac{3}{4}\partial_x^{-1}u_y,
\end{eqnarray}
and the noncommutative KP equation:
\begin{eqnarray}
\label{ncKP}
u_t+\frac{1}{4}u_{xxx}+\frac{3}{4}(u_x\star u+u\star u_x)
+\frac{3}{4}\partial_x^{-1}u_{yy}
+\frac{3}{4}[u,\partial_x^{-1} u_y]_\star=0,
\end{eqnarray}
where $\partial_x^{-1}f(x):=\int^x dx^\prime f(x^\prime),
~u_{xxx}=\partial^3 u/\partial x^3$ and so on.
This coincides with that in \cite{Paniak}.
There is seen to be a nontrivial deformed term $[u,\partial_x^{-1} u_y]_\star$
in the equation (\ref{ncKP}) which vanishes in the commutative limit.
In \cite{Paniak}, the multi-soliton solution is found
by the first order to small $\theta$ expansion,
which suggests that this equation would be considered as an 
integrable equation.

If we take the ansatz $T=\del_x^n L_{\scr\mbox{KP}}+T^\prime$,
we can get infinite number of the hierarchy equations.

\item The NC BCS equation:

This is obtained by following 
the same steps as in the commutative case.
The new equation is 
\begin{eqnarray}
&&u_t+\frac{1}{4}u_{xxy}+\frac{1}{2}(u_y\star u+u\star u_y)
+\frac{1}{4}u_x\star (\partial_x^{-1} u_y)\nonumber\\
&&+\frac{1}{4}(\partial_x^{-1}u_{y})\star u_x
+\frac{1}{4}[u,\partial_x^{-1}[u,\partial_x^{-1}u_{y}]_\star]_\star=0,
\end{eqnarray}
whose Lax pair and the ansatz are
\begin{eqnarray}
L_{\scriptsize\mbox{BCS}}&=&\partial_x^2+u(t,x,y),\nonumber\\
T&=&\partial_y L_{\scriptsize\mbox{BCS}}+T^\prime,\nonumber\\
T^\prime&=&A\del_x+B=\frac{1}{2}(\partial_x^{-1}u_y)\partial_x
-\frac{1}{4}u_y-\frac{1}{4}\partial_x^{-1}[u,\partial_x^{-1}u_y]_\star.
\end{eqnarray}
This time, a non-trivial term is found even in the operator $T$.

\end{itemize}

We can generate many other noncommutative Lax equations in the same way.
Furthermore if we introduce the noncommutativity into time coordinate
as $[t,x]=i\theta$,
we can construct noncommutative $(1+1)$-dimensional integrable equations.
Let us show some typical examples.

\begin{itemize}

\item The NC KdV equation:

The noncommutative KdV equation is simply obtained as
\begin{eqnarray}
u_t+\frac{3}{4}(u_x\star u +u\star u_x)+\frac{1}{4}u_{xxx}=0,
\end{eqnarray}
whose Lax pair and the ansatz are
\begin{eqnarray}
L_{\scr\mbox{KdV}}&=&\del_x^2+u(t,x),\nn
T&=&\del_x L_{\scr\mbox{KdV}}+T^\prime,\nn
T^\prime&=&A\del_x+B=\fr{1}{2}u\del_x+\fr{3}{4}u_x.
\end{eqnarray}

This coincides with that derived 
by using the bicomplex method \cite{DiMH2}
and by the reduction from noncommutative KP equation (\ref{ncKP})
setting the fields $y$-independent: ``$\partial_y  =0$.''
The bicomplex method guarantees the existence of
many conserved topological quantities,
which suggests that
noncommutative Lax equations would possess the integrability
Here we reintroduce the noncommutativity 
as $[t,x]=i\theta$.\footnote{We note that this reduction is formal
and the noncommutativity here contains subtle points in
the derivation from the $(2+2)$-dimensional
noncommutative ASD Yang-Mills equation by reduction
because the coordinates $(t,x,y)$
originate partially from the parameters 
in the gauge group of the noncommutative Yang-Mills theory \cite{AbCl, MaWo}.
We are grateful to T.~Ivanova for pointing out this point to us.}
We also find the noncommutative KdV hierarchy \cite{Toda},
by taking the ansatz $T=\del_x^n L_{\scr\mbox{KdV}}+T^\prime$.
It is interesting that for $n=2$, the hierarchy equation 
becomes trivial: $u_t=0$.

\item The NC Burgers equation \cite{HaTo2}:

As one of the important and new Lax equations, the noncommutative 
Burgers equation is obtained:
\begin{eqnarray}
u_t-\alpha u_{xx}+(1+\alpha-\beta)u_x\star u+(1-\alpha-\beta)u\star u_x=0,
\end{eqnarray}
whose Lax pair and the ansatz are
\begin{eqnarray}
L_{\scr\mbox{Burgers}}&=&\del_x+u(t,x),\nn
T&=&\del_x L_{\scr\mbox{Burgers}}+T^\prime,\nn
T^\prime&=&A\del_x+B=u\del_x+\alpha u_x+\beta u^2.
\end{eqnarray}

We can linearize it by the following two kind of
noncommutative Cole-Hopf transformations \cite{HaTo2}:
\begin{eqnarray}
\label{ch1}
u=\psi^{-1}\star \psi_x,
\end{eqnarray}
only when $1+\alpha-\beta=0$, and
\begin{eqnarray}
\label{ch2}
u=-\psi_x\star \psi^{-1},
\end{eqnarray}
only when $1-\alpha-\beta=0$.
The linearized equation is the noncommutative
diffusion equation
\begin{eqnarray}
\psi_t=\alpha\psi_{xx},
\end{eqnarray}
which is solvable via the Fourier transformation.
Hence the noncommutative Burgers equation
is really integrable.
The noncommutative Burgers hierarchy is also obtained
by taking the ansatz $T=\del_x^n L+T^\prime$ \cite{HaTo2}.

The transformations (\ref{ch1}) and (\ref{ch2})
are analogy of the commutative
Cole-Hopf transformation $u=\del_x \log \psi$ \cite{Cole, Hopf}.
This success makes us expect the possibility of
noncommutative extension of Hirota's bilinear forms \cite{Hirota},
tau-functions and Sato theory \cite{SaSa, DJM}. 

\end{itemize}


Let us here comment on the multi-soliton solutions.
First we note that 
if the field is holomorphic, that is, $f=f(x-vt)=f(z)$,
then the star product is reduced to the ordinary product:
\begin{eqnarray}
f(x-vt)\star g(x-vt)=f(x-vt)g(x-vt).
\end{eqnarray}
Hence the commutative
multi-soliton solutions where all the solitons
move at the same velocity always satisfy
the noncommutative version of the equations.
Of course, this does not mean that
the equations possess the integrability.

\subsection{Comments on the Noncommutative Ward Conjecture}

In commutative case, it is well known that 
many of integrable equations could be
derived from symmetry reductions of the four-dimensional
ASD Yang-Mills equation \cite{AbCl, MaWo},
which is first conjectured by R.Ward \cite{Ward}.

Even in noncommutative case,
the corresponding discussions would be possible and be interesting.
The noncommutative ASD Yang-Mills equations also have the Yang's 
forms \cite{Nekrasov2, Nekrasov} and many other similar
properties to commutative ones \cite{KKO}.
The simple reduction to three dimension
yields the noncommutative Bogomol'nyi equation
which has the exact monopole solutions and can be rewritten 
as the non-Abelian Toda lattice equation as
in Eq. (\ref{toda_lattice}) \cite{Nekrasov, GrNe}.
It is interesting that a discrete structure appears.
Furthermore M.Legar\'e \cite{Legare} succeeded in some reductions
of the $(2+2)$-dimensional noncommutative ASD Yang-Mills equations 
which coincide
with our results and those by using 
the bicomplex method \cite{DiMH, DiMH2},
which strongly suggests that the noncommutative deformation
would be unique and integrable
and the Ward conjecture would still hold on noncommutative spaces.

\newpage

\section{Conclusion and Discussion}

In the present thesis, 
we constructed various exact noncommutative solitons and 
discussed the corresponding D-brane dynamics.
We saw that ADHM/Nahm construction is very strong to
generate both commutative and noncommutative 
instantons/monopoles
and makes it possible to see the essential properties
of them clearly.
On noncommutative spaces, 
it was proved that resolutions of the singularities
actually occur and give various new physical objects. 
We could also see the equivalence between
the noncommutative deformation and 
the turning-on of the background magnetic ($B$-) fields
in gauge theories on D-branes.
Furthermore we found that ADHM construction naturally 
yields the ``solution generating technique''
which has been remarkably applied to
Sen's conjecture on
tachyon condensations in the context of string field theories.
The reason why the noncommutative descriptions could be successful 
is considered to be 
partially that the singular configuration becomes
smooth enough to be calculated due to the simple structure.
We constructed periodic instanton solutions 
and discussed the Fourier-transformations.
We saw that the transformed configuration satisfies 
the Bogomol'nyi equation and actually coincides with the fluxon,
which has perfectly consistent D-brane pictures 
Finally, we discussed noncommutative
extension of integrable systems
as a new study-area of them.
We proposed the strong way to generate
noncommutative Lax equations which are expected to be both 
integrable and obtained from the noncommutative Yang-Mills equation
by reductions.

\vs

There are many further directions following to these studies.

One of the expected directions is
the noncommutative extension of soliton theories and 
integrable systems in the lower-dimensions
which preserve the integrabilities
as is introduced in section 7.

In four-dimensional Yang-Mills theory,
the noncommutative deformation resolves the small instanton 
singularity of the (complete) instanton moduli space
and gives rise to a new physical object, the U(1) instanton.
Hence the noncommutative Ward conjecture would imply that
the noncommutative deformations of lower-dimensional integrable equations
might contain new physical objects
because of the deformations of the solution spaces in some case.

Now there are mainly three methods to yield noncommutative integrable
equations:
\begin{itemize}
\item Lax-pair generating technique
\item Bicomplex method
\item Reduction of the ASD Yang-Mills equation
\end{itemize}
The interesting point is that 
all the results are consistent
at least with the known noncommutative Lax equations, 
which suggests the existence and 
the uniqueness of the noncommutative deformations of integrable equations
which preserve the integrability.

Though we can get many new noncommutative Lax equations,
there need to be more discussions so that
such study should be fruitful as integrable systems.
First, we have to clarify whether the noncommutative Lax equations
are really good equations in the sense of integrability, that is,
the existence of many conserved quantities or
of multi-soliton solutions, and so on.
All of the previous studies including our works
strongly suggest that this would be true.
Second, we have to reveal the physical meaning of such equations.
If such integrable theories can be embedded in string
theories, there would be fruitful interactions
between the both theories,
just as between the (NC) ASD Yang-Mills equation and
D0-D4 brane system (in the background of NS-NS $B$-field).
There is a good string theory for this purpose:
$N=2$ string theory \cite{OoVa}.
The $(2+2)$-dimensional noncommutative ASD Yang-Mills equation and
some reductions of it can be embedded \cite{LPS, LePo}
in $N=2$ string theory,
which guarantees that such directions would have a physical
meaning and might be helpful to understand new aspects of 
the corresponding string theory.
This string theory has massless excitation modes only
and seems to make no problems in introducing
the noncommutativity in time direction
as $(1+1)$-dimensional noncommutative integrable equations.

\vs

The above direction is expected because of 
the success stories in 4-dimensional noncommutative Yang-Mills theories.
However even in this theories,
there are many problems to be solved.

The first one is the geometrical meaning of
the instanton charges for $G=U(1)$.
For $SU(2)$ part, there is an origin of the integer charge, 
the winding number: $\pi_3(SU(2))\simeq \pi_3(S^3)\in \Z$.
For U(1) part, however, $\pi_3(U(1))\simeq 0$
and at least the origin does not comes from the boundary of $\R^4$.
Crucial observations should be started with
the geometrical meaning of the shift operators.
There are several works in this direction,
for example, \cite{Furuuchi3, Matsuo, HaMo, IKS, Sako, STZ}.

Noncommutative monopoles also have many unclear points.
We have to clarify
whether the ``visible'' Dirac string which appears in the
exact solution of noncommutative 1-Dirac monopole are really
physical or not.
We should solve the inconsistency \cite{GrNe}
on the valuedness of the Higgs field
between exact noncommutative monopole (: single-valued) and 
exact nonlinear monopole (: multi-valued)
which should be equivalent to each other unless
the Seiberg-Witten's discussion \cite{SeWi} holds.
There are some discussions on these problems in \cite{Hamanaka2}.

\vs

In section 3-6, we saw that 
NC gauge theories could reveal
the corresponding D-brane dynamics
in some aspects.
It is natural to expect that
another D-brane system would be
analyzed by noncommutative gauge theories.

There is an origin of the duality of ADHM/Nahm construction,
that is, {\it Nahm transformation} which will be briefly introduced
in Appendix A.1.
This duality transformation is just related to the T-duality
transformation of the D0-D4 brane system where the coincident D4-branes
wrap on a four-torus.
The extension of Nahm transformation to even-dimensional tori
has been done by the author and H.~Kajiura \cite{HaKa}.
In commutative case, we have to suppose that
the number $k$ of the D0-branes is not zero.
On the other hand, the noncommutative extension of it is expected 
to admit the $k=0$ case because of the resolution of singularities,
which is just the T-duality transformation for
one kind of D-branes themselves.
Furthermore, there is some relationship between two
noncommutativities on torus and the dual torus \cite{KoSc}.
An attempt of the noncommutative extension of Nahm transformation
is found in \cite{ANS}, however, it tells nothing about the above point.
It is interesting to make it clear whether
the noncommutative Nahm transformation give the relationship of
the noncommutativities or not.
In order to study it concretely, 
we have to define the tensor product
of the modules on the product of torus and the dual torus
as is commented on in the conclusion of \cite{HaKa}.

The higher dimensional extension of ADHM construction is possible.
In fact, on the 8-dimensional Euclidean space,
there exist ``ASD'' configurations which satisfy
the 8-dimensional ``ASD'' equation \cite{CDFN,Ward} and
the ADHM construction of them in some special case \cite{CGK}.
Some works on the noncommutative 
extension of it have been done and the D-brane interpretations 
such as D0-D8 brane systems are presented for example 
in \cite{PaTe, Ohta_k, Hiraoka, BLP}.
In D0-D8 systems, there is seen to be a special behavior of D-branes
known as the {\it brane creation} \cite{HaWi}.
It is expected that (NC) higher-dimensional ADHM construction
might give gauge theoretical explanations of it
and some hints of new D-brane dynamics. 

\vs

There are mainly three aspects of noncommutative theories 
which show physical situations:
\begin{itemize}
\item the equivalence to physics in the presence of magnetic fields
\item a formulation of open string field theory \cite{Witten2}
\item a candidate for the geometry underlying quantum gravity
\end{itemize}
In this thesis, we focused on the first aspect and
applied it to the study of D-brane dynamics in the background $B$-field.
This approach is successful to some degree because of 
the simplicity. However the situation is rather restricted.

The second one is recently rewritten as NC-deformed
theories by I.~Bars et al. \cite{Bars}.
This direction is new and interesting.

The third one is more profound and very different from the
present discussions.
Very naively, quantum gravity might be formulated
in terms of noncommutative geometries
because the quantization processes usually introduce
the noncommutativity of the dynamical variables.
The quantization of gravities introduce
the noncommutativity of the metric (the gravitational field),
which would lead to noncommutative geometries.
There are several suggestions to justify the latter aspect,
for example, the {\it space-time uncertainty principle} 
proposed by T.~Yoneya \cite{Yoneya}.
We hope that such studies might shed light on
this challenging area.

\newpage

\begin{center}
{\large \bf Acknowledgments}
\end{center}

It is a great pleasure to thank Y.~Matsuo 
for advice and encouragement,
and 
Y.~Imaizumi, 
H.~Kajiura,
N.~Ohta,
S.~Terashima
and K.~Toda 
for collaboration.
In particular, K.~Toda has introduced me
to the wonderful world of integrable systems.
The author has visited in Korea, Kyoto (as an atom-type visitor), 
Netherlands, UK and Germany 
in order to make his knowledge related to the present thesis better
and is grateful to
all of the members at 
the Yukawa Institute for Theoretical Physics,
the Korea Institute for Advanced Study,
Sogang university,
York university,
Leiden university,
Amsterdam university,
Durham university,
Department of Applied Mathematics and Theoretical Physics 
at Cambridge university,
Queen Mary and Westfield college, university of London,
and Hannover university,
especially, 
C.~Chu,
E.~Corrigan,
J.~de Boer,
J.~Gauntlett,
O.~Lechtenfeld,
B.~Lee,
K.~Lee,
N.~Manton,
P.~van Baal
and 
R.~Ward
for hospitality, hearty support and fruitful discussions
during his stay.
He has had many chances to have talks related to this thesis
at the conferences, workshops and seminars at the universities or institutes.
He would also like to express his gratitude to
the organizers and the audiences for hospitality and discussions.
Thanks are also due to
T.~Eguchi,
H.~Fuji,
K.~Fujikawa,
K.~Furuuchi,
G.~Gibbons,
R.~Goto,
K.~Hashimoto,
Y.~Hashimoto,
T.~Hirayama,
K.~Ichikawa,
Y.~Imamura,
N.~Inoue,
M.~Ishibashi,
K.~Izawa,
M.~Jinzenji,
A.~Kato,
M.~Kato,
H.~Kanno,
T.~Kawano,
Y.~Kazama,
I.~Kishimoto,
Y.~Konishi,
H.~Konno,
Y.~Matsumoto,
S.~Moriyama,
A.~Mukherjee,
H.~Nakajima,
K.~Ohta,
H.~Ooguri,
R.~Sasaki,
A.~Schwarz,
J.~Shiraishi,
Y.~Sugawara,
T.~Takayanagi,
G.~'t Hooft,
I.~Tsutsui,
T.~Uesugi,
M.~Wadati,
S.~Watamura,
T.~Watari,
T.~Yanagida,
and all other members of his group
for enjoyable discussions and encouragements.
In particular,
M.~Asano,
K.~Hosomichi,
and S.~Terashima
have spent a lot of their time on education
and the author would like to express his special thanks to them.

Thanks to all colleagues around the author,
he could spend a stimulating and happy life,
which leaves him plenty of golden memories.
Finally he would like to thank 
his parents and his brother for continuous encouragement.

This work was supported in part by
the Japan Scholarship Foundation and 
the Japan Securities Scholarship Foundation (\#12-3-0403)

\newpage

\appendix


\section{ADHM/Nahm Construction}

In this appendix, we review foundation
of ADHM/Nahm construction of instantons/monopoles
on commutative spaces and presents our conventions.

ADHM/Nahm construction is one of the strongest methods
to generate all instantons/monopoles.
Instantons and monopoles have (anti-)self-dual and
stable configurations
and play important roles in revealing non-perturbative
aspects of Yang-Mills theories. 
ADHM/Nahm construction is based on the
one-to-one correspondence between
instanton/monopole moduli space and 
the moduli space of ADHM/Nahm data 
and can be applied to the instanton calculus and so on.
(For a review, see \cite{DHKM}.)

ADHM construction is a descendent
of the twistor theory \cite{Penrose}. (For reviews, see
\cite{MRS, MaWo, PeRi, WaWe}.)
In 1977, R.~Ward applied the twistor theory to
instantons and 
replaced the self-duality of the gauge fields on $S^4$
with the holomorphy of the vector bundles on $\C P_3$ \cite{Ward}.
The problem on the holomorphy of the vector bundle
is reduced to algebraic problems from algebro-geometric idea.
There are two treatments of it:
the method of algebraic curves and the method of monads.

M.~Atiyah and R.~Ward developed the former treatment and 
showed that an ansatz ({\it Atiyah-Ward ansatz})
gives rise to instantons \cite{AtWa}.
This idea has a close relationship to 
the inverse scattering methods (or B\"acklund transformations)
in soliton theories \cite{BeZa, CFGY, Yang}
and has made much progress with integrable systems \cite{WaWe, MaWo}.

On the other hand, Atiyah, Drinfeld, Hitchin and Manin
developed the latter treatment and found
the strong algebraic method to generate {\it all}
instanton solutions on $S^4$,
which is just the ADHM construction \cite{ADHM}.
(In this thesis, we treat instantons on $\R^4$
which is proved to be equivalent to instantons on $S^4$
from the conformal invariance and Uhlenbeck's theorem 
\cite{Uhlenbeck}.)
The idea of ADHM construction was applied to
the construction of monopoles by W.~Nahm \cite{Nahm}-\cite{Nahm6},
which is called {\it ADHMN construction} or {\it Nahm construction}.
Furthermore the duality in Nahm construction
which is like Fourier-transformation
was extracted into as a profound duality
of instantons on four-torus by
Schenk \cite{Schenk}, Braam and van Baal \cite{BrvB}.
This is called {\it Nahm transformation}
and has close relationship to
Fourier-Mukai transformation \cite{Mukai} in
algebraic geometry and T-duality in string theory.
(For a review on T-duality, see \cite{GPR}.)
Hence the duality on Nahm transformation is often called
{\it Fourier-Mukai-Nahm duality}.

In this appendix, we begin with the Fourier-Mukai-Nahm duality
and derive ADHM/Nahm duality from it intuitively.
Then we introduce the detailed discussion
on ADHM/Nahm construction on commutative spaces
and present our conventions which is used in main parts of 
the present thesis.

\noindent
\unl{\bf Notations and Comments in the Appendix}
\begin{itemize}
\item The size of a $m\ti n$ matrix $M$
is denoted by $M_{[m]\ti[n]}$.
In particular $m \ti m$ diagonal matrices are sometimes denoted by $M_{[m]}$.
\item The Lie algebra of a Lie group $G$
is represented as the corresponding calligraph symbols ``$\cG$,''
where the element $g$ of the Lie group
and that $X$ of the Lie algebra have the relation: $g=e^{X}$.
\item ``$\ap$'' means ``asymptotically equal to at spatial 
infinity'' $r:=\vert x\vert\rar\infty$. 
\item  Usually the trace symbols $\Tr$ and  $\tr$
are taken with respect to the color indices of the gauge
group and the spinor indices, respectively.
\item The convention of the indices can be summarized up as follows:
\begin{eqnarray*}
\mbox{4-dimensional space indices~}[4]&:&1\leq\mu,\nu,\rho,\cdots\leq 4\\
\mbox{3-dimensional space indices~}[3]&:&1\leq i,j,k,\cdots\leq 3\\
\mbox{Color indices~}[N]&:&1\leq u,v,w,\cdots\leq N\\
\mbox{Instanton number indices~}[k]&:&1\leq p,q,r,\cdots\leq k\\
\mbox{Spinor indices~}[2]&:&1\leq \alpha,\beta,\gamma,\cdots,\leq2
\end{eqnarray*}
\end{itemize}

\subsection{A Derivation of ADHM/Nahm construction from Nahm Transformation}

ADHM/Nahm construction looks very complicated,
however, is simple and beautiful in fact.
In order to explain this points clearly,
we introduce the beautiful duality transformation,
Nahm transformation \cite{Schenk, BrvB}
as the background of ADHM/Nahm construction.

Nahm transformation is a duality transformation 
(one-to-one mapping) between
the instanton moduli space on a four-torus $T^4$ with $G=U(N),~C_2=k$
and that on the dual torus $\wht^4$ with $\widehat{G}=U(k),~C_2=N$.
This situation is realized as D0-D4 brane systems where
the D4-branes wrap on $T^4$.
We can take T-duality transformation in the four directions where
the D4-brane lie, which is just the Nahm transformation.
In this subsection, we review the Nahm transformation briefly
and discuss a derivation of ADHM/Nahm construction
by taking some limits.

\vs
\unl{\bf Poincar\'e Line Bundle}
\vs

Let us set up the stage first.
We introduce the Poincar\'e Line Bundle.

Let us suppose that $\Lambda$  denotes 
the rank-four lattice of $\R^4$.
Then a four-torus $T^4$ and the dual torus $\wht^4$ are
given as follows;
\begin{eqnarray}
T^4:=\R^4/\Lambda,~~~\wht^4:=\R^{4*}/2\pi\Lambda^*,
\end{eqnarray}
where $\R^{4*}$ is the dual vector space of $\R^4$
and $\Lambda^*$ is the dual lattice of $\Lambda$:
\begin{eqnarray}
\Lambda^*:=\left\{\mu\in\R^{4*}~\vert~ \mu\cdot\lambda\in\Z,\all\lambda\in
\Lambda\right\}.
\end{eqnarray}
In this subsection, the dot ``$\cdot$ '' denotes 
the inner product of the elements of $\R^4$ and $\R^{4*}$.
Hence roughly speaking, the torus and the dual torus have
the opposite size to each other : $(\mbox{vol}\,T^4)
\cdot(\mbox{vol}\,\wht^4)=(2\pi)^4$.
The coordinates of $\R^4$ and $\R^{4*}$
are represented as $x^\mu$ and $\x_\mu$, respectively. 

Next let us introduce the trivial bundle $\cL=T^4\ti \C
\rar T^4$ on $T^4$ and pull it back onto $T^4\ti \R^{4*}$
by the projection $\pi:T^4\ti \R^{4*}\rar T^4$.
The gauge group of the bundle is U(1). 
On the trivial line bundle $\pi^*\cL\rar T^4\ti\R^{4*}$
which is the pull-back bundle of $\cL$ by the projection $\pi$,
the natural gauge field can be defined as
\begin{eqnarray}
\omega (x,\x)=i\x_\mu dx^\mu,
\end{eqnarray}
which is considered as that on $\pi^*\cL\rar T^4\ti\wht^4$.
In fact, the gauge field $\omega(x,\x)$ is 
equivalent to $\omega(x,\x+2\pi\mu)$
and connected by the following gauge transformation:
\begin{eqnarray}
\lab{equiv-u1}
\omega(x,\x+2\pi\mu)= g^{-1}\omega(x,\x)g+g^{-1}dg,~~~~~~~
^{\exists}g(x)=e^{2\pi i\mu\cdot x}\in U(1),~~~\mu\in \Lambda^*.
\end{eqnarray}
This gauge-equivalent relation define the line bundle
on $T^4\ti\wht^4$ which is called
{\it Poincar\'e line bundle} and is denoted by $\cP\rar T^4\ti \wht^4$.
The curvature $\Omega(x,\x)$ of the Poincar\'e line bundle is
\begin{eqnarray}
\Omega(x,\x)=id\x_\mu\wedge dx^\mu.
\end{eqnarray}

The dual Poincar\'e line bundle $\widehat{\cP}\rar T^4\ti \wht^4$
is also constructed from the trivial 
line bundle $\widehat{\cL}=\wht^4\ti \C
\rar \wht^4$ on the dual torus $\wht^4$
and the gauge field is given by $\omega^\pr (x,\x) = ix^\mu d\x_\mu$.
The gauge field $\om(x,\x) = i \x_\mu dx^\mu$
is mapped to $\omega^\pr (x,\x) = -ix^\mu d\x_\mu$
by the gauge transformation $\exp(- i\x\cdot x)$
on $\R^4\ti\R^{4*}$:
\begin{eqnarray}
 \om(x,\x)= i \x_\mu dx^\mu ~~\longr
~~\omega^\pr
(x,\x) = \om(x,\x) +e^{i\x \cdot x}de^{-i\x \cdot x}
=-i x^\mu d\x_\mu,
\end{eqnarray}
which shows that $\widehat{\cP}$ is the complex conjugate of $\cP$.

The Poincar\'e line bundle yields the Fourier-transformation like
duality in Nahm transformation.

Let us summarize on Poincar\'e line bundle:
\begin{eqnarray*}
\ba{ccccc}
~&~&\cP&~&~\\
~&~&\dar&~&~\\
\cL &~&T^4\ti\wht^4&~&\wha{\cL} \\
\dar&\st{\pi}{\swarrow}&~&\st{\hat{\pi}}{\searrow}&\dar\\
T^4&~&~&~&\wht^4
\ea
\end{eqnarray*}

\vs
\unl{\bf Nahm Transformation}
\vs

Now let us define Nahm transformation $\cN:(E,A)\map(\widehat{E},\whA)$,
where $E$ is the $N$-dimensional complex vector bundle on $T^4$
with Hermitian metric and $G=U(N),~C_2=k$. 
First we pull the bundle $E$ back by the projection $\pi$.
The gauge field on $\cP\ot \pi^*E\vert_{T^4\ti\left\{\x\right\}}$
is defined by $A_\x:=A\ot 1_{\scr \cL}
+1_{[N]}\ot i\x_\mu dx^\mu$. 
The field strength $F_\x$ from $A_\x$ equals to $F$ from $A$. 
The covariant derivative from $A_\x$ is denoted by $D[A_\x]:=d+A_\x$. 

Next let us define Dirac operator.
Suppose that  $S^{\pm}\rar T^4$ is the spinor bundle on $T^4$.
The Dirac operator acting on 
the section $\Gamma(T^4,S^\pm\ot E\ot \cP)$ is given by
\begin{eqnarray}
\cD [A_\x]&:=&e^\mu\ot D[A_\x]=e_\mu \ot(\del_\mu+A_\mu+i\x_\mu),\nn
\cDb [A_\x]&:=&\eb^\mu \ot D[A_\x]=\eb_\mu\ot (\del_\mu+A_\mu+i\x_\mu).
\end{eqnarray}
Exactly speaking, we should call them {\it Weyl operators}
rather than Dirac operators.
Here, however, we use the word ``Dirac operator'' for simplicity,
which makes no confusion, we hope.

\begin{figure}[htbn]
\begin{center}
\epsfxsize=100mm
\hspace{0cm}
\epsffile{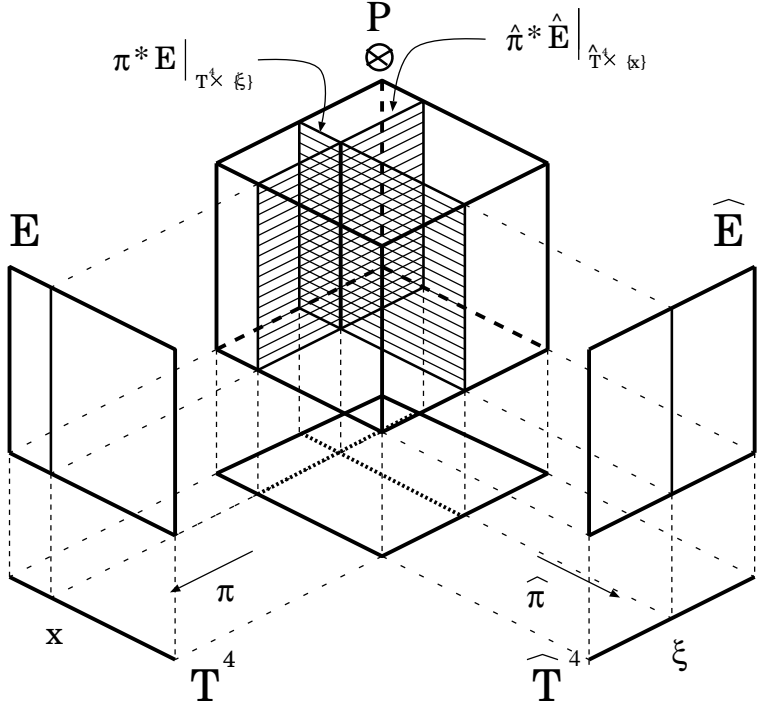}
\caption{The stage of Nahm transformation}
\label{Poin}
\end{center}
\end{figure}

Here let us construct the dual vector bundle $\widehat{E}$
on $\wht^4$ by using the Dirac zero-mode $\psi_\x^p(x),~p=1,\cdots,k$.
Concretely we take $\Ker \cDb[A_\x]$ as the fiber $\widehat{E}_\x$.
Atiyah-Singer family index theorem says $\dim \Ker \cDb [A_\x]=k$. 
Suppose $\wha{H}\rar \wht^4$ as infinite-dim trivial vector bundle
whose fiber is $\wha{H}_\x:=L^2(T^4,S^+\ot E\ot
\cP\vert_{T^4\ti\left\{\x\right\}})$,
the bundle $\widehat{E}_\x=\Ker \cDb [A_\x]$
is sub-bundle of $\wha{H}_\x$ and  $\widehat{E}$ is sub-bundle of 
$\wha{H}$. (See Fig. \ref{Poin}.) 

\begin{eqnarray}
\ba{ccccccc}
~&~&~&~&\pi^*\cF\ot\cP&~&~\\
~&~&~&~&\dar&~&~\\
(\cF,A)\rar (\whf,\whA)~:&~&\cF
 &~&T^4\ti\wht^4&~&\hat{\pi}_*(\pi^*\cF\ot\cP)\\
~&~&\dar&\st{\pi}{\swarrow}&~&\st{\hat{\pi}}{\searrow}&\dar\\
~&~&T^4&~&~&~&\wht^4
\ea
\end{eqnarray}
\begin{eqnarray}
\ba{ccccccc}
~&~&~&~&\cP\ot\hat{\pi}^*\whf&~&~\\
~&~&~&~&\dar&~&~\\
(\cF,A)\lar (\whf,\whA)~:&~&\pi_*(\cP\ot\hat{\pi}^*\whf)
 &~&T^4\ti\wht^4&~&\wha{\cF} \\
~&~&\dar&\st{\pi}{\swarrow}&~&\st{\hat{\pi}}{\searrow}&\dar\\
~&~&T^4&~&~&~&\wht^4
\ea
\end{eqnarray}

Here we introduce the projection 
\begin{eqnarray}
P : \wha{H} \rar \widehat{E}
\end{eqnarray}
and define the covariant derivative as follows
\begin{eqnarray}
\lab{covder}
\wha{D}=P\widehat{d}:
\Gamma(\wht^4,\wha{E})\rar\Gamma(\wht^4,\Lambda^1\ot\widehat{E}),
\end{eqnarray}
which specifies the gauge field $\wha{A}$ on $\widehat{E}$.
This is the Nahm transformation 
(mapping): $\cN:(E,A)\map (\widehat{E},\wha{A})$.
The concrete representation of the dual gauge fields are given by
\begin{eqnarray}
\whA_\mu^{pq}=\int_{T^4}d^4 x~\psi^{\dagger p}\fr{\del}
{\del \x^\mu}\psi^q,
\end{eqnarray}
where $\psi^p~(p=1,2,\ldots,k)$ is the $k$ normalizable Dirac zero-modes. 

The similar argument is possible from $\wht^4$
which specifies the inverse 
transformation: $\wha{\cN}:(\widehat{E},\wha{A})\map(E,A)$.
Then the dual Dirac operator is defined by
\begin{eqnarray}
\widehat{\cD} {[\wha{A}_x]}
&:=&e_\mu \ot (\widehat{\del}^\mu+\wha{A}^\mu-ix^\mu),\nn
\widehat{\cDb} {[\wha{A}_x]}
&:=&\eb_\mu\ot (\widehat{\del}^\mu+\wha{A}^\mu- ix^\mu).
\end{eqnarray}

Furthermore we can prove that Nahm transformation is one-to-one,
that is, $\cN\wha{\cN}=$id. and  $\wha{\cN}\cN=$id.

Summary is the following:

\vs
\unl{\bf Nahm transformation}
\begin{eqnarray*}
\lab{summary}
\ba{ccc}
E&~&\widehat{E}\\
\dar&~&\dar\\
T^4&~&\wht^4\\
G=U(N)&~&\wha{G}=U(k)\\
k\mbox{-instanton}&\st{1~:~1}{\longlr}&N\mbox{-instanton}\\
~&~&~\\
~&\mbox{massless Dirac eq.}&~\\
~&\cDb \psi =0&~\\
\mbox{instanton}~:~A_{\mu[N]}&\st{k \mbox{ solutions} :~\psi({\x},x)}{\longr}&
\dis\whA_{\mu[k]}=\int_{T^4}d^4 x~\psi^\dagger\fr{\del}{\del \x^\mu}
\psi\\
~&~&~\\
~&\mbox{massless Dirac eq.}&~\\
~&\widehat{\cDb}v=0&~\\
\dis A_{\mu[N]}=\int_{\wht^4}d^4 \x~v^\dagger\fr{\del}{\del x^\mu}
v
&\st{N \mbox{ solutions}: 
v(x,\x)}{\longl}&\mbox{instanton}~:~\whA_{\mu[k]}\\
\ea
\end{eqnarray*}

\begin{center}
\begin{figure}[htbn]
\epsfxsize=120mm
\hspace{2.5cm} 
\epsffile{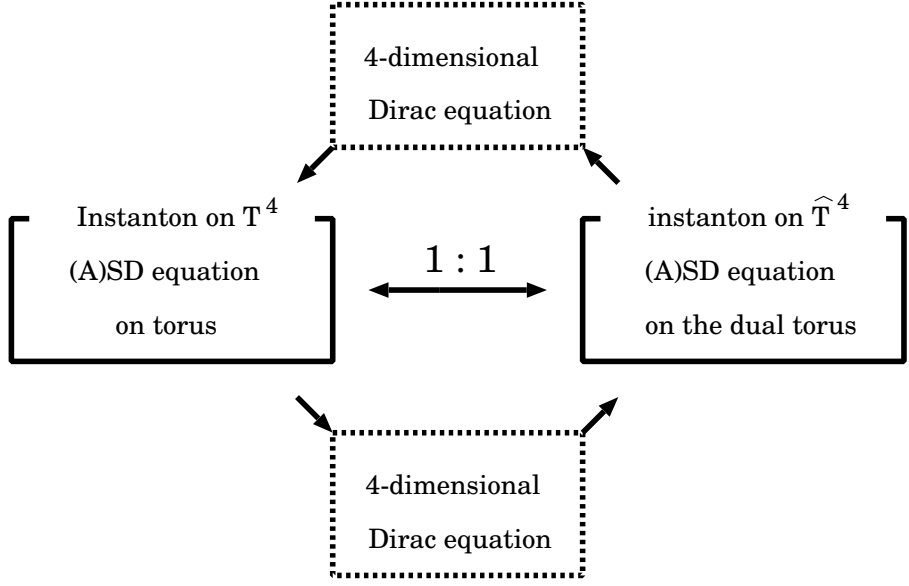}
\caption{Nahm Transformation}
\label{caloron}
\end{figure}
\end{center}

\vs
\unl{\bf Examples}
\vs

Let us transform concrete solutions \cite{HaKa}. 
There is known 
to be $G=U(N^2)(\simeq U(N)\ot U(N)),~k^2$-instanton solutions:
\begin{eqnarray}
A_1=0,~~~A_2=-\frac{i}{2\pi}\frac{k}{N}x_1\otimes 1_{[N]},~~~
A_3=0,~~~A_4=1_{[N]}\otimes\frac{i}{2\pi}\frac{k}{N}x_3,
\end{eqnarray}
which actually satisfies ASD eq.
and the instanton number is calculated as $-k^2$:
\begin{eqnarray}
F_{12}=-F_{34}=-\fr{i}{2\pi}\fr{k}{N}1_{[N]}\otimes 1_{[N]}.
\end{eqnarray}
By solving the Dirac equation in the background of 
the instantons, we find the Dirac zero-mode:
\begin{eqnarray}
\label{zero_torus}
\psi_{uu^\prime}^{pp^\prime}(\xi,x)
&=&\left(\frac{N}{2\pi k}\right)^{\half}
\sum_{s,t\in \Z} e^{ix_1(\frac{k}{N}(\frac{x_2}{2\pi}+u+Ns)+p)}
e^{2\pi i\xi_2 (\fr{x_2}{2\pi}+u+Ns+\frac{N}{k}(\xi_1+p))}
e^{-\frac{\pi k}{N}(\frac{x_2}{2\pi}+u+Ns+\fr{N}{k}(\xi_1+p))^2}\nonumber\\
&&\times e^{-ix_3(\frac{k}{N}(\frac{x_4}{2\pi}+u^\prime+Nt)+p^\prime)}
e^{-2\pi i\xi_4(\fr{x_4}{2\pi}+u^\prime+Nt+\frac{N}{k}(\xi_3+p^\prime))}
e^{-\frac{\pi k}{N}(\frac{x_4}{2\pi}+u^\prime+Nt+\fr{N}{k}(\xi_3+p^\prime))^2}
\end{eqnarray}
Then we can calculate the dual gauge field
in usual manner:
\begin{eqnarray}
\widehat{A}_1=-2\pi i\frac{N}{k}\xi_2\otimes 1_{[k]},~~~\widehat{A}_2=0,~~~
\widehat{A}_3=1_{[k]}\otimes 2\pi i\frac{N}{k}\xi_4,~~~\widehat{A}_4=0.
\end{eqnarray}
This trivially solves the ASD equation and 
is proved to be $\widehat{G}=U(k^2),~N^2$-instanton.
We can calculate the Green function substituting this into (\ref{zero_torus}).

\vs
\noindent
{\bf Note}
\vs
\begin{itemize}
\item The extension of Nahm transformation to even-dimensional 
tori are discussed in \cite{HaKa}.
\item The D-brane interpretations of Nahm transformation
and extension to other gauge groups are discussed in \cite{Hori}.
\end{itemize}

\vs
\unl{\bf A Derivation of ADHM/Nahm Construction}
\vs

Though we saw a beautiful duality in Nahm transformation.
this is no use for constructing explicit instanton
solutions because we have to make two steps 
to get explicit instanton solutions on torus,
that is, solving dual ASD equation and Dirac equation on the dual torus,
which spend more effort than solving ASD equation directly.

If we want to make the duality useful,
we often take some limit with respect to
the parameters in the theory.
This time there are good parameters, radius of torus $r_\mu$.
Now we take some limit of the parameters and derive
non-trivial duality in the extreme situations,
which is found to be just ADHM/Nahm construction. 

\begin{itemize}
\item Taking all four radii infinity $\Rightarrow$ ADHM construction

Then the radii of the dual torus become zero.
Hence the dual torus shrink into one point and
the derivative becomes meaning less because
the derivative measures the difference between two points.
As the result, all the derivatives in the dual ASD equation and 
the dual massless Dirac equation drop out naively and 
the differential equations becomes matrix equations.
This degeneration of the Nahm transformation leads to
the non-trivial results:
we can construct instanton solutions on ${\bf R}^4$ 
(=infinite-size torus) by solving matrix equations,
which is just ADHM construction ($T_\mu=\wha{A}_\mu$).
For more detailed discussion, see \cite{vanBaal}.

\item Taking three radii infinity and the other radius zero 
$\Rightarrow$ Nahm construction

Then the torus and the dual torus become $\R^3$ and $\R$,
respectively.
In similar way, the differential equations on dual side
become ordinary differential equations because 
the derivative only in one direction survives,
which concludes that 
we can construct BPS monopole solutions 
(=ASD configuration on ``$\R^3$'') 
by solving the ordinary differential equations,
which is just Nahm construction. 

\end{itemize}

\subsection{ADHM Construction of Instantons on $\R^4$}

In this subsection,
we review the ordinary ADHM construction of 
instantons on commutative space based on
Corrigan-Goddard's paper \cite{CoGo3}
and my review \cite{Hamanaka4}.

The most fundamental object in ADHM construction
is the Dirac operator.
The important equations such as ASD equation and so on
can be understood from the viewpoint of Dirac operators.

Here we impose on this point and discuss the
duality in ADHM construction.
At the same time, we set up the notations.
The outline of the review is the following as Nahm transformation:

\begin{center}
\begin{figure}[htbn]
\epsfxsize=120mm
\hspace{2.5cm} 
\epsffile{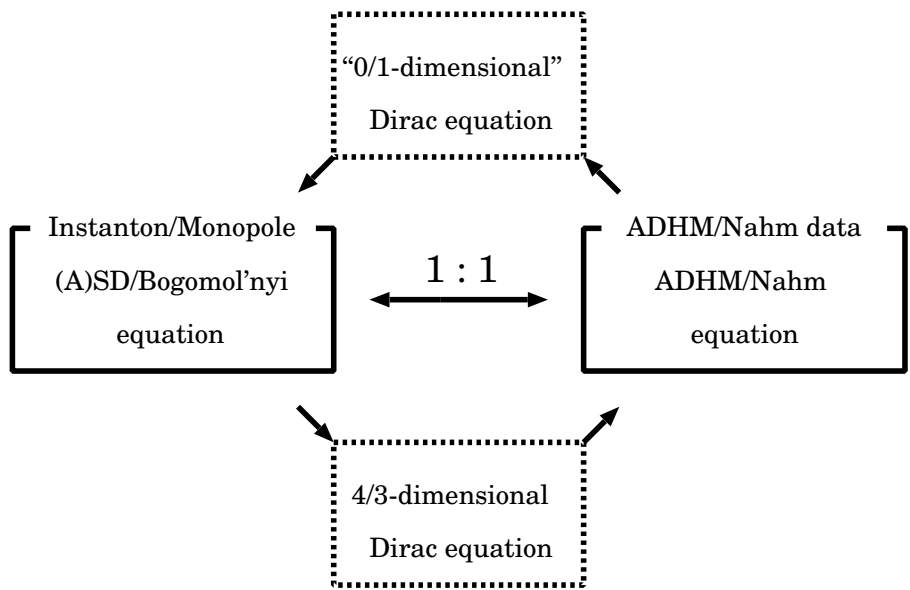}
\caption{ADHM/Nahm Construction}
\label{caloron}
\end{figure}
\end{center}

As we comment in the previous subsection,
(A)SD / Bogomol'nyi / Nahm / ADHM equation is basically
considered as 4 / 3 / 1 / 0-dimensional ASD equation.

In order to discuss the duality,
we first present instantons and ADHM data,
and then define the duality mapping
and finally comment on the one-to-one
correspondence without proofs.

\vs
\unl{\bf (Instanton)}
\vs

Let us explain what instantons are.
For simplicity,
suppose that the gauge $G$ is $SU(N),~N\geq 2$.
(There is no difference whether $G=U(N)$ or $G=SU(N)$.)
We can fix the self-duality of instantons ASD 
without loss of generality.
Those who know the basic notion of instantons
may skip this part except for
the representation of ASD equation
from the viewpoint of Dirac operators.

Instantons on four-dimensional commutative Euclidean  space
are the configuration of the gauge fields which
satisfies ASD equation and make the Yang-Mills action
minimize and be finite.

Let us define the Dirac equation
which is the most fundamental:
\begin{itemize}
\item Dirac operator
\begin{eqnarray}
\cD_x:=e^\mu\ot D_\mu=e^\mu\ot (\del_\mu +A_\mu),
~~~\cDb_x:=\eb_\mu\ot D_\mu=-\cD^\dagger.
\end{eqnarray}
\end{itemize}
Here $D_\mu$ is an ordinary covariant operator and $e_\mu$
is the two-dimensional representation matrix of quartanion $(i,j,k,1)$
(Euclidean 4-dimensional Pauli matrix):
\begin{eqnarray}
e_\mu:=(-i\sigma_i,1),~~~\eb_\mu:=\eb_\mu=(i\sigma_i,1),
\end{eqnarray}
which satisfies
\begin{eqnarray}
\label{edage}
\eb_\mu e_\nu=\delta_\mn +i\eta_\mn^{(+)}
=\delta_\mn +i\eta_\mn^{i(+)}\sigma_i,~~~
e_\mu \eb_\nu=\delta_\mn +i\eta_\mn^{(-)}
=\delta_\mn +i\eta_\mn^{i(-)}\sigma_i.
\label{eedag}
\end{eqnarray}
The symbol $\eta_\mn^{i(\pm)}$ is called 't Hooft's eta symbol
\cite{tHooft, tHooft3} and is concretely represented
\begin{eqnarray}
\lab{th}
\eta^{i(\pm)}_\mn=\epsilon_{i\mn 4}\pm\delta_{i\mu}\delta_{\nu 4}
\mp\delta_{i\nu}\delta_{\mu 4},
\end{eqnarray}
which is anti-symmetric and (A)SD
with respect to $\mu,~\nu$:
\begin{eqnarray}
\eta^{i(\pm)}_\mn=\pm *\eta^{i(\pm)}_\mn,
\end{eqnarray}
where $*$ is Hodge operator, and defined by $*X_\mn:=(1/2)
\epsilon_{\mn\rho\sigma}X^{\rho\sigma}$.
(For example, $*X_{12}=X_{34},~*X_{13}=X_{42},~\ldots$...)

Some formula on $e_\mu, \eta_\mn^{i(\pm)}$ are as follows:
\begin{eqnarray}
&&e_\mu \eb^{}_\nu+e_\nu \eb^{}_\mu=
\eb_\mu e_\nu+\eb_\nu e_\mu=2\de_{\mu\nu}\\
&&e_\mu \eb^{}_\nu e_\mu=-2e_\nu,~~~e_\mu e_\nu e_\mu=-2\eb_\nu\\
&&e_2 e_\mu e_2 = -\eb_\mu^{\scr\mbox{t}}\\
&&\lab{tr}
\tr (e_\mu \eb^{}_\nu)=\tr (\eb_\mu e_\nu)=2\de_{\mu\nu},\\
&&\lab{th1}
 \y^{i(+)}_{\mu\nu}=-\fr{i}{2}\tr (\si^i\eb^{}_\mu e_\nu),~~~
\y^{i(-)}_{\mu\nu}=-\fr{i}{2}\tr (\si^ie_\mu \eb_\nu)\\
&&\lab{th2}
\y^{i(+)}_{\mu\nu}\y^{j(+)}_{\mu\nu}
=\y^{i(-)}_{\mu\nu}\y^{j(-)}_{\mu\nu}=4\de^{ij}.
\end{eqnarray}
{}From now on, we often omit the symbol of the tensor product $\ot$.

Let us define the ASD equation
by using the Dirac operator,
which is based on the following observation: 
\begin{eqnarray*}
\begin{array}{ccc}
\mbox{Gauge fields are ASD.}& \Lra &
\begin{array}{c}
\mbox{The ``square'' of the Dirac operator $\cDb \cD$}\\
\mbox{commutes with Pauli matrices.}
\end{array}
\end{array}
\end{eqnarray*}
In fact the ``square'' of the Dirac operator $\cDb \cD$ is
\begin{eqnarray}
\cDb \cD=\eb^{\mu} \ot D_\mu e^\nu \ot D_\nu&=&1_{[2]}\ot D^2
+\fr{i}{2}\y^{(+)i\mu\nu}\sigma_i\ot [D_\mu,D_\nu]\nn
&=&1_{[2]}\ot D^2 +\fr{i}{2}\y^{(+)i\mu\nu}\sigma_i\ot F_{\mu\nu},
\end{eqnarray}
which gives the proof of the observation\footnote{If we treat 
SD instantons, then we have only to replace $e_\mu$ with $\eb_\mu$.}.
The condition $F_\mn=F_\mn^{(-)}$ is the ASD equation
and concretely represented as:
\begin{itemize}
\item The ASD equation ($\Lra [\cDb \cD, \sigma_i]=0$)
\begin{eqnarray}
&&F_{12}+F_{34}=0,~~~F_{13}-F_{24}=0,~~~F_{14}+F_{23}=0.
~~~(\mbox{real rep.})\\
&&\Lra F_{z_1\zb_1}+F_{z_2\zb_2}=0,~~~F_{z_1z_2}=0.
~~~~~~~~~~~~~~~~~(\mbox{complex rep.})\\
&&\Lra F_\mn+*F_\mn=0.
\label{asd}
\end{eqnarray}
\end{itemize}
The ASD equation gives the minimum of 
the Yang-Mills action:
\begin{eqnarray}
\lab{action-inst}
I_{\scr\mbox{YM}}&=&-\fr{1}{2g_{\mbox{\scr YM}}^2} 
\int d^4x~ \Tr (F_{\mu\nu}F^{\mu\nu})
=-\fr{1}{4g_{\mbox{\scr YM}}^2} \int d^4x~ \Tr (F_{\mu\nu}F^{\mu\nu}
+*F_{\mu\nu}\, *F^{\mu\nu})\nn
&=&-\fr{1}{4g_{\mbox{\scr YM}}^2} 
\int d^4x~ \Tr \left( (F_{\mu\nu}\pm *F^{\mu\nu})^2
\mp 2F_{\mu\nu}\, *F^{\mu\nu}\right)\nn
&=&-\fr{1}{4g_{\mbox{\scr YM}}^2} 
\int d^4x~ \Tr (F_{\mu\nu}\mp *F^{\mu\nu})^2
\pm\fr{8\pi^2}{g_{\mbox{\scr YM}}^2}\unb{\left[\fr{-1}{16\pi^2}
\int d^4x~ \Tr  (F_{\mu\nu}\, *F^{\mu\nu})\right]}_{=:\nu[A_\mu]}
\end{eqnarray}
The condition that the square part in the final line should
be zero is just the same as the ASD equation. 
The second term $\nu[A_\mu]$ in the RHS takes an integer.
The gauge field should be pure-gauge at infinity,
that is, $A_\mu\ap g^{-1}\del_\mu g,~^\exists g\in SU(N)$.
(then $F_\mn\ap 0$.)
Then the integer $\nu[A_\mu]$ is called the {\it instanton number}.
Here we consider the instantons whose instanton number is $-k$.
($k$ ASD instantons):
\begin{itemize}
\item instanton number (the gauge field behaves at infinity as
pure gauge: $A_\mu\ap g^{-1}\del_\mu g,~^\exists g\in SU(N)$)
\begin{eqnarray}
\lab{instantonsuu}
\nu[A_\mu]&:=&-\fr{1}{16\pi^2}\int d^4x~ \Tr (F_{\mu\nu}\, *F^{\mu\nu})
~=~-\fr{1}{8\pi^2}\int \Tr (F\we F)\nonumber\\
&=&-\fr{1}{8\pi^2}\int d\Tr (A\we dA+\frac{2}{3}A\we A\we A)\nn
&\st{\scr\mbox{Stokes}}{=}&-\fr{1}{8\pi^2}\int_{S^3} \Tr (\unb{A\we dA
+\frac{2}{3}A\we A\we A}_{=A\we F-\frac{1}{3}A\we A\we A})\nonumber\\
&=&\fr{1}{24\pi^2}\int_{S^3} \Tr ((g^{-1}dg)\we (g^{-1}dg)\we (g^{-1}dg))
\in \Z\nonumber\\
&=&-k.
\end{eqnarray}
\end{itemize}
Furthermore we need the condition
that $D^2$ has the inverse:
\begin{itemize}
\item $D^2$ is invertible (there exists the  Green function $G(x,y)$
of $D^2$.):
\begin{eqnarray}
D^2_x ~^\exists G(x,y)=-\delta(x-y),~~~G(x,y) \simeq \cO(r^{-2}).
\end{eqnarray}
\end{itemize}
Exactly speaking, more detailed conditions are needed,
which is written in \cite{DoKr}.

The gauge transformation is defined as usual:
\begin{itemize} 
\item  gauge transformation
\begin{eqnarray}
A_\mu\rar g^{-1}A_\mu g + g^{-1}\del_\mu g,~~~g(x)\in SU(N).
\end{eqnarray}
\end{itemize}

Instantons whose instanton number is $-k$
is specified by finite parameters up to the freedom of the 
gauge transformation
The space of the parameters is represented by $\cM_{N,k}^{\mbox{inst}}$

Let us summarize instantons:
\vs

\fbox{{\bf Instantons}}
\begin{eqnarray}
\cM_{N,k}^{\mbox{inst}}&=&
\fr{\left\{
\ba{c|l}
A_\mu^{(N,k)}&
\ba{l}{\mbox{ASD equation}}
\\
A_\mu~:~N\ti N \mbox{ anti-Hermite matrices}\\
\dis \nu[A_\mu]
=-k\\
\cDb\cD {\mbox{ : invertible}}
\ea
\ea\right\}}{(A_\mu\sim g^{-1}A_\mu g+g^{-1}\del_\mu g,
~~~g(x)\in SU(N))}\nonumber\\
\dim\cM_{N,k}^{\mbox{inst}}&=&\left\{\ba{ll}4Nk-N^2+1&~~~N\leq
2k\\4k^2+1&~~~N> 2k \ea \right.
\end{eqnarray}

The dimension of instanton moduli space $\dim\cM_{N,k}^{\mbox{inst}}$
is calculated by using the results of 
Atiyah-Singer index theorem
\begin{eqnarray}
\lab{instdim}
\dim\cM_k^{\mbox{inst}}=4hk-\fr{\chi+\si}{2}\dim G,
\end{eqnarray}
where $h$, $\chi$ and $\si$
are the dual Coxeter number of the gauge group $G$,
Euler number of the base manifold and signature of the
base manifold. (See Diagram 1, 2.) 

\begin{center}
Diagram 1: simply-connected compact simple Lie group\\
\begin{tabular}{|c|c|c|c|} \hline
Lie group $G$&rank&dimension&the dual Coxeter number $h$\\ \hline\hline
$SU(N)~(N\geq 2)$&$N-1$&$ N^2 -1$ &$N$\\ \hline
$SO(N)~(N\geq 2)$&$\dis\left[\frac{N}{2}\right]$&$\dis\frac{1}{2}N(N-1)$
&$N-2~(N\geq 4)$\\ \hline
$Sp(N)_{[2N]\ti [2N]}$&$N$&$N(2N+1)$&$N+1$\\ \hline
$G_2$&2&14&4\\ \hline
$F_4$&4&52&9\\ \hline
$E_6$&6&78&12\\ \hline
$E_7$&7&133&18\\ \hline
$E_8$&8&248&30\\ \hline
\end{tabular}\\
\end{center}
\vs
\begin{center}
Diagram 2: Euler numbers $\chi$ and signatures $\si$ of four-manifolds\\
\begin{tabular}{|c|c|c|} \hline
four-manifold&Euler number $\chi$&signature $\si$\\\hline\hline
$T^4$&0&0\\\hline  
$S^4 $&2&0\\\hline
$\C P_2$&3&$-$1\\\hline
$S^2\ti S^2$&4&0\\\hline
$K3$&24&$-$16\\\hline
\end{tabular}\\
\end{center}

\vs
\unl{\bf (ADHM)}
\vs

Next let us define ADHM data
which is the dual of instantons on ``0-dimensional space''
as we mention in the end of the previous subsection.
That is why ADHM side contains no derivative.

Let us define the (dual) ``0-dimensional Dirac operator'' $\na$ 
as follows:
\begin{eqnarray}
\label{cxd}
\nabla(x):=Cx-D,
\end{eqnarray}
where
\begin{eqnarray}
x:=x^\mu \ot e_\mu=\left(\ba{cc}x^4-ix^3&
-(x^2+ix^1)\\x^2-ix^1&x^4+ix^3\ea\right)
=\left(\ba{cc}\zb_2&-z_1\\
\zb_1&z_2\ea\right)
\end{eqnarray}
and $x^\mu$ or $z_{1,2}$ represents the coordinates of $\R^4$ or $\C^2$,
respectively.
Here the symbol $x$ in Eq. (\ref{cxd}) means 
precisely $x\ot 1_{[k]}$. This kind of omission 
is sometimes used in this appendix.
The matrix $C$ is $(N+2k)\ti 2k$ constant matrix:
\begin{eqnarray}
\lab{cd}
C=\left(\ba{c}0_{[N]\ti [2k]}\\1_{[2k]\ti [2k]}\ea\right)_{[N+2k]\ti[2k].}
\end{eqnarray}
Hence the matrix $D$ has all the information and 
is called {\it ADHM data}
and is represented in various ways:
\begin{eqnarray}
D&=&\left(\ba{c}-S_{[N]\ti [2k]}\\
          T_{[2k]\ti [2k]}
          \ea\right)_{[N+2k]\ti[2k]}
=\left(\ba{c}-S_{[N]\ti [2k]}\\e_{\mu [2]\ti[2]}\ot T^\mu_{[k]\ti[k]} 
       \ea\right)_{[N+2k]\ti[2k]}\nonumber\\
&=&\left(\ba{cc}-I^\dagger&-J\\
                T^4-iT^3&-(T^2+iT^1)\\
                T^2-iT^1&T^4+iT^3\ea\right)_{[N+2k]\ti[2k]}
=\left(\ba{cc}-I^\dagger &-J\\
          B_2^\dagger&-B_1\\
          B_1^\dagger&B_2
          \ea\right)_{[N+2k]\ti[2k],}
\end{eqnarray}
where the matrices $I,J, B_{1,2}$ are $k\ti N, N\ti k, k\ti k$
complex matrices and  $B_{1,2}$ is the complex representation
of $T_\mu$ ($k\ti k$ Hermitian matrix).
(Please do not confuse
the matrix $D$ in eq. (\ref{cxd}) with the covariant derivative $D_\mu$
in (instanton).)

Then the ``0-dimensional Dirac operator'' can be rewritten as
\begin{eqnarray}
\na (x)&=&\left(\ba{c}S\\
            e_\mu\ot (x^\mu-T^\mu)\ea\right)
=\left(\ba{cc}I^\dagger &J\\
          \zb_2-B_2^\dagger&-(z_1-B_1)\\
          \zb_1-B_1^\dagger&z_2-B_2
          \ea\right),\nn
\na (x)^\dagger&=&\left(\ba{cc}S^\dagger&
            \eb_\mu\ot (x^\mu-T^\mu)\ea\right)
=\left(\ba{ccc}I&z_2-B_2&z_1-B_1\\
          J^\dagger&-(\zb_1-B_1^\dagger)&\zb_2-B^\dagger_2
          \ea\right).
\label{Dir_op}
\end{eqnarray} 

Now let us introduce the (dual) ``0-dimensional ASD equation,''
in the similar way as instantons.
We take the condition ``$\nabla^\dagger \nabla$ should commutes
with Pauli matrices'' as the dual ASD equation. 
This is concretely written down as:
\begin{itemize} 
\item ADHM equation (``0-dimensional ASD equation''):
\begin{eqnarray}
&&\left\{\ba{l}~\dis [T_1,T_2]+[T_3,T_4]-\fr{i}{2}
             (I^\dagger I-JJ^\dagger)=0,\\
             ~\dis [T_1,T_3]-[T_2,T_4]-\fr{1}{2}(IJ+J^\dagger J^\dagger)
             =0,\\
             ~\dis [T_1,T_4]+[T_2,T_3]-\fr{i}{2}(IJ-J^\dagger J^\dagger)
             =0.\ea\right.~~~~~~(\mbox{real rep.})\label{adhm}\\
&&\Lra\left\{\ba{l}~\dis 
(\mu_{\scr{\R}} :=)~~~[B_1,B_1^\dagger]+[B_2,B_2^\dagger]
+II^\dagger-J^\dagger J=0,\\
~\dis(\mu_{\scr{\C}} :=)~~~[B_1,B_2]+IJ=0.
~~~~~~~~~~~~~(\mbox{complex rep.}) \ea\right.
\label{adhm2}\\
&&\Lra\tr (\si^i( S^\da S+T^\da T))=0.~~~~~~~~~~~~~~~~~~~~~( ^\forall i=1,2,3)
\label{adhm3}
\end{eqnarray} 
\end{itemize}
The LHS in the complex representation is
often represented as $\mu_{\scr{\R}},\dis \mu_{\scr{\C}}$
in the context of hyperK\"ahler quotient \cite{HKLR}.
Here we note that ADHM data $T^\mu, B_{1,2}$ always appear
in pair with the coordinates $x^\mu,z_{1,2}$
and therefore the existence of the commutators of ADHM data
implies that of the coordinates,
such as $\mu_{\scr{\R}}=-[z_1,\zb_1]-[z_2,\zb_2]$.
The commutator of the coordinate is zero on commutative space,
of course, however, on noncommutative spaces this causes various
important results.

Now we get
\begin{eqnarray}
\lab{adhm1}
(\nabla(x)^\da\nabla(x))&=&
\left(\ba{cc}
\Box&0_{[k]}\\
0_{[k]}&\Box\ea\right)_{[2k]\ti [2k],}\\
\Box(x)_{[k]}&=&\half \tr(D^\dagger D)+2T_\mu x^\mu +\vert x\vert^2.
\nonumber
\label{box}
\end{eqnarray}

As in instanton case, there needs to be the following condition:
\begin{itemize}
\item $\Box$ is invertible (The existence of the inverse matrix $f$)
\begin{eqnarray}
\label{0green}
\Box ^{\exists}f=1~~~\Lra ~~~f(x)_{[k]}=\Box^{-1}\simeq \cO(r^2).
\end{eqnarray}
\end{itemize}

\noindent
There exists the transformation
which leaves 
ADHM equation and the constant matrix $C$
and is called the ``gauge transformation'' of ADHM data:
\begin{itemize}
\item``gauge transformation'' of ADHM data
\begin{eqnarray}
\lab{equiv}
I\rar R^\dagger I Q^\dagger,~~~J\rar QJR,~~~T_\mu\rar R^\dagger T_\mu R,
~~~~~~~Q\in SU(N),~R \in U(k)
\end{eqnarray}
\end{itemize}

Let us consider the quotient space of the ADHM data
by the equivalent relation (\ref{equiv})
and represent it as $\cM_{k,N}^{\mbox{\scr{ADHM}}}$,
which is called the moduli space of ADHM data.
The dimension of the moduli space $\dim\cM_{k,N}^{\mbox{\scr{ADHM}}}$
can be easily calculated from the constraints:
\begin{itemize}
\item For $N\leq 2k$
\begin{eqnarray}
\dim\cM_{k,N}^{\mbox{\scr{ADHM}}}&=&\unb{2\cdot
2k(N+2k)}_{D}-\unb{3k^2}_{(\ref{adhm})}-\unb{4k^2}_{T_\mu^\dagger=T_\mu}
-\unb{(N^2-1)}_{Q}-\unb{k^2}_{R}\nn
&=&4Nk-N^2+1.
\end{eqnarray}
\item For $N> 2k$\\The same calculation as $N\leq 2k$ case 
over-subtracts
the degree of freedom of $U(N-2k)$ $N\leq 2k$,
\begin{eqnarray}
\dim\cM_{k,N}^{\mbox{\scr{ADHM}}}&=&4Nk-N^2+1+(N-2k)^2=4k^2+1.
\end{eqnarray}
\end{itemize}
This shows the beautiful coincident:
$\dim\cM_{k,N}^{\mbox{\scr{ADHM}}}
=\dim\cM_{N,k}^{\mbox{inst}}$.

Let us summarize on the ADHM data:
\vs

\fbox{{\bf ADHM data}}
\begin{eqnarray}
\cM_{k,N}^{\mbox{\scr{ADHM}}}&=&
\fr{\left\{\ba{l|l}D^{(k,N)}=\left(\ba{c}-S^{(k,N)}\\
e_\mu\ot
T^{\mu(k)}\ea\right)&
\ba{l}\mbox{ADHM equation}
\\
T^\mu~:~k\ti k~\mbox{Hermite matrix}\\
S~:~N\ti 2k \mbox{ complex matrices}\\
\nabla^\da\nabla \mbox{ is invertible.}\ea \ea\right\}}
{(I\sim R^\dagger I Q^\dagger,~
J\sim QJR,~T_\mu\sim R^\da T_\mu R,~~~Q\in SU(N),
~R\in U(k))}\nonumber\\
\dim\cM_{k,N}^{\mbox{\scr{ADHM}}}&=&\left\{\ba{ll}
4Nk-N^2+1&~~~N\leq2k\\4k^2+1&~~~N> 2k \ea \right.
\end{eqnarray}

The goal of this subsection is to outline the proof of 
\begin{eqnarray}
\cM_{N,k}^{\mbox{inst}}
\st{1:1}{=}\cM_{k,N}^{\mbox{\scr{ADHM}}}.
\end{eqnarray}

For simplicity, let us take $N\leq2k$ case. 

\vs
\unl{\bf (ADHM)$\longr$(Instanton)}
\vs

Now we show the detailed discussion of
the main part of ADHM construction:
From given ADHM data $S^{(k,N)},T_\mu^{(k)}$ to
instantons $A_\mu=A_\mu(S,T)$.
We present how to construct the gauge field
from the ADHM data and then check that the gauge field
satisfies all of the properties on instantons.

First let us consider the following ``0-dimensional Dirac equation'':
\begin{eqnarray}
\lab{0dirac}
\nabla^\da V=0,
\end{eqnarray}
where $V$ is called the ``0-dimensional Dirac zero-mode.''
The number of the normalized zero-mode $V$ is
$(N+2k-2k=)~N$ and we can arrange the independent $N$ solution
at each row  and consider $V=V_{[N+2k]\ti [N]}$
The normalization condition is 
\begin{eqnarray}
\lab{0norm}
V^\da V&=&1_{[N]}.
\end{eqnarray}
Taking ``0-dimensional Dirac equation,'' normalization condition and 
 and the invertibility of $\na^\dagger\na$ into account,
we get the following relation:
\begin{eqnarray}
\lab{0comp}
VV^\da =1_{[N+2k]}-\nabla f \nabla^\dagger.
\end{eqnarray}
In order to prove it, let us introduce the convenient matrix $W$ as
\begin{eqnarray}
W:=\left(\ba{cc}\na&V\ea\right)_{[N+2k]\ti[N+2k].}
\end{eqnarray}
From Eqs. (\ref{0green}), (\ref{0dirac}), (\ref{0norm}),
the $(N+2k)$ rows of the matrix $W$ is 
independent to each other and there exists the inverse
of $W$. Hence
\begin{eqnarray}
W(W^\dagger W)^{-1}W^\dagger\equiv 1~
\Lra~ V(\unb{V^\dagger V}_{=1})^{-1}V^\dagger 
+ \nabla (\nabla^\dagger \nabla)^{-1}\na^\dagger=1,
\end{eqnarray}
which implies (\ref{0comp}).

The condition (\ref{0comp}) shows the completeness of
the each rows of $W$ in $(N+2k)$-dimensional vector space
and is called the {\it completeness condition}.

The matrix $W$ simplifies the relations:
\begin{eqnarray}
W^\dagger W \equiv \left(\ba{cc}
\na^\dagger\na&\na^\dagger V\\
V^\dagger \na&V^\dagger V
\ea
\right) = \left(\ba{cc}
1_{[2]}\ot \Box_{[k]}&O\\
O&1_{[N]}
\ea
\right).
\end{eqnarray}

Here let us introduce 
\begin{eqnarray}
P&:=&VV^\dagger,\\
V&=&\left(\ba{c}u_{[N]\ti[N]}\\v_{\,[2k]\ti[N]}\ea\right)=
\left(\ba{c}u_{[N]\ti[N]}\\v_{1\,[k]\ti[N]}\\v_{2\,[k]\ti[N]}\ea\right),\\
v&=&C^\dagger V,
\end{eqnarray}
where $P$ is the projection in $(N+2k)$-dimensional space onto
the $N$-dimensional subspace.

From the zero-mode $V$, we can construct the gauge field $A_\mu$ as
\begin{eqnarray}
\lab{4gauge}
A_\mu=V^\da \del_\mu V\ap\cO(r^{-1}).
\end{eqnarray}
The normalization condition (\ref{0norm}) shows $A^{\da}_\mu=-A_\mu$ 
(anti-Hermitian) and $G=U(N)$. 

The geometrical meaning of (\ref{4gauge}) is as follows.
The covariant derivative on the 
$N$-dimensional subspace spanned by $V_u~(u=1,\ldots,N)$
could be defined from
the natural, trivial covariant derivative $\del_\mu$
on $(N+2k)$-dimensional space as the projection onto the 
$N$-dimensional space: $D_\mu:=P\del_\mu$.
By acting the covariant derivative to
the function $s(x)$ restricted on
the subspace, which is spanned by $V^u s_u(x)$,
we get
\begin{eqnarray}
D_\mu(V^u s_u)=P\del_\mu (V^v s_v)
&=&V^uV_u^\dagger(V^v(\del_\mu s_v)+(\del_\mu V^v)s_v)\nn
&=&V^u(\delta_{uv}\del_\mu+(V_u^\dagger\del_\mu V^v))s_v.
\end{eqnarray}
Here the second term of the RHS $V_u^\dagger\del_\mu V^v$
should be just the gauge field $A_{\mu u}^{~~~v}$
which is consistent with (\ref{4gauge}).
The important point here is that
we take the Dirac zero-mode as the basis of 
the subspace.

Here we present some important relations:
\begin{eqnarray}
\label{delf}
\del_\mu f&=&-f(\del_\mu f^{-1}) f,\\
\lab{jyuuyou}
e_\mu\nabla^\da Ce_\mu &=&-2C^\da \nabla,\\
D_\mu V^\da&=&V^\da \del_\mu(VV^\da)=-V^\da Cf e_\mu \nabla^\da,\lab{dv}\\
\lab{ddv}
D^2V^\da&=&-4V^\da CfC^\da,\\
\lab{lapv0}
D^2u^\dagger&=&0,\\
\lab{logdet}
\Tr (F_{\mu\nu}F^{\mu\nu})&=&-\del^2\del^2\log\det f.
\end{eqnarray}
Eqs. (\ref{delf})-(\ref{ddv}) holds even when $C$
is not the canonical form
(\ref{cd}).
The proof of (\ref{logdet}) is found
in \cite{DHKM}. (See also \cite{CGOT, Osborn2}.)

So far we define how to construct the gauge field from the ADHM
data via ``0-dimensional Dirac equation.''
Next let us check this gauge field is the $G=SU(N)$, $k$-instanton.

First we check the anti-self-duality 
by calculating the field strength $F_{\mu\nu}$ 
from $A_\mu=V^\dagger \partial_\mu V$:
\begin{eqnarray}
\lab{curv1}
F&=&dA+A\wedge A\nonumber\\
&=&dV^\dagger \wedge dV+V^\dagger dV\wedge V^\dagger dV
=dV^\dagger \wedge dV-dV^\dagger V\wedge V^\dagger dV\nonumber\\
&=&dV^\dagger(1-VV^\dagger)\wedge dV
\st{(\ref{0comp})}{=}dV^\dagger\nabla f\nabla^\dagger \wedge dV\nonumber\\
&\st{(\ref{0dirac})}{=}&V^\dagger (d\nabla) f\wedge (d\nabla^\dagger) V
=V^\da Ce_\mu dx^\mu f\wedge dx^\nu \eb^{}_\nu C^\dagger V\nonumber\\
&\st{(\ref{box})^{-1}}{=}&V^\da C dx^\mu f\wedge dx^\nu e_\mu 
\eb^{}_\nu
 C^\dagger V
\st{(\ref{eedag})}{=}
iV^\da Cf\unb{\eta^{(-)}_{\mu\nu}}_{\scr\mbox{ASD}}
C^\da V dx^\mu \wedge dx^\nu,\\
 F_{\mu\nu}&=&2iV^\da Cf\eta^{(-)}_{\mu\nu}C^\da V
=2iv^\dagger f\eta^{(-)}_{\mu\nu} v.
\end{eqnarray}

Next in order to show the gauge field $A_\mu$ behaves at infinity
as pure gauge, let us examine the behavior at infinity. 
At the region $\vert x\vert\rar \infty$, ``0-dimensional Dirac equation''
(\ref{0dirac}) becomes $x^\dagger C^\dagger V\ap 0$, and hence $v\ap 0$. 
Then the normalization condition (\ref{0norm}) 
shows $u\ap {}^\exists g(x)\in U(N)$ and 
\begin{eqnarray}
A_\mu\ap g^{-1}\del_\mu g.
\end{eqnarray}
Multiplying the both hands of (\ref{0dirac}) by $x$, 
we get
\begin{eqnarray}
\lab{vpr}
V^\da C=\fr{V^\da Dx^\da}{\vert x\vert^2}
\end{eqnarray}
Hence the behavior of $V$ at infinity is summarized as
\begin{eqnarray}
V_x=\left(\ba{c}u_x\\v_x\ea\right)\ap\left(\ba{c}\cO(1)
\\\cO(r^{-1})\ea\right).
\end{eqnarray}

Instanton number is calculated by using Eq. (\ref{logdet}):
\begin{eqnarray}
\nu[A_\mu]&=&-\fr{1}{16\pi^2}\int d^4x~\Tr( F_{\mu\nu}\, *F^{\mu\nu})
=-\fr{1}{16\pi^2}\int d^4x~\del^2\del^2\log\det f\nonumber\\
&=&-\fr{1}{16\pi^2}\int dS^\mu_x\del_\mu\del^2\Trk\log \unb{f}_
{\ap\vert x\vert^{-2}}=-\fr{8}{16\pi^2}\int d\Omega_x~\Trk 1_{[k]}~=~-k,
\end{eqnarray}
where $d\Om_x $ denotes surface element of $x$-space whose radius is 1
and $\dis\int d\Om_x=2\pi^2$.
(The surface area $S^{n-1}$ of the $n-1$-dimensional 
sphere with the radius $r$ is
Vol$(S^{n-1}_r)=\dis\fr{2\pi^{\fr{n}{2}}}{\Gamma(\fr{n}{2})}r^{n-1}$, 
where $\Gamma(1)=1,~\Gamma(\half)=\sqrt{\pi}$.)

The invertivilities of $D^2$ is proved from the existence of 
the Green function of $D^2$ which is concretely represented as \cite{CFGT}:
\begin{eqnarray}
\lab{green4}
G(x,y)=\fr{1}{4\pi^2}\fr{V_x^\dagger V_y}{\vert x-y\vert^2},
\end{eqnarray}
which satisfies $D^2G(x,y)=-\de(x-y)$.

In order to prove it, let us calculate the LHS first:
\begin{eqnarray}
\lab{ddg}
&&D^2G(x,y)\nn
&&=\fr{1}{4\pi^2}\left\{\unb{\del_x^2
\left(\fr{1}{\vert x-y\vert^2}\right)
}_{-4\pi^2\de(x-y)}
V_x^\dagger V_y+2\del_{x\mu}\left(\fr{1}{\vert x-y\vert^2}\right)
D^\mu_x V_x^\dagger
V_y+\fr{1}{\vert x-y\vert^2}D_x^2V_x^\dagger V_y\right\}.~~~~~~~~~~
\end{eqnarray}
Here let us discuss both in $x=y$ case and in $x\neq y$ case.
\begin{itemize}
\item When $x=y$, using Eq. (\ref{dv}), 
\begin{eqnarray}
D^\mu_x V_x^\dagger V_y=-V^\dagger Cfe^\mu \unb{\nabla_{x=y}^\dagger V_y}_{=0}.
\end{eqnarray}
Hence the second and third terms in Eq. (\ref{ddg}) vanish.
Therefore, 
\begin{eqnarray}
D^2G(x,y)=-\de(x-y)\unb{V_{x=y}^\dagger V_y}_{=1}=-\de(x-y).
\end{eqnarray}
\item When $x\neq y$, by using Eq. (\ref{ddv}), 
\begin{eqnarray}
D^2G(x,y)&=&-\unb{\de(x-y)}_{=0}V_x^\dagger V_y+\fr{1}{4\pi^2}\left\{
2\del_{x\mu}\left(\fr{1}{\vert x-y\vert^2}\right)D^\mu_x V_x^\dagger
V_y+\fr{1}{\vert x-y\vert^2}D_x^2V_x^\dagger V_y\right\}\nonumber\\
&=&\fr{1}{4\pi^2}\left\{\fr{(x-y)^\mu}{\vert x-y\vert^4}
(V_x^\dagger Cfe_\mu\nabla_x^\dagger)V_y
+\fr{1}{\vert x-y\vert^2}(-4V_x^\dagger CfC^\dagger)V_y\right\}\nonumber\\
&=&\fr{1}{\pi^2\vert x-y\vert^2}V_x^\dagger Cf
\left\{\fr{(x-y)^\mu}{\vert x-y\vert^2}e_\mu
(\nabla_y^\dagger+(x-y)^\dagger C^\dagger)-C^\dagger \right\}V_y\nonumber\\
&=&\fr{1}{\pi^2\vert x-y\vert^2}V_x^\dagger Cf
\left\{\unb{\fr{(x-y)^\mu}{\vert x-y\vert^2}(x-y)^\nu
(\de_\mn+i\y^{(-)}_{\mn})}_{=1+0}-1\right\}C^\dagger V_y~=~0.\nn
\end{eqnarray}
\end{itemize}
Now Eq. (\ref{green4}) is proved. 
 
The transformation of $V$
\begin{eqnarray}
V\rar Vg,~~~g(x)\in SU(N)
\end{eqnarray}
preserves Eqs. (\ref{0dirac})-(\ref{0comp}) and 
is equal to the gauge transformation of $A_\mu$:
\begin{eqnarray}
A_\mu\longr A^{\pr}_\mu&=&(Vg)^\da\del_\mu (Vg)
=g^{-1}\unb{(V^\da \del_\mu V)}_{A_\mu}g+g^{-1}\del_\mu g.
\end{eqnarray}

We note that this discussion (ADHM) $\longr$ (Instanton)
holds even when the matrix $C$ is not the canonical form
(\ref{cd}) but a general complex matrix.
This restriction does not lose generality
because it is always taken by the following
degree of freedom.
Now let us suppose that  $C$ is a general complex matrix
and consider the following transformation:
\begin{eqnarray}
\lab{amb}
&&D\rar D^\pr=\cU D\cR,~~~C\rar C^\pr=\cU C\cR,
~~~V\rar V^\pr=\cU V\nonumber\\
&&\cU \in U(N+2k),~~\cR\in GL(k;\C)\ot 1_{[2]}.
\end{eqnarray}
This preserve the Eqs. (\ref{adhm}), (\ref{0dirac})-(\ref{0comp}).
By using this degree of freedom, we can set $C$
the canonical form (\ref{cd}).

\vs
\unl{\bf (Instanton)$\longr$(ADHM)}
\vs

Here we discuss the inverse construction 
(Instanton)$\longr$(ADHM), that is,
we construct the ADHM data $S=S(A),~T_\mu=T_\mu(A)$
from given $SU(N)$, $k$-instantons $A_\mu^{(N,k)}$.
We have to show that the $S,T_\mu$ have 
all the properties of ADHM data.

First let us consider the massless Dirac equation
in the background of instanton $A_\mu$:
\begin{eqnarray}
\lab{4dirac}
\cDb \psi&=&0.
\end{eqnarray}
The solution $\psi$ is called Dirac zero-mode and 
it is shown that there are independent $k$ solutions by
Atiyah-Singer index theorem.
Hence we can consider $\psi$ as $2N\ti k$ matrix 
whose $k$ rows are consist of the normalized $k$ zero-mode
and the normalization condition is
\begin{eqnarray}
\lab{4norm}
\int d^4 x~\psi^\dagger\psi=1_{[k]}.
\end{eqnarray}
The completeness condition is
\begin{eqnarray}
\label{4comp}
\psi(x)\psi^\dagger(y)=\de(x-y)+\cD G(x,y)\st{\lar}{\cDb},
\end{eqnarray}
where $G(x,y)$ is Green function of $D^2$.
This condition is guaranteed by the normalizability of $\psi$
and the invertibility of $D^2$ as in (ADHM) $\rar$ (instanton).

Here we introduce the following symbol on the spinor index of $\psi$:
\begin{eqnarray}
\wtps:=\psi^t \cdot e_2,
\end{eqnarray}
where $\psi^t$ is the transposed matrix of $\psi$ w.r.t. spinor indices
and is considered as $N\ti 2k$ matrix.

From the zero-mode $\psi$, we can construct ADHM data $S,T$ as
\begin{eqnarray}
\lab{adhms}
\wtps&\ap&
-\fr{g^\dagger Sx^\da}{\pi\vert x\vert^4}+\cO(r^{-4}),\\
\lab{adhmt}
T_\mu&=&\int d^4x~\psi^\da x_\mu\psi,
\end{eqnarray}
where $g$ is just the $N\times N$ matrix which appears
in the asymptotic behavior of $A_\mu$: $A_\mu\ap g^{-1}\del_\mu g$.
The matrices $S,T_\mu$ are actually $N\ti 2k,k\ti k$.
We can easily show that $T_\mu$ is Hermitian.

Let us check that the data (\ref{adhms}) and (\ref{adhmt})
satisfies ADHM equation (\ref{adhm3}).
In order to do so, we calculate first
\begin{eqnarray}
\lab{com}
T^\mu T^\nu=\int d^4x ~x^\mu \psi^\dagger(x) \psi(x) \int d^4y
~y^\nu \psi^\dagger(y) \psi(y).
\end{eqnarray}
Substituting the completeness condition (\ref{4comp})
into Eq. (\ref{com}), we get 
\begin{eqnarray}
\lab{tt}
T^\mu T^\nu=\int d^4x~x^\mu x^\nu \psi^\dagger(x)\psi(x)+\int d^4x
d^4y~x^\mu y^\nu \psi^\dagger(x) e^\rho \eb^{\si}
D_\rho G(x,y)\st{\lar}{D}_\si \psi(y).
\end{eqnarray}

The explanation of the integrals are done by restricting
the integral region within $\vert x\vert\leq R_x,~\vert y\vert\leq
R_y$ and taking the limit $R_x\rar \infty,~R_y\rar \infty$.
This integral contains diverse parts which are
dropped out in the contraction by 't Hooft's eta symbol
and cause no problem.
The twice integration of the second term of (\ref{tt})
leads to
\begin{eqnarray}
\lab{term2}
{\mbox{The second term of }}(\ref{tt})
&=&\int x^\mu d^4x~y^\nu d^4y~\tr \left(\eb^{\rho} \wtps^\dagger(x)
D_\rho G(x,y)\st{\lar}{D}_\si \wtps(y)e^\si\right)\nonumber\\
&=&-\int x^\mu dS_x^\rho~ y^\nu dS_y^\si~ \tr \left(\eb^{\rho}
\wtps^\dagger(x) G(x,y)\wtps(y)e^\si\right)\nonumber\\
&&+\int  d^4x ~y^\nu dS_y^\si~ \tr \left(\eb^{\mu}
\wtps^\dagger(x) G(x,y)\wtps(y)e^\si\right)\nonumber\\
&&+\int x^\mu dS_x^\rho ~d^4y ~\tr \left(\eb^{\rho}
\wtps^\dagger(x) G(x,y)\wtps(y)e^\nu\right)\nonumber\\
&&-\int  d^4x~ d^4y~ \tr \left(\eb^{\mu}
\wtps^\dagger(x) G(x,y)\wtps(y)e^\nu\right),
\end{eqnarray}
where the volume integral and the surface integral are taken within
the region $\vert x\vert\leq R_x,~\vert y\vert\leq R_y$
and within $\vert x\vert= R_x,~\vert y\vert= R_y$, respectively.

Here let us take $R_y\rar\infty$ first.
Then the first and second terms of (\ref{term2}) become
\begin{eqnarray}
\int \unb{y^\nu dS_y^\si}_{\cO(R_y^4)}~\unb{G(x,y)}_{\cO(R_y^{-2})}
\unb{\wtps(y)}_{\cO(R_y^{-3})}\ap
\cO(R_y^{-1})\st{R_y\rar\infty}{\longr}0.
\end{eqnarray}
The third term of (\ref{term2}) behaves
\begin{eqnarray}
\int \unb{d^4 y}_{\cO(R_y^4)}~\unb{G(x,y)}_{\cO(R_y^{-2})}
\unb{\wtps(y)}_{\cO(R_y^{-3})}\ap
\cO(R_y^{-1}),
\end{eqnarray}
which shows that the integration converses.
In order to evaluate the integral,
let consider the following differential equation:
\begin{eqnarray}
\lab{chi}
D^2\wtc (x)=-4\pi\wtps(x),~~~\wtc (x)\ap0.
\end{eqnarray}
From Eq. (\ref{adhms}), we can see
\begin{eqnarray}
\label{chibc}
\wtc (x)\ap -\fr{g^\dagger Sx^\da}{\vert x\vert^2}.
\end{eqnarray}
Eq. (\ref{chi}) is equivalent to 
\begin{eqnarray}
\int d^4 y~G(x,y)\wtps(y)=\fr{1}{4\pi}\wtc (x),
\end{eqnarray}
and the third term of (\ref{term2}) becomes
\begin{eqnarray}
&&\int x^\mu dS_x^\rho ~d^4y ~\tr \left(\eb^{\rho}
\wtps^\dagger(x) G(x,y)\wtps(y)e^\nu\right)\nonumber\\
&&\st{R_y\rar\infty}{\longr}\fr{1}{4\pi}\int x^\mu dS_x^\rho ~\tr
\left(\eb^{}_\rho\wtps^\dagger(x) \wtc (x)e^\nu\right)
=\fr{1}{4\pi^2}\int x^\mu\fr{x^\rho}{\vert x\vert}\vert x\vert^3
d\Om_x ~\tr \left(\fr{\eb^{}_\rho xS^\da Sx^\da e^\nu}
{\vert x\vert^6}\right)\nonumber\\
&&\st{R_x\rar\infty}{\longr}\fr{1}{4\pi^2}\int \fr{x^\mu}{\vert
x\vert^2}d\Om_x ~\tr \left(S^\da Sx^\da e^\nu\right)
=\fr{1}{8}\tr \left(S^\da S\eb^{\mu} e^\nu\right).
\end{eqnarray}

Now let us contract the both side of Eq. (\ref{tt}) 
by $\y^{(+)}_{\mu\nu}$. Though the first term of Eq. (\ref{tt})
and the fourth term (\ref{term2}) diverse,
they drop out by the contraction by $\y^{(+)}_{\mu\nu}$
which is SD and anti-symmetric w.r.t. $\mu\lra\nu$.
The fourth term of (\ref{term2}) is ASD
because $\eb^{}$ moves to the right side of $e$
through the spinor trace. Then we get
\begin{eqnarray}
\y^{(+)}_{\mu\nu}\left(T^\mu T^\nu-\fr{1}{8}\tr
\left(S^\da S\eb^{\mu} e^\nu \right)\right)=0.
\end{eqnarray}
By using the relations on 't Hooft's eta symbol
(\ref{th1}), (\ref{th2}), we obtain ADHM equation:
\begin{eqnarray}
\tr \left(\si^i(S^\da S+T^\da T)\right)=0.
\end{eqnarray}

We can also check the invertibility 
of $\nabla^\da\nabla$ 
basically showing $f\sim (\del^2)^{-1} \psi^\dagger\psi$
as Eq. (\ref{green0}),
which shows the existence of the inverse $f$ of $\nabla^\da\nabla$.

The transformation for $g,\psi$
\begin{eqnarray}
g\rar Q^\dagger g,~~~\psi\rar\psi R,~~~~~~Q\in SU(N),~~~R\in U(k)
\end{eqnarray}
preserves Eqs. (\ref{4dirac})-(\ref{4comp}) 
and $A_\mu\ap g^{-1} \del_\mu g$
and hence is ``the gauge transformation'' 
for $S,T_\mu$.

\vs
\unl{\bf Completeness:  (ADHM)$\longr$(Instanton)$\longr$(ADHM)}
\vs

In this section, we prove the completeness, that is,
the composite transformation: ADHM construction and 
the inverse construction should be identity.
We start with a given ADHM data $ S^{(k,N)}, ~T_\mu^{(k)}$
and construct the instantons $A_\mu= A_\mu (S,T) $
in ADHM construction and 
reconstruct from the instantons 
ADHM data $S^{\pr(k^\pr,N^\pr)}=S^{\pr(k^\pr,N^\pr)}(A (S,T) ),
~T^{\pr(k^\pr)}=T^{\pr(k^\pr)}(A(S,T))$.
We show that the reconstructed ADHM data coincides with the 
original ones $S^{(k,N)}, ~T_\mu^{(k)}$
($k^\pr=k, N^\pr=N,S^\pr =S,T_\mu^\pr =T_\mu$).

The solution $\psi$ of the Dirac equation (\ref{4dirac})
can be represented by the ADHM data $D$ and the descendents $V,f$ as
\begin{eqnarray}
\lab{psiv}
\wtps=\fr{1}{\pi}V^\da Cf=\fr{1}{\pi}v^\dagger f,
\end{eqnarray}
which is proved by $\cDb \psi=0 \Lra D_\mu\wtps e^\mu=0$ and 
\begin{eqnarray}
\lab{4zero}
\pi D_\mu\wtps e^\mu&=&D_\mu(V^\dagger Ce^\mu f)
=\left\{\del_\mu V^\dagger +(V^\dagger \del_\mu V)V^\dagger \right\}
Ce^\mu f+V^\dagger Ce^\mu \del_\mu f\nonumber\\
&=&\del_\mu V^\dagger(1-VV^\dagger)Ce^\mu f-V^\dagger Ce^\mu f \del_\mu
(\nabla^\dagger\nabla)f\nonumber\\
&=&(\del_\mu V^\dagger)\nabla f\nabla^\dagger Ce^\mu f-V^\dagger Ce^\mu f
(\eb_\mu C^\dagger\nabla +\nabla^\dagger Ce_\mu)f\nonumber\\
&\st{(\ref{jyuuyou})}{=}&-V^\dagger(C\unb{e_\mu f\nabla^\dagger Ce^\mu}_
{-2fC^\dagger \nabla }
+4CfC^\dagger \nabla -2CfC^\dagger \nabla)f=0.
\end{eqnarray}
There is an important relation between $\psi$ and $f$:
\begin{eqnarray}
\lab{green0}
\psi^\da\psi=-\fr{1}{4\pi^2}\del^2 f.
\end{eqnarray}
The proof is straightforward in similar way.
Eq. (\ref{box}) implies
\begin{eqnarray}
\label{asymp_f}
f&=&\Box^{-1}=\fr{1}{\vert x\vert^2}\left(1_{[k]}-\fr{2T_\mu x^\mu}{\vert
x\vert^2}+\fr
{ \tr(D^\dagger D)}{2\vert x\vert^2}
\right)^{-1}\\
&=&\fr{1_{[k]}}{\vert x\vert^2}+\fr{2T_\mu x^\mu}{\vert x\vert ^4}
-\fr{\tr (D^\da D)}{2\vert x\vert^4}+\fr{4(T_\mu x^\mu)^2}{\vert x\vert^6}
+\fr{2T_\mu x^\mu \tr (D^\da D)}{\vert x\vert^6}
+\fr{(\tr (D^\dagger D))^2}{4\vert x\vert^6}+\cdots\nonumber\\
\lab{leading}
\psi^\da\psi&=&\de^4(x)\cdot 1_{[k]}+\fr{\tr (S^\da S)}
{\pi^2\vert x\vert^6}-\fr
{9\tr (D^\dagger D)T_\mu
x^\mu}{4\pi^2\vert x\vert^8}-\fr{3(\tr (D^\dagger D))^2}{2\pi^2\vert
x\vert^8}
+\cdots,
\end{eqnarray}
which gives the proof of the normalization condition of $\psi$:
\begin{eqnarray}
\int d^4x~\psi^\da \psi&=&1_{[k]}.
\end{eqnarray}
Eq. (\ref{adhmt}) gives rise to new ADHM data as
\begin{eqnarray}
\lab{T1}
(T^{\pr\mu}=)~\int d^4x~\psi^\da x^\mu  \psi
&\st{(\ref{green0})}{=}&
-\fr{1}{4\pi^2}\int dS^\nu\unb{(x^\mu\del_\nu-\de_\nu^{~\mu})f}
_{\cO(r^{-4})\mbox{\scr{ part vanishes}}}\nonumber\\
&\st{(\ref{asymp_f})}{=}&
-\fr{1}{4\pi^2}\int dS^\nu~\left\{x^\mu\del_\nu\left(\fr{-2T_\rho x^\rho}
{\vert x\vert^4}\right)+\fr{2T_\rho x^\rho}{\vert x\vert^4}
\de_\nu^{~\mu}\right\}\nonumber\\
&=&-\fr{T_\rho}{2\pi^2}\int \Big{\{}
\unb{\left(\fr{x^\rho}{\vert x\vert}dS^\mu-\fr{x^\mu}{\vert x\vert}dS^\rho
\right)}_{=0}\fr{1}{\vert x\vert^3}
-\fr{x_\nu}{\vert x\vert^6}\unb{4x^\mu x^\rho dS^\nu}_{=\de^{\mu\rho}
\vert x\vert^2}
\Big{\}}\nonumber\\&=&T^\mu,
\end{eqnarray}
which just coincides with the original one!
In order to get new ADHM data $S$, let us examine the 
behavior of  $\wtps$ at infinity 
as Eq. (\ref{adhms}).
Substituting Eq. (\ref{vpr}) into it and 
using the asymptotic form of $V_x$: ``$V_x\ap g$''
and of $f$: (\ref{asymp_f}), we get
\begin{eqnarray}
\wtps=\fr{V^\da Dx^\da f}{\pi\vert x\vert^2}\ap
-\fr{g^\dagger Sx^\da}{\pi\vert x\vert^4},
\end{eqnarray}
which shows that the reconstructed ADHM data $S$
also coincides with the original one.
This result is consistent 
with the asymptotic behavior of $\psi^\da \psi$ (\ref{leading}).

\vs
\unl{\bf Uniqueness: (Instanton)$\longr$(ADHM)$\longr$(Instanton)}
\vs

The opposite discussion of (ADHM) $\longr$ (instanton) $\longr$ (ADHM)
is possible and we can show that
the new instanton just coincides with the original ones.
A key formula is 
\begin{eqnarray}
\lab{dirac3}
D_\mu V^\da&=&-\pi\wtps e_\mu\nabla^\da,
\end{eqnarray}
which shows very beautiful duality.
This is actually the composit of  Eqs. (\ref{dv}) and (\ref{psiv}).

In this way, we can show the one-to-one correspondence
between the instanton moduli space and the moduli space of ADHM data,
which makes the practical calculation on instantons
very easy to treat.

\vs
\noindent
{\bf Note}
\begin{itemize}
\item ADHM constructions for other gauge groups are 
discussed in \cite{GSW}.
\item ADHM constructions on the ALE spaces are discussed in \cite{BFMR, KrNa}
and their D-brane interpretations are presented in \cite{DoMo}.
\end{itemize}

\subsection{Nahm Construction of Monopoles on $\R^3$}

In this subsection,
we review the application of ADHM construction to 
monopoles (Nahm construction) \cite{Nahm}-\cite{Nahm6}. 
The proof of one-to-one correspondence between
monopole moduli space and the moduli space of Nahm data
is similar to ADHM construction.
Hence here we just set up the notation and
give a brief discussion pointing out the
similarities and the differences.

\vs
\unl{\bf (Monopole)}
\vs

(BPS) monopoles are defined the translational invariant instantons
which live on $\R^3$ whose coordinates are $x^1,x^2,x^3$.
For simplicity,
suppose that $G=SU(2)$ and the self-duality is ASD.

As in instanton case,
we have to define the 
``3-dimensional Dirac operator'' first:
\begin{itemize}
\item`` 3-dimensional Dirac operator'' 
\begin{eqnarray}
\cD_{\scr{\bf x}}(\xi)&:=&1_{[2]}\ot i(\xi-\Phi)+e^i\ot D_i,\nn
\cDb_{\scr{\bf x}}(\xi)&:=&1_{[2]}\ot -i(\xi-\Phi)+e^i\ot D_i=-\cD,
\end{eqnarray}
\end{itemize}
which can be interpreted to be obtained 
by replacing $\del_4, A_4$ with $i\xi, -i\Phi$
in the 4-dimensional Dirac operator in instanton case.

Let us present the conditions similar to instantons:
\begin{itemize}
\item Bogomol'nyi equation (``3-dim ASD equation'')
\begin{eqnarray}
B_i=-[D_i,\Phi],
\end{eqnarray}
where $B_i:=(i/2)\epsilon_{ijk}F^{jk}$ are magnetic fields.
This equation comes from the condition 
that $\cDb \cD$ commutes with matrices.
\end{itemize}

Bogomol'nyi equation represents the condition
that the energy functional of $(3+1)$-dimensional Yang-Mills-Higgs theory
should take the minimum:
\begin{eqnarray}
E&=&\frac{1}{4}\int d^3x
{\mbox{Tr}}\left[F_{ij}F^{ij}+2D_i\Phi D^i\Phi\right]\nn
&=&\frac{1}{2}
\int d^3x {\mbox{Tr}}(B_i\mp D_i\Phi)^2
\pm 2\pi a
\unb{\left[\fr{1}{2\pi a}\int d^3x
{\mbox{Tr}}\partial_i(B_i\Phi)\right]}
_{=:\nu[\Phi,A_i]}.
\end{eqnarray}
The second term in the RHS $\nu[\Phi,A_i]$
is just the definition of the monopole charge.
If the behavior of the Higgs field at infinity 
is as follows up to degree of gauge freedom, 
the magnetic charge $\nu[\Phi,A_i]$
becomes $-k$:
\begin{eqnarray}
\Phi\ap\left(\fr{a}{2}-\fr{k}{2r}\right)\sigma_3+\cO(r^{-2}).
\label{bc_higgs}
\end{eqnarray}
The vacuum expectation value of the Higgs is $a/2$. Then
\begin{itemize}
\item magnetic charge
\begin{eqnarray}
\nu[\Phi,A_i]&=&\fr{1}{2\pi a}\int_{S^2} dS_i~{\mbox{Tr}}\, (B_i\Phi)
=\fr{1}{2\pi}\int_{S^2} dS_i~B^{a=3}_i= -k.
\end{eqnarray}
\end{itemize}
We need the following condition:
\begin{itemize}
\item $\cDb \cD$ is invertible:
\begin{eqnarray}
\cDb \cD ^{\exists}G(\xi;{\bf x},{\bf y})=-\delta({\bf x}-{\bf y}).
\end{eqnarray}
\end{itemize}

The monopole moduli space is denoted by $\cM_{2,k}^{\mbox{mono}}$ 
and parameterized by finite number of parameters.
We summarize the $SU(2),~k$-monopole:

\vs
\fbox{{\bf Monopoles}}
\begin{eqnarray}
\cM_{2,k}^{\mbox{mono}}&=&
\fr{\left\{\ba{c|l}
(\Phi^{(2,k)},A_i^{(2,k)})&
\ba{l}
\mbox{Bogomol'nyi equation}
\\
A_\mu:=(-i\Phi,A_i)~:~N\ti N {\mbox{ anti-Hermitian matrices}}\\
\mbox{The b.c. of the Higgs field }(\ref{bc_higgs})\\
\cDb \cD {\mbox{ : invertible}}
\ea
\ea\right\}}{(A_\mu\sim g^{-1}A_\mu g+g^{-1}\del_\mu g,
~~~g({\bf x})\in SU(2))}\nonumber\\
\dim\cM_{2,k}^{\mbox{mono}}&=&4k-1
\end{eqnarray}

The dimension of the moduli space $\dim\cM_{2,k}^{\mbox{mono}}$
is calculated by the index theorem \cite{Weinberg, CoGo2,
Taubes}.
The degree contains that of center of mass of the monopoles.

\vs
\unl{\bf (Nahm)}
\vs

Next we define Nahm data. 

First we define the ``1-dimensional Dirac operator''
by using $k\ti k$ Hermitian matrices $T_i(\xi)$:
\begin{itemize} 
\item ``1-dimensional Dirac operator''
\begin{eqnarray}
\lab{ndata}
\nabla_\xi({\bf x})=i\fr{d}{d\xi}+e_i(x^i-T^i),~~~
\nabla_\xi({\bf x})^\dagger=i\fr{d}{d\xi}+\eb_i(x^i-T^i),
\end{eqnarray}
where $x^i$ denotes the coordinates of $\R^3$
and $\xi$ is an element of the interval $(-(a/2),a/2)$ 
for $G=SU(2)$.
The region of $\xi$ depends on the gauge group
and the way of the breaking. For example,
in $G=U(2)$ case, the region is a finite interval ($a_-,a_+$)
and in $G=U(1)$ case, it becomes semi-infinite.
\item Nahm equation (``1-dim ASD 
equation''$\Lra~\nabla^\dagger\nabla$ commutes with Pauli matrices):
\begin{eqnarray}
\lab{nahm}
\fr{dT_i}{d\xi}&=&i\ep_{ijl}T_j T_l 
\end{eqnarray}
\item The boundary condition of $T_i(\xi)$
\begin{eqnarray}
\label{bc_nahm}
T_i(\xi)&\st{\xi\rar\pm a/2}{\longr}&\fr{\tau_i}{\dis\xi\mp \fr{a}{2}}
+({\mbox{regular terms w.r.t. }}\xi)\\
{\mbox{where}}&&\tau_i~:~k{\mbox{-dimensional 
irreducible rep. of $\cS\cU(2)$}}
~~~[\tau_i,\tau_j]=i\ep_{ijl}\tau_l.\nonumber
\end{eqnarray}
\end{itemize}

The space of Nahm data up to gauge degree of freedom
is denoted by $\cM^{\mbox{\scr Nahm}}_{k,2}$
and called the moduli space of Nahm data,
which is summarized as follows:
\vs

\fbox{{\bf Nahm data}}
\begin{eqnarray}
\cM_{k,2}^{\mbox{\scr{Nahm}}}&=&
\fr{\left\{\ba{c|l}T_i^{(k,2)}&
\ba{l}\mbox{Nahm equation}
\\
T_i~:~k\ti k {\mbox{ Hermitian matrices}}\\
\mbox{The b.c. of Nahm data }(\ref{bc_nahm})\\
\na^\dagger\na {\mbox{ : invertible}}
\ea \ea\right\}}
{(T_i\sim R^{-1}T_i R,~~~R(\xi)\in U(k))}\nonumber\\
\dim\cM_{k,2}^{\mbox{\scr{Nahm}}}&=&4k-1.
\end{eqnarray}
The dimension is calculated directly from Nahm data \cite{Bowman2}.

There is a duality:
\begin{eqnarray}
\cM_{2,k}^{\mbox{mono}}\st{1:1}{=}\cM_{k,2}^{\mbox{\scr{Nahm}}},
\end{eqnarray}
which is proved as in ADHM construction
\cite{CoGo3, Hitchin2, Nahm4, Nakajima}.

\vs
\unl{\bf (Nahm)$\longr$(Monopole)}
\vs

We give the way to construct monopole solution $\Phi=\Phi(T),A_i=A_i(T)$
from given Nahm data $T_i^{(k)}$. 

First we solve the ``1-dimensional Dirac equation'':
\begin{eqnarray}
\lab{1dirac}
\nabla_{{\bf x}}(\xi)^\dagger v=i\left(
\ba{cc}
\dis\del_\xi+x^3-T^3&x^1-ix^2-T^1+iT^2\\
x^1+ix^2-T^1-iT^2&\dis\del_\xi-x^3+T^3
\ea
\right)
\left(
\ba{c}
v_1\\v_2
\ea
\right)
=0,
\end{eqnarray}
where $v$ is the $2k\ti 2$ matrix whose rows are
the independent normalized two solutions of (\ref{1dirac}):
\begin{eqnarray}
\lab{1norm}
\int d\xi ~v^\da v&=&1_{[2]}.
\end{eqnarray}
The completeness condition is also held:
\begin{eqnarray}
\lab{1comp}
 v(\xi)v(\xi^\pr)^\dagger&=&\de(\xi-\xi^\pr)-\nabla(\xi) f(\xi,\xi^\pr)
\st{\lar}{\nabla}(\xi^\pr)^\dagger.
\end{eqnarray}

We can construct the Higgs field $\Phi$
and gauge fields $A_i$ from the zero-mode $v$
as like instantons: 
\begin{eqnarray}
\label{mono}
\Phi=\int d\xi~v^\dagger\xi  v,~~~A_i=\int d\xi~v^\dagger \del_i v.
\end{eqnarray}
Here $A_i$ is a $2\ti 2$ matrix and  $A_i^\dagger=-A_i$
which implies $G=U(2)$.

We can show that the Higgs field and the gauge fields
is the $k$-monopole solution in similar way to ADHM
and the transformation for $v$: $v\rar vg,~g({\bf x})\in SU(2)$
preserves Eqs. (\ref{1dirac}) and (\ref{1norm}) and 
becomes the gauge transformation of $A_\mu$. 

\vs
\unl{\bf (Monopole)$\longr$(Nahm)}
\vs

We can construct Nahm data from given monopoles
as in ADHM case.
The steps are all similar to ADHM.
First we solve the massless 3-dimensional  Dirac equation
in the background of the given monopoles $\Phi, A_i$:
\begin{eqnarray}
\lab{3df2}
&&\cDb_{{\bf x}}(\xi)\psi(\xi,{\bf x})=0,\\
\lab{nor3f}
&&\int d^3x~\psi_\xi^\da \psi_\xi=1_{[k]}.
\end{eqnarray}
Then we can construct Nahm data  $T_i$
from the Dirac zero-mode $\psi_\xi$ ($2N\ti k$ matrix \cite{Callias})
\begin{eqnarray}
\lab{nahmt0}
\lab{nahmt}
T_i=\int d^3x~\psi_\xi^\da x_i\psi_\xi.
\end{eqnarray}
The data $T_i$ are actually $k\ti k$ Hermitian matrices.
We can show that these data satisfy Nahm equation.
The diagonal components of $T_i$ represent the
positions of $k$ monopoles.

Furthermore we can show the completeness and the 
uniqueness on Nahm construction,
which prove the one-to-one correspondence between the 
monopole moduli space and the moduli space of Nahm data.

\vs
\noindent
{\bf Note}
\begin{itemize}
\item General proofs of Nahm construction for other gauge groups
are summarized in \cite{HuMu}.
\item Explicit construction of spherically symmetric monopole
solutions for $G=SU(N)$ are presented in \cite{BCGPS}.

\end{itemize}

\end{document}